\documentclass[12pt,preprint]{aastex}

\pdfoutput=1
\usepackage{amsmath}
\usepackage{amssymb}
\usepackage{amsmath}
\usepackage{mathrsfs}
\usepackage{natbib}
\usepackage[pdftex,bookmarks,naturalnames]{hyperref}

\shorttitle{Poloidal Flux Transport}
\shortauthors{Beckwith et al.}

\begin{document}

\title{Transport of Large Scale Poloidal Flux in Black Hole Accretion}

\author{Kris Beckwith} 
\affil{Institute of Astronomy\\
University of Cambridge\\
Madingley Road\\
Cambridge CB3 0HA\\
United Kingdom}

\email{krcb@ast.cam.ac.uk}

\author{John F. Hawley}
\affil{Astronomy Department\\
University of Virginia\\ 
P.O. Box 400325\\
Charlottesville, VA 22904-4325}

\email{jh8h@virginia.edu}

\and
\author{Julian H. Krolik}
\affil{Department of Physics and Astronomy\\
Johns Hopkins University\\
Baltimore, MD 21218}

\email{jhk@pha.jhu.edu}

\begin{abstract} 

We report on a global, three-dimensional GRMHD simulation of an accretion
torus embedded in a large scale vertical magnetic field orbiting a
Schwarzschild black hole.  This simulation investigates how a large
scale vertical field evolves within a turbulent accretion disk and
whether global magnetic field configurations suitable for launching
jets and winds can develop.  We find that a ``coronal mechanism'' of
magnetic flux motion, which operates largely outside the disk body,
dominates global flux evolution.  In this mechanism, magnetic stresses
driven by orbital shear create large-scale half-loops of magnetic field
that stretch radially inward and then reconnect, leading to discontinuous
jumps in the location of magnetic flux.  In contrast, little or no flux
is brought in directly by accretion within the disk itself.  The coronal
mechanism establishes a dipole magnetic field in the evacuated funnel
around the orbital axis with a field intensity regulated by a combination
of the magnetic and gas pressures in the inner disk.  These results prompt
a reevaluation of previous descriptions of magnetic flux motion associated
with accretion.  Local pictures are undercut by the intrinsically global
character of magnetic flux.  Formulations in terms of an ``effective
viscosity" competing with an ``effective resistivity" are undermined by
the nonlinearity of of the magnetic dynamics and the fact that the same
turbulence driving mass motion (traditionally identified as ``viscosity'')
can alter magnetic topology.

\end{abstract}


\section{Introduction}\label{intro}

Astrophysical jets are seen in such a wide range of astrophysical systems
that their creation must not require particularly unusual conditions.
A tentative consensus has emerged that jets are a natural consequence of
accretion, rotation and magnetic fields \cite[for a summary of this position,
see][]{Livio:2000}.  Although the presence of accretion and
rotation are a given in an accreting system, the presence of magnetic
fields has long been considered to be less certain.  However, since
magnetohydrodynamic (MHD) turbulence driven by the magnetorotational
instability \cite[MRI; see][]{Balbus:1991,Balbus:1998} accounts for the
internal stresses that drive accretion, one is at least assured that the
presence of accretion itself implies the existence of a tangled magnetic
field of some reasonable strength.

But is that enough?  Most jet launching mechanisms that have been
developed depend on the existence of a large-scale, organized poloidal
field, following in general terms the scenarios outlined analytically in
the influential papers of \cite{Blandford:1977} and \cite{Blandford:1982}.
In the Blandford-Payne model, a large-scale poloidal magnetic field is
anchored in and rotates with the disk.  If the fieldlines are angled
outward sufficiently with respect to the disk, there can be a net outward
force on the matter.  As matter is accelerated along the rotating
fieldlines, its angular momentum increases still further, 
accelerating and driving an outflow.  In this model the power
for the resulting jet comes from the rotation of the accretion disk.
In the Blandford-Znajek model, the jet is powered by the rotating
space-time of a black hole.  (In the case of jets from systems with
stars rather than black holes, stellar rotation can play a similar
role.)  Radial magnetic field lines lie along the hole's rotation
axis and are anchored in the event horizon.  Fieldline rotation is
created by frame-dragging and this drives an out-going Poynting flux.
Most simulations that have demonstrated jet-launching have {\it assumed}
the existence of a large-scale poloidal (mainly vertical) field in the
initial conditions \cite[see, e.g., the review by][]{Pudritz:2007}.
Although such simulations have provided considerable evidence that
magnetic fields can be effective in powering jets, they cannot account
for the origin of those fields.

If, as it seems, some sort of large-scale poloidal field is required
for jet production, how can such a field be established in the near
hole region?   One possibility is an accretion disk dynamo, presumably
working with the turbulent fields generated naturally by the MRI.
A dynamo is an attractive possibility because it has the potential to
be ubiquitous.  A number of models have been put forward suggesting how
some form of inverse cascade might occur within the MHD turbulence to
generate a large scale field \cite[e.g.,][]{Tout:1996,Uzdensky:2008},
but these scenarios remain speculative and nothing definitive has been
seen in global disk simulations done to date.  In any case, the strict
conservation of flux within a volume means that a disk dynamo could
produce net flux only through interactions with boundaries, in this case
either the black hole, or by expelling flux to large radius.

The global advection of net flux is an important process, therefore,
whether the disk functions as a dynamo or net poloidal field is simply
carried in to the near hole region from large radius.  Even without
any dynamo action, a weak net field might become locally strong it
were concentrated by the accretion process.  In the absence of a
single field polarity at large distance, a near-hole net field could
nevertheless be built up due to a random walk process as field is
accreted \citep[e.g.,][]{Thorne:1986}.  Whether or not net field can be
so accreted, however, is a matter of some uncertainty. The main concern
is that the field would not be accreted if, as seems intuitively likely,
the field intensity declines outward and the rate of diffusion of the
field through the matter exceeds the rate at which matter accretes
\citep{van-Ballegooijen:1989,Lubow:1994,Heyvaerts:1996,Ogilvie:2001}.
This competition between inward advection and outward diffusion is
typically described in terms of an effective turbulent accretion
viscosity $\nu_t$ determining the accretion rate, and an effective
magnetic diffusivity, $\eta_t$ that sets the field diffusion rate.
The fate of the field is then determined by the effective magnetic
Prandtl number $\mathcal{P}_m = \nu_t/\eta_t$.
In this picture, it is argued that the relevant scale for effective
resistivity is $\sim H$ whereas the relevant scale for effective
viscosity is $\sim R$; consequently, field can be brought inward only
if ${\cal P}_m > R/H$.

It is far from clear, however, that field transport within
MHD turbulence can be described in such terms.  The picture just
described assumes that the fluid motion is independent of the field,
yet in Nature the dynamical coupling between fluid and field assures
that even weak fields are amplified, and that the effective viscosity
$\nu_t$ is, in fact, the result of magnetic stresses.  Thus, there
is no kinematic regime in which the Lorentz forces
can be neglected.  The resulting MRI-driven turbulence is inherently
three-dimensional; indeed, three-dimensional studies are essential to
generating long-term sustained MHD turbulence due to the fundamental
restriction of the anti-dynamo theorem. By contrast, the framework of
effective turbulent viscosity and diffusivity is cast in axisymmetry,
in the hope that suitable time- and space-averaging of the field
transport process in a fully turbulent, three-dimensional disk can lead
to some value of $\eta_t$ in much the same way that suitable averaging of
correlations in the magnetic field, normalized to some suitable pressure
term (either gas, magnetic or total pressure) leads to a value of $\nu_t$.
But the primary mechanism by which magnetic fields change the fluid
elements to which they are attached is through magnetic reconnection
driven by the fluid turbulence, not gradual slippage via resistivity.
The effective viscosity and magnetic diffusivity are derived from making
the {\it ansatz} that stress and reconnection can be {\it modeled} as a
viscosity or resistivity, i.e., a constant coefficient multiplying the
gradient in orbital rotation rate or magnetic field strength.  Whether
or not such a model is appropriate to accretion flows remains to be
determined.  Finally, net flux and field topology are global concepts;
a purely local description of field motion in terms of transport 
coefficients may not be sufficient.

Numerical simulations offer a promising approach to investigate these
and other issues related to global field evolution within accretion flows.
Of course, simulations entail their own set of difficulties:
global simulations require adequately fine resolution over a wide range
of radii in the disk body and throughout a good deal of the surrounding
coronal region as well.  The numerical grid boundaries should be
sufficiently remote so as to minimize artificial effects on the disk.
Although axisymmetric simulations can be of interest, the problem
ultimately must be done in three dimensions because any axisymmetric
calculation fails to describe properly essential properties of both the
fluid turbulence and the magnetic field evolution.  Finally, because
we need to follow the disk long enough to observe net accretion in an
approximate inflow equilibrium, the duration of the simulation must
be long compared to typical local dynamical times in the disk.

Work along these lines is ongoing.  Global, three dimensional black
hole accretion simulations without initial fields threading the disk
were carried out and described in a series of papers beginning with
\cite{De-Villiers:2003b}.  In these simulations initial weak dipole
loops are contained within an isolated gas torus.  The subsequent
MRI-driven accretion flow carries the inner half of the initial field
loop (with a single sign for the net vertical field component) down
to the black hole, creating radial field as the loop is stretched.
Differential rotation along that radial field leads to rapid growth of
the toroidal field, especially within the plunging region of the flow.
This growing toroidal field causes the ejection of a magnetic tower
into the low density funnel region and the establishment of a global
dipole field anchored in the black hole.  This field can then produce
a Poynting flux jet if the black hole is rotating.

\cite{Beckwith:2008a} examined several alternative initial field
configurations in comparison to the dipolar loop.  Quadrupolar initial
field loops produce a much weaker funnel field.  A similar result
was found by \cite{McKinney:2004} for axisymmetric simulations, and
\cite{McKinney:2009} in a three-dimensional model.  A variation proposed
by \cite{Hawley:2006} and further elaborated upon by \cite{Beckwith:2008a}
is a series of dipolar loops contained within the accretion flow.   Such
loops can lead to the creation of a {\it temporary} net sense of vertical
field that can support a significant field in a relativistic outflow along
the rotation axis for a sizable period of time.  \cite{Beckwith:2008a}
also computed a model that began with a purely toroidal field.  In this
case the MRI creates turbulence, field and angular momentum transport,
but without creating any overall organization to the poloidal field.
As a consequence, no funnel field is generated.

In this paper we continue our study of the influence of field topology
on accretion and potential jet formation through a three-dimensional
general relativistic MHD simulation with an initial field configuration
consisting of a net vertical field passing through an orbiting ring of
gas at modestly large radius.  We will investigate the evolution of
the disk in the presence of a net field while determining if such an
initial condition can lead to the establishment of field configurations
conducive to jet formation.  From this study we hope to gain more insight
into whether or not large scale fields can be advected along with the
accreting gas and to lay the groundwork for further work.

The rest of this paper is structured as follows:  In \S\ref{numerics},
we describe the methodology and the initial conditions for our
simulations. In \S\ref{flux_evolve} we discuss their results as they
pertain to magnetic flux motion.  Finally, in \S\ref{conclusion},
we summarize our results, compare them with current models from
the literature,  and explain their implications for observations of
relativistic jets in Nature.

\section{Numerical Details}\label{numerics}

For this work we employ the general relativistic MHD code 
GRMHD developed in
\cite{De-Villiers:2003,De-Villiers:2006}. ÊGRMHD has already 
been used in several studies to simulate black hole accretion in three 
spatial dimensions \citep{De-Villiers:2003b,Hirose:2004, 
De-Villiers:2005,Krolik:2005,Hawley:2006,Beckwith:2008a,Beckwith:2008b}. 
GRMHD solves the equations of ideal non-radiative MHD in the static Kerr 
metric of a rotating black hole using Boyer-Lindquist coordinates. Ê
Values are expressed in gravitational units $(G = M = c = 1)$ with 
line element $ds^{2} = g_{tt} dt^{2} + 2g_{t \phi} dt d\phi + g_{rr} 
dr^{2} + g_{\theta \theta} d \theta^{2} + g_{\phi \phi} d\phi^{2}$ and 
signature $(-,+,+,+)$. The relativistic fluid at each grid zone is 
described by its density $\rho$, specific internal energy $\epsilon$, 
$4$-velocity $U^{\mu}$, and isotropic pressure $P$. The relativistic 
enthalpy is $h = 1 + \epsilon + P / \rho$. The pressure is related to 
$\rho$ and $\epsilon$ via the equation of state for an ideal gas, $P = 
\rho \epsilon ( \Gamma - 1)$. The magnetic field is described by two 
sets of variables. ÊThe first is the constrained transport
\citep[CT;][]{Evans:1988} magnetic 
field $\mathcal{B}^{i} = [ijk] F_{jk}$, where $[ijk]$ is the completely
anti-symmetric symbol, and $F_{jk}$ are the spatial components of the
electromagnetic field strength tensor. From the 
CT magnetic field components, we derive the magnetic field
four-vector, $(4\pi)^{1/2}b^{\mu} = *F^{\mu \nu} U_{\nu}$, and the
magnetic field scalar, $||b^{2}|| = b^{\mu} b_{\mu}$. The
electromagnetic component of the stress-energy tensor is $T^{\mu
\nu}_{\mathrm{(EM)}} = \frac{1}{2}g^{\mu \nu} ||b||^{2} + U^{\mu}
U^{\nu}||b||^{2} - b^{\mu} b^{\nu}$.

\subsection{Initial and Boundary Conditions}

In this study we carry out both axisymmetric
and fully three dimensional simulations around a
Schwarzschild black hole. ÊThe initial condition, as in previous works
\citep{De-Villiers:2003b,Hirose:2004,De-Villiers:2005,Krolik:2005,Hawley:2006,Beckwith:2008a,Beckwith:2008b},
consists of an isolated gas torus upon which a weak magnetic field is
imposed. ÊWe use the torus model described by \cite{De-Villiers:2003b}
with an adiabatic index of $\Gamma = 5/3$. ÊThe initial angular momentum
distribution is slightly sub-Keplerian, with a specific angular momentum
at the inner edge of the torus (located at $r=30$M) $\ell_{\rm in}
\equiv U_\phi/U_t = 6.10$. ÊFor this choice of $\ell_{\rm in}$, the
pressure maximum is at $r \approx 40M$ and the outer edge of the torus is
located at $r \approx 60M$. ÊThe angular period at the pressure maximum is
$\Omega^{-1} \approx 250M$. ÊWith this inner edge radius, the resulting
accretion flow should have sufficient radial range to permit a detailed
study of the advection of vertical flux. ÊThe choice of angular momentum
parameter at the inner edge of the torus, $\ell_{\rm in}$, was made so
that the initial torus (and hence the steady state accretion flow) would
have a scale height similar to that of previous simulations \cite[see
e.g.][]{Beckwith:2008b}. ÊThe torus is initially surrounded by zones
filled with zero-velocity gas set to a constant value of density and
pressure. ÊThe ratios of these ``vacuum'' values to the torus maximum are
$2\times 10^{-4}$ for the pressure, and $5\times 10^{-6}$ for the density.
Because the code is relativistic, the timestep is always limited by the
time for light to cross the smallest cell; high Alfv\'en speeds in the
low-density coronal region do not create any unique difficulty.

The initial magnetic field is homogeneous and vertical,
filling the annular cylinder $35M \le R \le 55M$ (where $R$
is the cylindrical radius). This configuration differs from
the vertical field used in some previous studies \cite[see
e.g.][]{Koide:1999,McKinney:2004,De-Villiers:2006} which fills all space,
not just through the initial torus. By isolating the field in this
manner, we know that any vertical flux that
ends up crossing the black hole event horizon was brought in from field
initially passing through the torus at large radius, rather than already
being present at the horizon in the initial conditions. ÊAlso, because
equilibrium field strengths scale strongly with radius, a global vertical
field that is strong enough to be interesting at one radius is likely
to be overwhelming at some radii while negligible at others. We prefer
to see the global field strength emerge as an outcome of the evolution.
In our configuration the initial vertical field is subthermal within the
torus and is unstable to the MRI. ÊIt should be noted, though, that this
initial field configuration is not in dynamic equilibrium outside of
the torus. ÊAlso, this is but one very specific realization of a torus
embedded in large scale field. ÊFuture experiments will need to explore
a wider variety of initial configurations.

\begin{figure}
\leavevmode
\begin{center}
\includegraphics[width=0.49\textwidth]{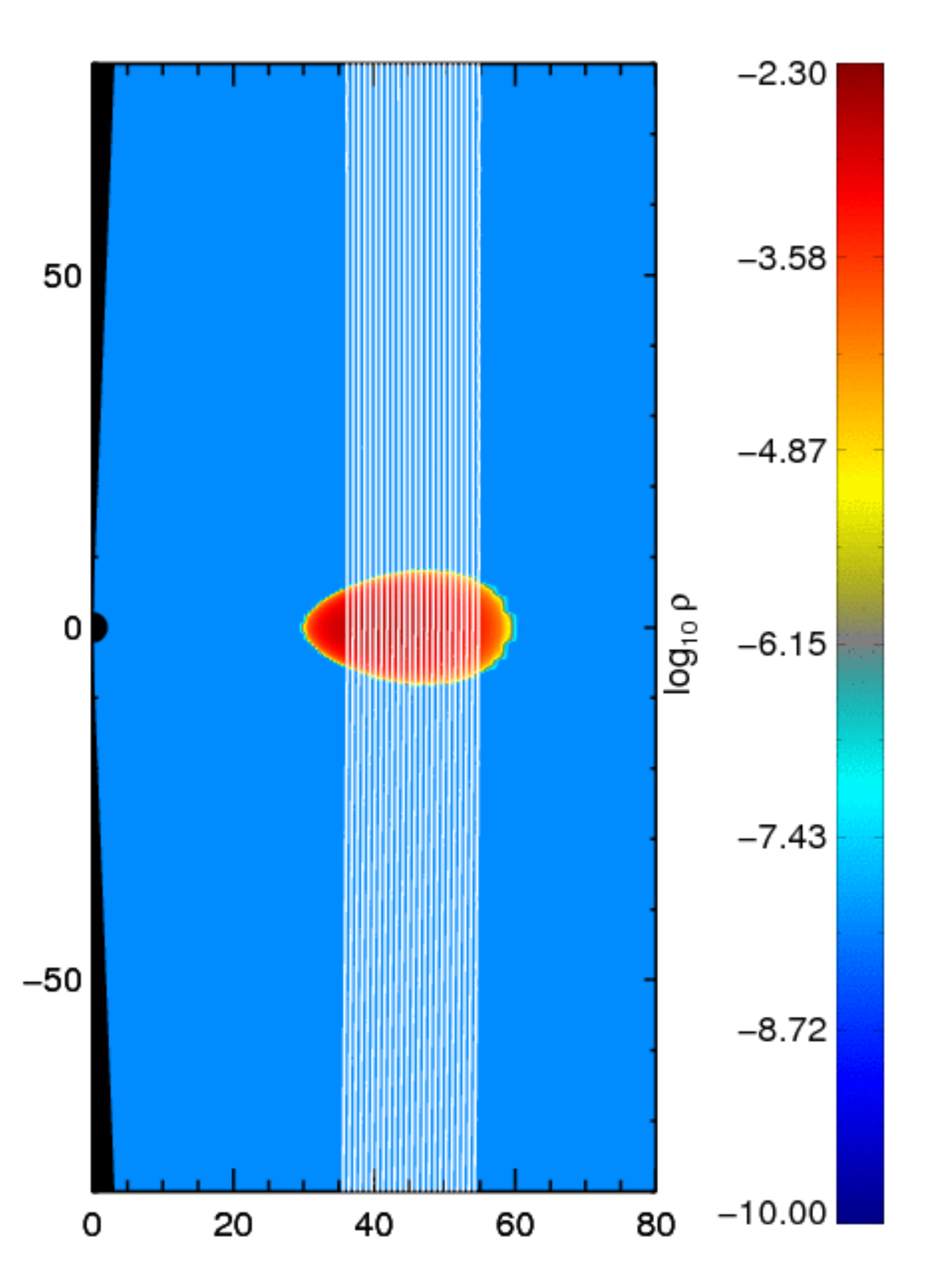}
\includegraphics[width=0.49\textwidth]{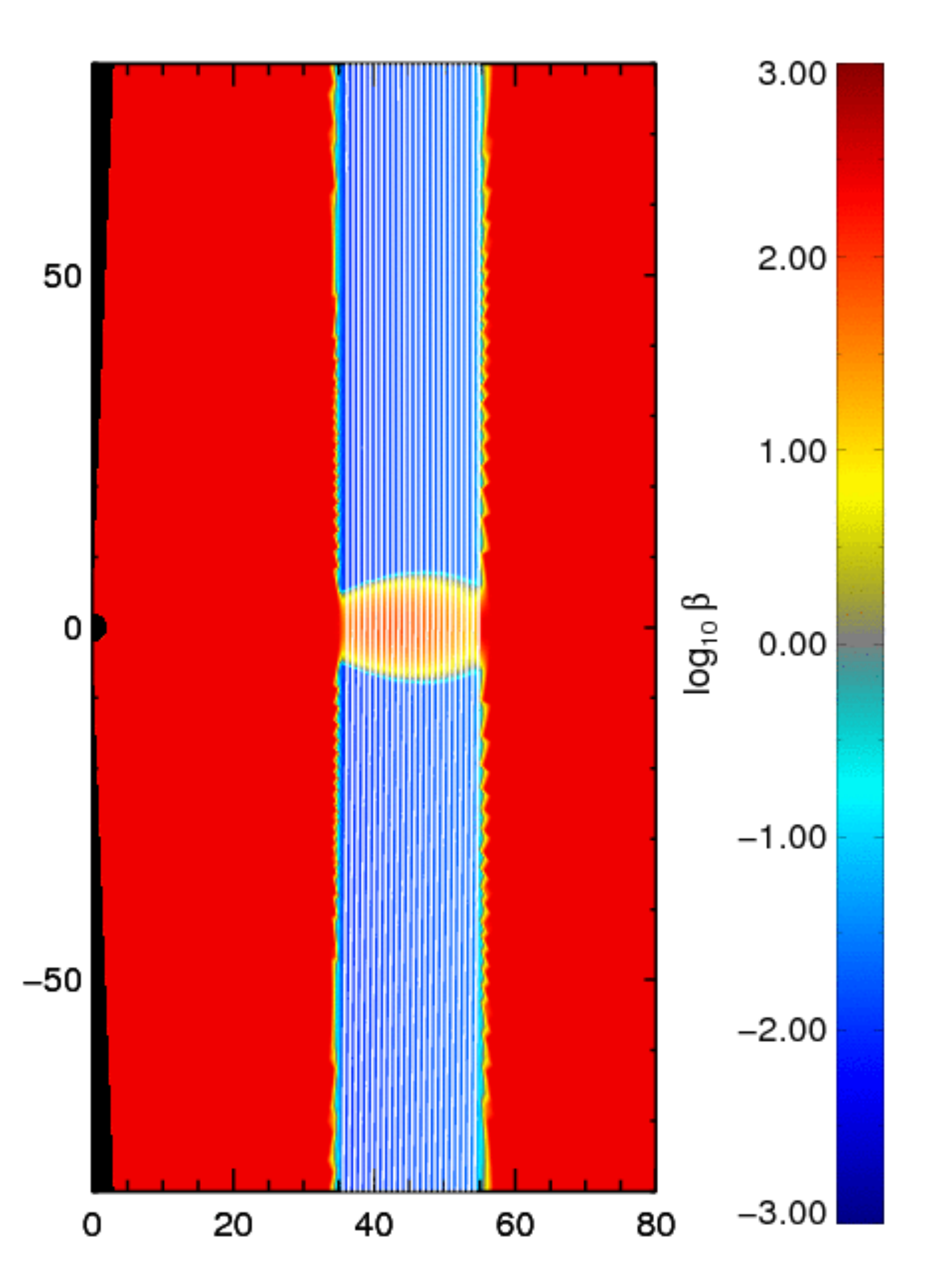}
\end{center}
\caption[]{Initial torus and field configuration.  White contours denote
magnetic field lines, color contours density distribution (left panel) and
the gas $\beta$ parameter (right panel).}
\label{initialstate}
\end{figure}

This initial field is computed from the curl of the four-vector potential,
i.e.,
\begin{equation}\label{eqn:curlA}
F_{\alpha \beta}
= \partial_\alpha A_\beta - \partial_\beta A_\alpha
= A_{\beta, \alpha} - A_{\alpha, \beta} .
\end{equation}
with only $A_\phi \neq 0$. The vector
potential corresponding to a uniform (Newtonian) vertical field geometry
is proportional to the cylindrical radius, i.e.,
\begin{equation}
A_{\phi} \propto r \sin \theta .
\end{equation}
The vector potential for the initial field is
\begin{equation}
A_{\phi} = A_0 \left\{ \begin{array}{lll}
R_{in} & \quad \mbox{$R < R_{in}$} \\
r \sin \theta & \mbox{$R_{in} \le R \le R_{out}$} \\ R_{out} & \quad
\mbox{$R > R_{out}$} \end{array} \right. 
\end{equation} 
where $R_{in}$ and $R_{out}$ are $35M$ and $55M$. ÊOutside of those
cylindrical radii the field is zero. $A_{0}$ was specified so that
the ratio of the average thermal to 
average magnetic pressure
inside the torus is $ \beta=100$. At the center of the torus $\beta =
275$ and $\beta$ declines smoothly toward 1 near the edge of the
torus.  In the surrounding corona,
$\beta=0.038$. The initial mass, field and $\beta$
distribution is shown in Figure \ref{initialstate}.

The simulation we will focus on is three-dimensional with $256 \times
256 \times 64$ zones in $(r,\theta,\phi)$. ÊThe simulation is labeled
``VD0m3d.'' ÊGiven the challenge of adequately resolving such models
and of doing anything like a convergence study, we also carried
out two-dimensional simulations at two different resolutions. ÊIn two
dimensions we used $256 \times 256$ and $512 \times 512$ zones in
$(r,\theta)$. ÊThe two-dimensional simulations are labeled ``VD0m2d''
and ``VD0h2d'' where the ``m" designates medium resolution and the ``h''
designates the higher resolution case.

For all models the radial grid extends from an inner boundary at
$r_{in}=2.104$ to an outer boundary located at $r_{out} = 500 M$. The
radial grid was graded using a logarithmic distribution that concentrates
zones near the inner boundary. An outflow condition is applied at both
the inner and outer radial boundaries. ÊIn contrast to the zero net flux
simulations that we have performed previously, magnetic fields pierce
the outer boundary in the initial state. This net flux is kept fixed
throughout the simulation, but otherwise the field at the boundary can
evolve. ÊWhen the fluid velocity normal to the boundary is directed
outward we perform a simple copy of fluid variables into the
ghost zones that lie outside of the boundary.
In the case where the fluid velocity
is inward, the ghost zones are filled with a cold, low density, zero
momentum fluid at the vacuum value so that only cold, very low density
gas can enter the computational grid. To minimize the influence of the
outer boundary it is located as far away from the region of interest as
is feasible. Any torques that arise from the interaction of the outer
boundary with the large scale vertical field should have minimal influence
on the turbulent disk itself.

The $\theta$-grid spans the range $0.01 \pi \le \theta \le 0.99 \pi$.
This creates a conical cutout that surrounds the coordinate axis.
Such a cutout will prevent flow through the coordinate axis, but it is
computationally advantageous to avoid the coordinate singularity.
The size of this $\theta$-cutout is determined by two considerations.
First, light travel times across narrow $\phi$-zones near
the axis sets the timestep, so the cutout should not be too small.  In 
this simulation, $\Delta t = 2.53
\times 10^{-3}M$. The $\theta$ cutout must not be too large;
initial vertical field should not cross the $\theta$-cutout before
reaching the outer radial boundary at $r=500M$. A reflecting boundary
condition is used along the $\theta$-cutout. The $\theta$ zones were
distributed logarithmically so as to concentrate zones around the equator.

The $\phi$-grid spans the quarter plane, $0 \le \phi \le \pi / 2$.
The $\phi$ grid is uniform and periodic boundary conditions are
applied. The use of this restricted angular domain significantly reduces
the computational requirements of the simulation.
\cite{Schnittman:2006} examined the variance in surface density as a
function of $\phi$ and found the characteristic size of perturbations
to be around $25^\circ$.  Thus, while some global features may be
lost by using a restricted domain, the character of the local 
MRI turbulence is captured.

\subsection{Magnetic Flux}

In analyzing the evolution of the large scale poloidal flux we will
make use of the poloidal flux function, which 
corresponds to the $\phi$-component of the magnetic vector potential 
\citep{van-Ballegooijen:1989,Heyvaerts:1996,Reynolds:2006}, which we
must calculate from the simulation data. In GRMHD,
we directly evolve the components of the Faraday tensor, $F_{\mu \nu}$
using the alternative form of the induction equation \citep[for further details
see][]{De-Villiers:2003}:
\begin{equation}
\partial_\delta F_{\alpha \beta} + \partial_\alpha F_{\beta \delta}
+ \partial_\beta F_{\delta \alpha} = 0
\end{equation}
The space-space components of $F_{\mu \nu}$ are identified with the 
constrained-transport (CT) magnetic fields via $\mathcal{B}^{i} = [ijk] F_{jk}$:
\begin{equation}
{\cal B}^r = F_{\phi \theta}; \;\;
{\cal B}^\theta = F_{r \phi}; \;\;
{\cal B}^\phi = F_{\theta r}.
\end{equation}
This identification allows us to write the induction equation in the familiar form
\begin{equation}
\partial_t {\cal B}^{i} - \partial_j \left( V^{i} {\cal B}^j - {\cal
B}^i V^{j} \right) = 0 ,
\end{equation}
where $V^i = U^i/U^t$.
The tensor $F_{\mu,\nu}$ is related to the
magnetic induction in the fluid frame, $B^\alpha$ by the relation
\begin{equation}
\label{fmunudef}
F_{\mu\,\nu} = \epsilon_{\alpha\,\beta\,\mu\,\nu}\,B^{\alpha}\,U^\beta ,
\end{equation}
where
$\epsilon_{\mu\,\nu\,\delta\,\gamma}=\sqrt{-g}\,[\mu\,\nu\,\delta\,\gamma]$.

The Faraday tensor can also be written in terms of a vector potential:
\begin{equation}
F_{\alpha \beta}
= \partial_\alpha A_\beta - \partial_\beta A_\alpha
= A_{\beta, \alpha} - A_{\alpha, \beta} .
\end{equation}
It is convenient to define an azimuthally-averaged
poloidal flux function, $A_\phi$.  This can be derived from the azimuthally-averaged
CT field data by noting that the total (spatial) derivative of $A_\phi$ is
\begin{equation}
dA_\phi = \frac{\partial A_\phi}{\partial r} dr
+ \frac{\partial A_\phi}{\partial \theta} d\theta .
\end{equation}
>From the definition of the Faraday tensor, we have that
\begin{equation}
F_{\phi \theta} = A_{\theta,\phi} - A_{\phi,\theta} ; \;\;
F_{r \phi} = A_{\phi,r} - A_{r,\phi} .
\end{equation}
After azimuthal averaging, this reduces to
\begin{equation}
F_{\phi \theta} = - A_{\phi,\theta} ; \;\;
F_{r \phi} = A_{\phi,r} .
\end{equation}
The space-space components of $F_{\mu \nu}$ are identified with the 
CT magnetic fields via $\mathcal{B}^{i} = [ijk] F_{jk}$ so that
\begin{equation}
{\cal B}^r = F_{\phi \theta} = -A_{\phi,\theta} ; \;\;
{\cal B}^\theta = F_{r \phi} = A_{\phi,r} .
\end{equation}
We therefore have that
\begin{equation}
dA_\phi = F_{r \phi} dr
- F_{\phi \theta} d\theta
= {\cal B}^\theta dr - {\cal B}^r d\theta .
\end{equation}

To find the flux from this differential, we could in principle integrate
over any poloidal surface. For
the radial motion of vertical flux through the disk, 
the surface of greatest significance
is the equatorial plane. Ê
Because this surface is interrupted by the event horizon, we must
stretch it over half of the event horizon. ÊWe
will therefore define a flux function that is the sum of two parts:
the radial flux through the inner horizon boundary,
\begin{equation}\label{eqn:psidefn}
\Psi(\theta)|_{r=r_{in}} = \int^{\theta}_{0} \, d\theta^\prime \,
{\cal B}^r (r_{in},\theta^\prime),
\end{equation}
and the vertical flux through the equator,
\begin{equation}\label{eqn:phidefn}
\Phi(r)|_{\theta=\pi/2} = \int^r_{r_{in}} \, dr^\prime \,
{\cal B}^\theta (r^\prime,\theta=\pi/2).
\end{equation}
Using these functions, we can determine how the accumulated flux
through the top half of the event horizon depends on polar angle,
$\Psi(\theta)$, and how much total vertical flux has been brought
to within radius $r$. The full flux function along the total path is
\begin{equation}\label{eqn:fluxdefn} 
\mathcal{A} (r,\theta=\pi/2) = 
\Psi(\pi/2) - \Phi(r). 
\end{equation} 
This corresponds to the net flux piercing the surface covering the
top hemisphere of the black hole horizon and the equatorial plane
out to radius $r$. The full vector potential component $A_\phi
(r,\theta)$ is obtained by integrating the CT variables throughout
the computational domain starting from $r=r_{in}$ and $\theta
= 0$. $\mathcal{A} (r,\theta=\pi/2)$ is therefore identical to
$A_\phi (r,\theta=\pi/2)$.  For the three-dimensional simulation,
VD0m3d, ${\cal B}^r$ and ${\cal B}^\theta$
are integrated over over the $\phi$-dimension prior to computing
$A_\phi (r,\theta)$.  Given the periodic  nature of the
$\phi$ direction this is sufficient to compute the total flux piercing
the volume bounded by the equatorial plane and the black hole horizon.
We note that this procedure is
consistent with the approach adopted in the calculations of
\cite{van-Ballegooijen:1989},
\cite{Heyvaerts:1996}, and \cite{Reynolds:2006}.

\section{Global Flux Transport}
\label{flux_evolve}

\subsection{Overall evolution}
\label{overall_evolve}

\begin{figure}
\leavevmode
\begin{center}
\includegraphics[width=0.24\textwidth, viewport=40 30 360 620,clip] {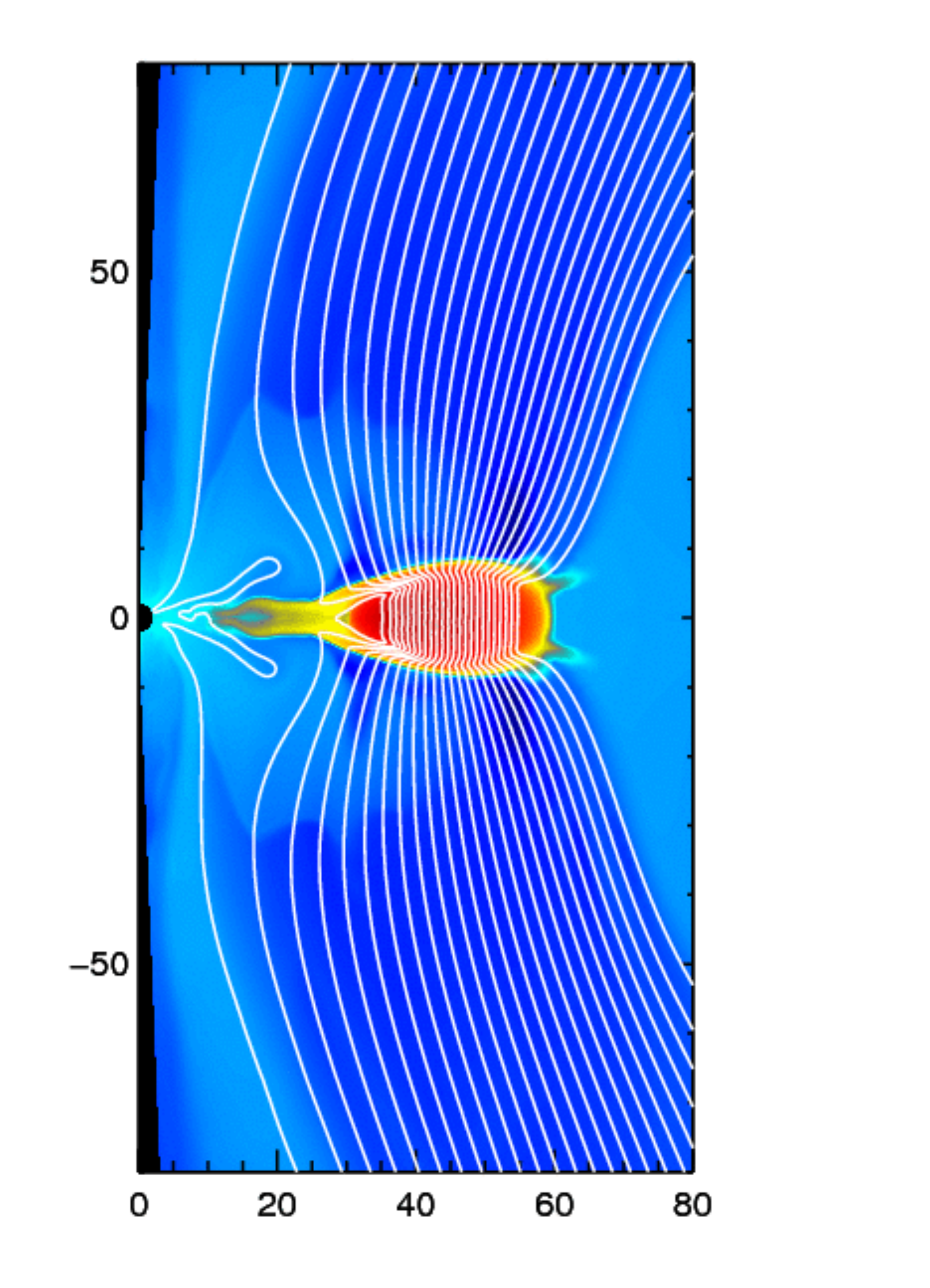}
\includegraphics[width=0.24\textwidth, viewport=40 30 360 620,clip] {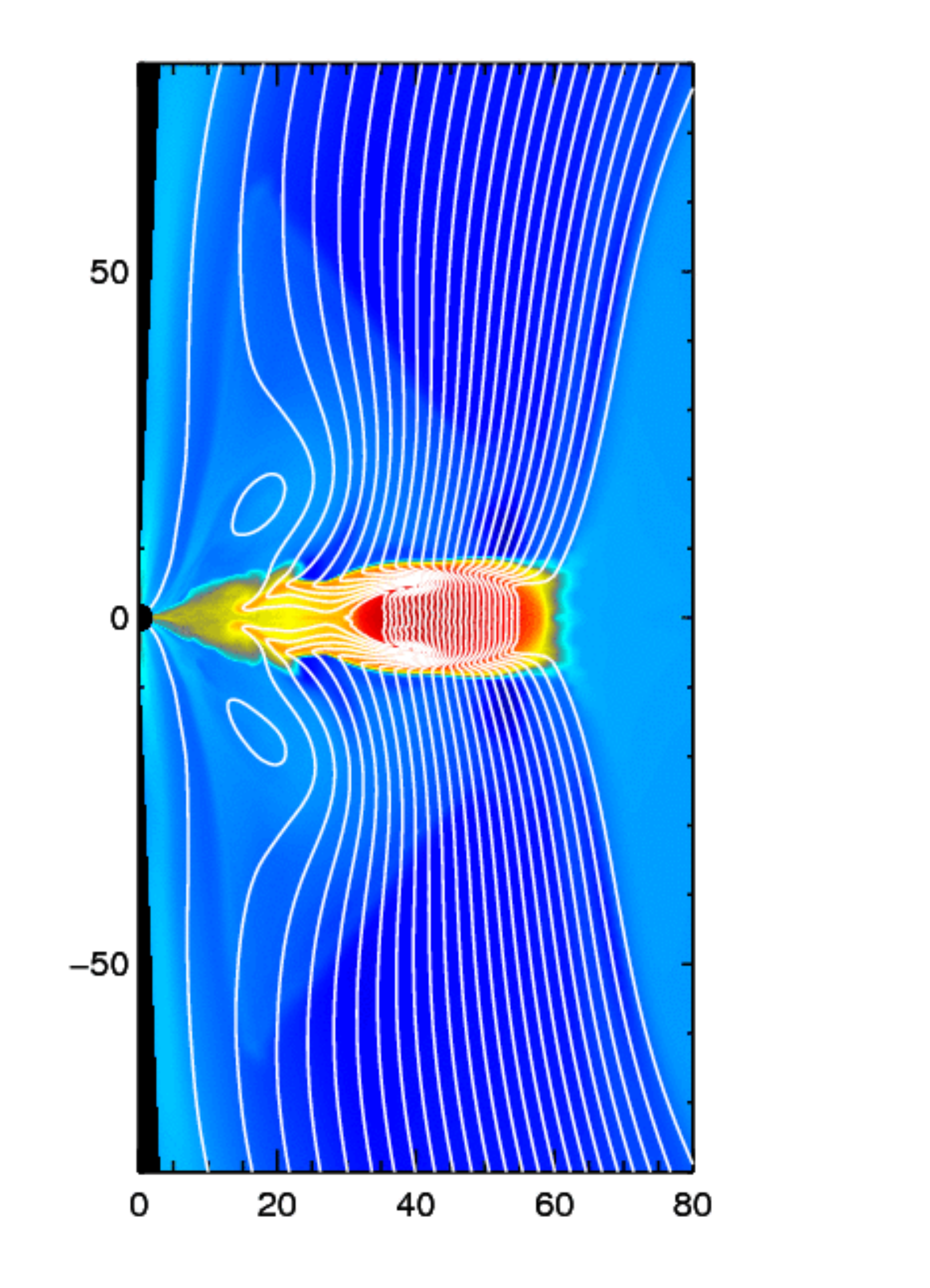}
\includegraphics[width=0.24\textwidth, viewport=40 30 360 620,clip] {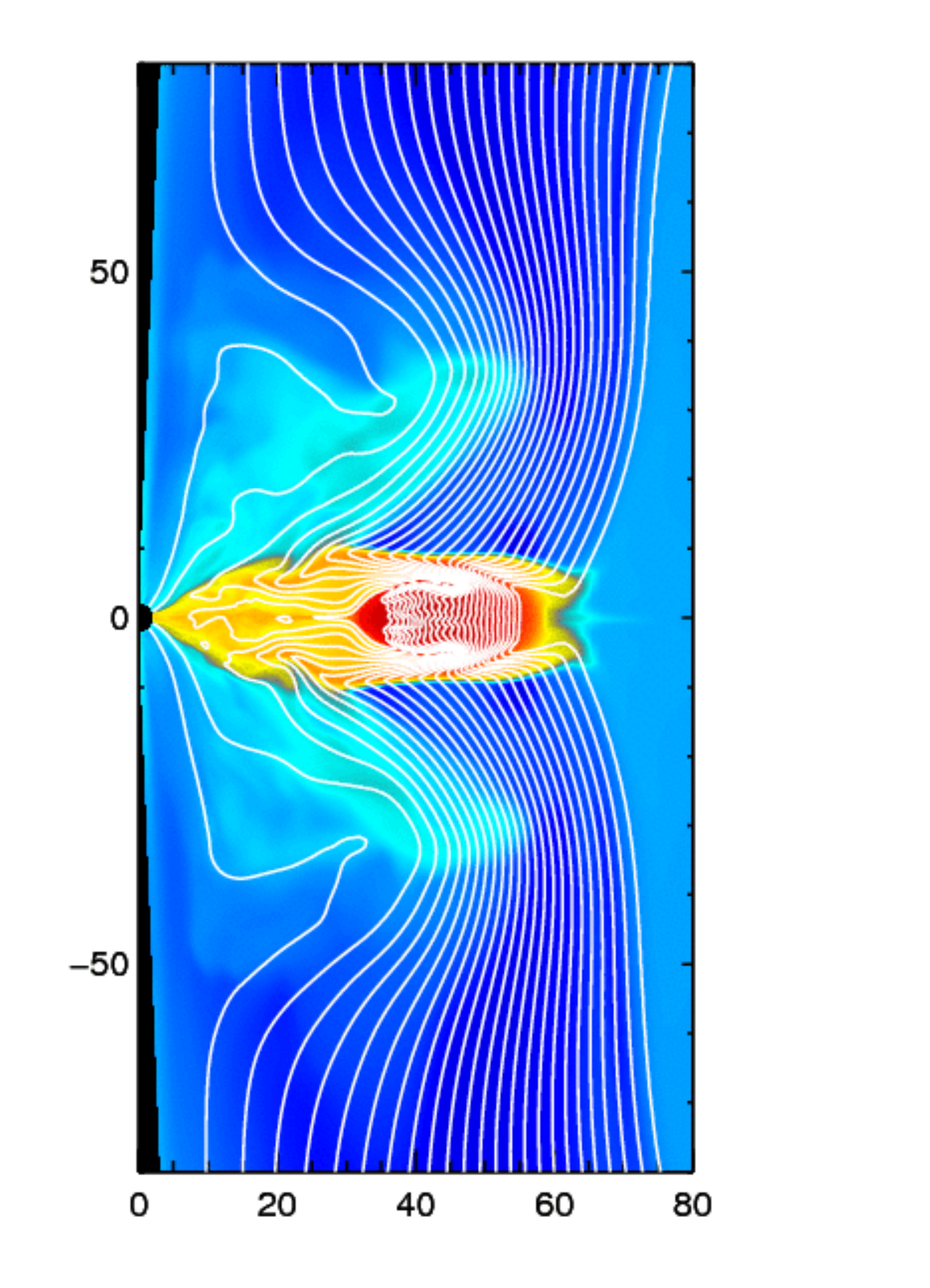}
\includegraphics[width=0.24\textwidth, viewport=40 30 360 620,clip] {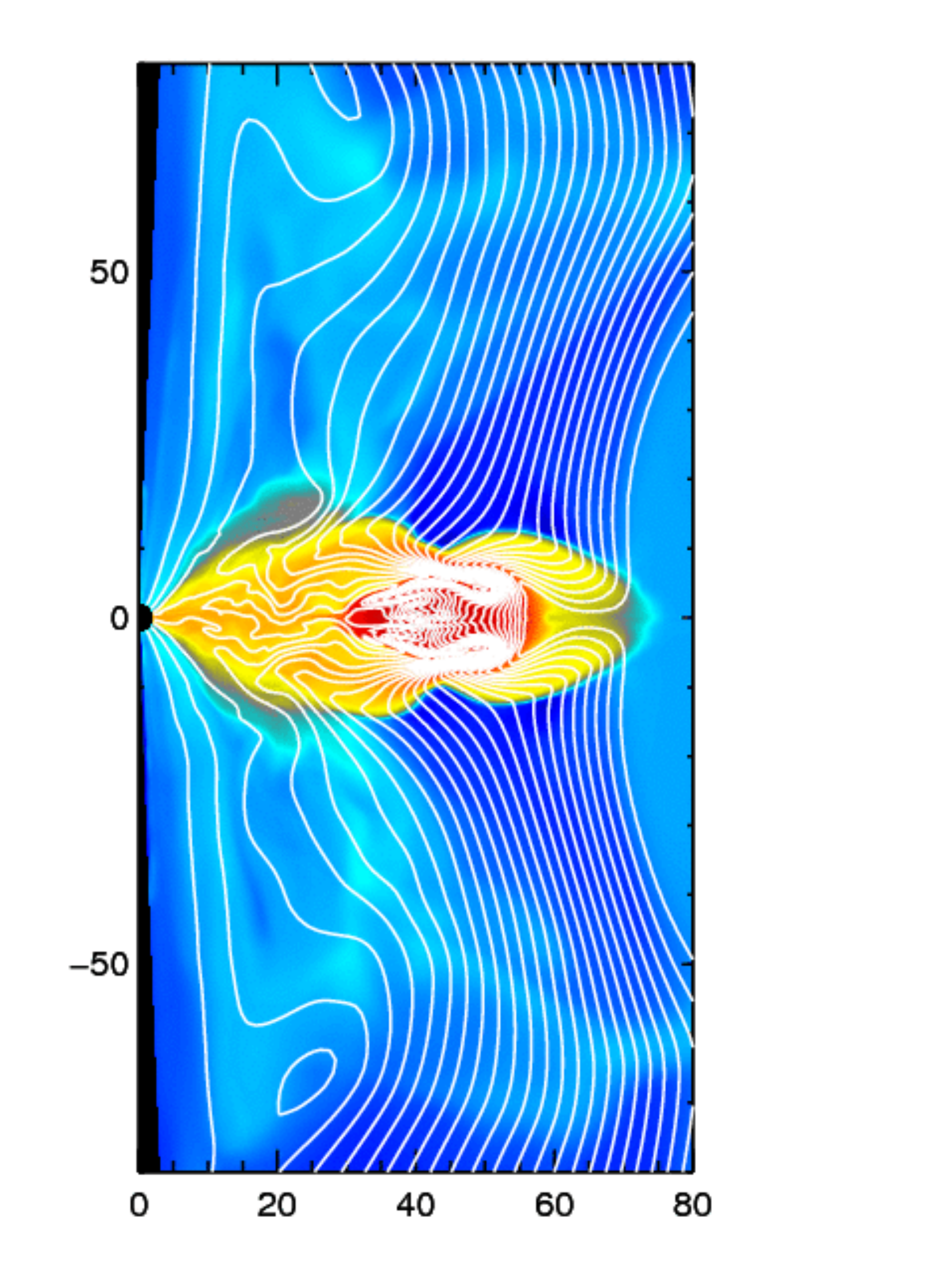}
\includegraphics[width=0.6\textwidth]{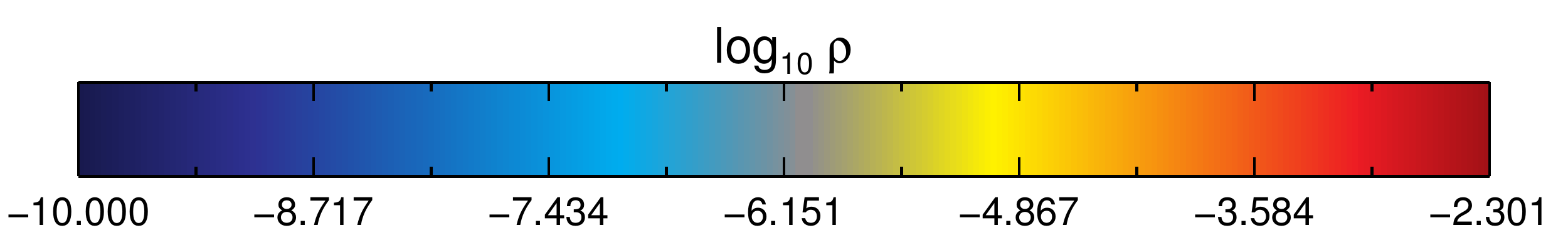}
\end{center}
\caption[]{Initial evolution of the flow in simulation VD0m3d showing 
the advection of the vertical field from the initial location of the 
torus to the black hole. White contours are magnetic field lines, 
color contours show the density distribution. From left to right, the 
panels correspond to times $t=1000,\,1500,\,2000,\,2500M$. The 
color scale is the same as in Figure \ref{initialstate}. By 
$t=2500M$, the funnel region has become strongly magnetic pressure 
dominated and two large scale channel modes have formed within the 
initial torus.}
\label{initial_evolution} 
\end{figure}

The simulation begins with an isolated torus threaded with a column
of vertical field.  The first part of the simulation is marked by
relatively rapid evolution of the coronal field and the surface layers
of the torus.  Figure \ref{initial_evolution} shows the evolution of the
poloidal magnetic flux and the density distribution over the time period
$1000M\le t \le 2500M$.  This process, along with the subsequent evolution
of the flow is also shown in an animation included in the online edition of this work.
We encourage the reader to refer to this material as it clarifies many of the issues that
we shall discuss at length in the remainder of the text.  As noted above, the initial state is not in
equilibrium within the corona.  There is differential rotation between the
coronal gas and the torus, and the magnetic pressure within the vertical
field column is not balanced by pressure within the initial atmosphere.
The field expands radially, and stresses at the boundary of the disk
begin to drive low density gas from the torus outward along field lines.
In the inner half of the torus, low density surface layers move inward in
two thin accretion streams above and below the equatorial plane, dragging
poloidal field lines down to the black hole.  This behavior was dubbed an
``avalanche flow'' when it was observed in the axisymmetric MHD accretion
torus simulations of \cite{Matsumoto:1996}, and it is a common feature
to accretion models that begin with vertical fields threading the disk
\citep[e.g.,][]{Koide:1999}.  These surface features develop well before
field deeper within the torus is significantly amplified by the MRI.
The surface layers in the outer half of the torus also evolve, but here
mass flows radially outward along the field lines, carrying the field
along as it does so.

By $t=2000M$ net flux has begun to build up on the black hole horizon,
well before any significant mass accretion has begun.  This flux arrives
there by a process we call the ``coronal mechanism,'' which we discuss
briefly here and will discuss again in greater detail in \S~\ref{coronal_flux}.
A typical field line participating in the coronal mechanism may be traced from
the outer boundary at high latitude more or less vertically down
to the upper surface of the torus, where it bends sharply toward the hole
near the boundary of the denser region.
It goes in some distance before doubling back
(forming a ``hairpin" shape) and returning to the torus.  There it passes
through the equator, and turns inward
to mirror its behavior on the other side of the equator.  A short time
later, the apex of each hairpin moves inward and closer to the equator,
where,
in the early stages of this simulation when there is little mass
at small radii, it can encounter an oppositely-oriented hairpin from the
other hemisphere.  Reconnection ensues, causing the magnetic
flux distribution to change essentially instantaneously.

The four panels of Fig.~\ref{initial_evolution} show these structures in
differing ways.  In the first panel, an isolated hairpin field line can
be clearly seen approaching very close to the horizon.  By the time of
the second panel, the bends in this field line have reconnected, attaching
it to the event horizon and liberating a pair of closed loops.  Also
in the second panel, new hairpins
can be seen forming along the top and bottom surfaces of the disk in the
region $20M < r < 30M$.  These hairpins evolve further and become embroiled
in the beginnings of disk turbulence (the channel modes) in the third and
fourth panels.

Magnetic field enters the extremely low-density funnel around the rotation
axis by an offshoot of the coronal mechanism just described.  There is
initially no field within the funnel.
Once horizontal field lines have been brought close to the black hole, their
radial, near-horizontal components become unstable to the formation
of ballooning half-loops that rise upward into the funnel.  These loops
are the initial source of field for the funnel.   Only the inner half of the
rising loop enters the funnel and, as a result, the field direction changes
sign abruptly at the funnel's outer edge along the centrifugal barrier.

The coronal process also occurs at radii exterior to the initial 
pressure maximum of the torus.  There it acts to move net flux 
progressively outward. Above the disk surface, the field is stretched 
out radially, but subsequently bends back down toward the equator, where 
reconnection can occur.  This transfers the net flux from field lines 
going through the torus (those lines now close in large loops) to 
vertical field lines at large radius.

Thus, in the early phase of the evolution the global topology of the 
field is rapidly rearranged through coronal motions without any 
significant evolution within the main disk.  Although these features 
result from the initial conditions, the process illustrates how (in 
principle) coherent large scale poloidal flux can be rapidly moved over 
large radial distances.

\begin{figure}
\leavevmode
\begin{center}
\includegraphics[width=0.24\textwidth, viewport=40 30 360 620,clip] {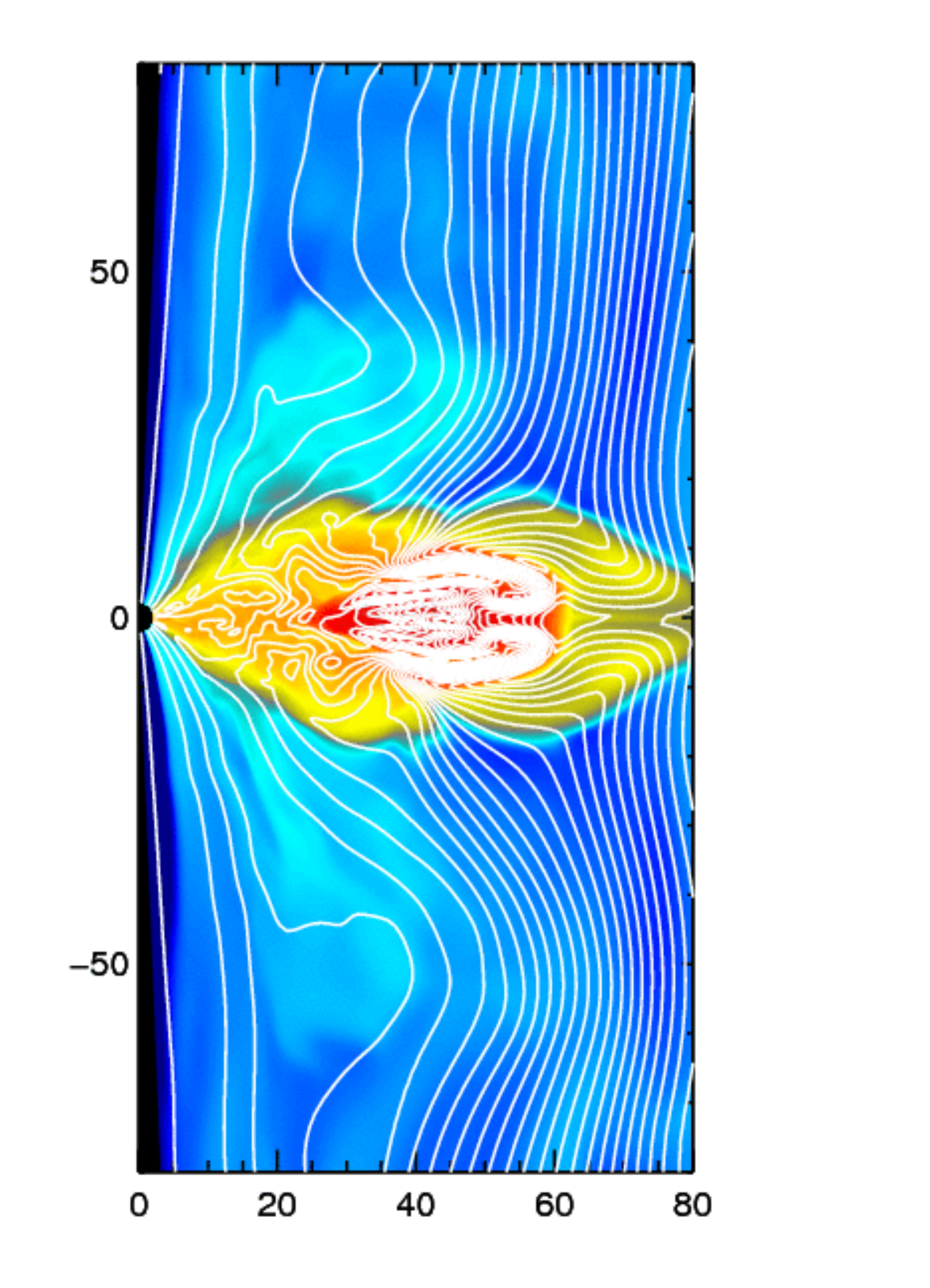}
\includegraphics[width=0.24\textwidth, viewport=40 30 360 620,clip] {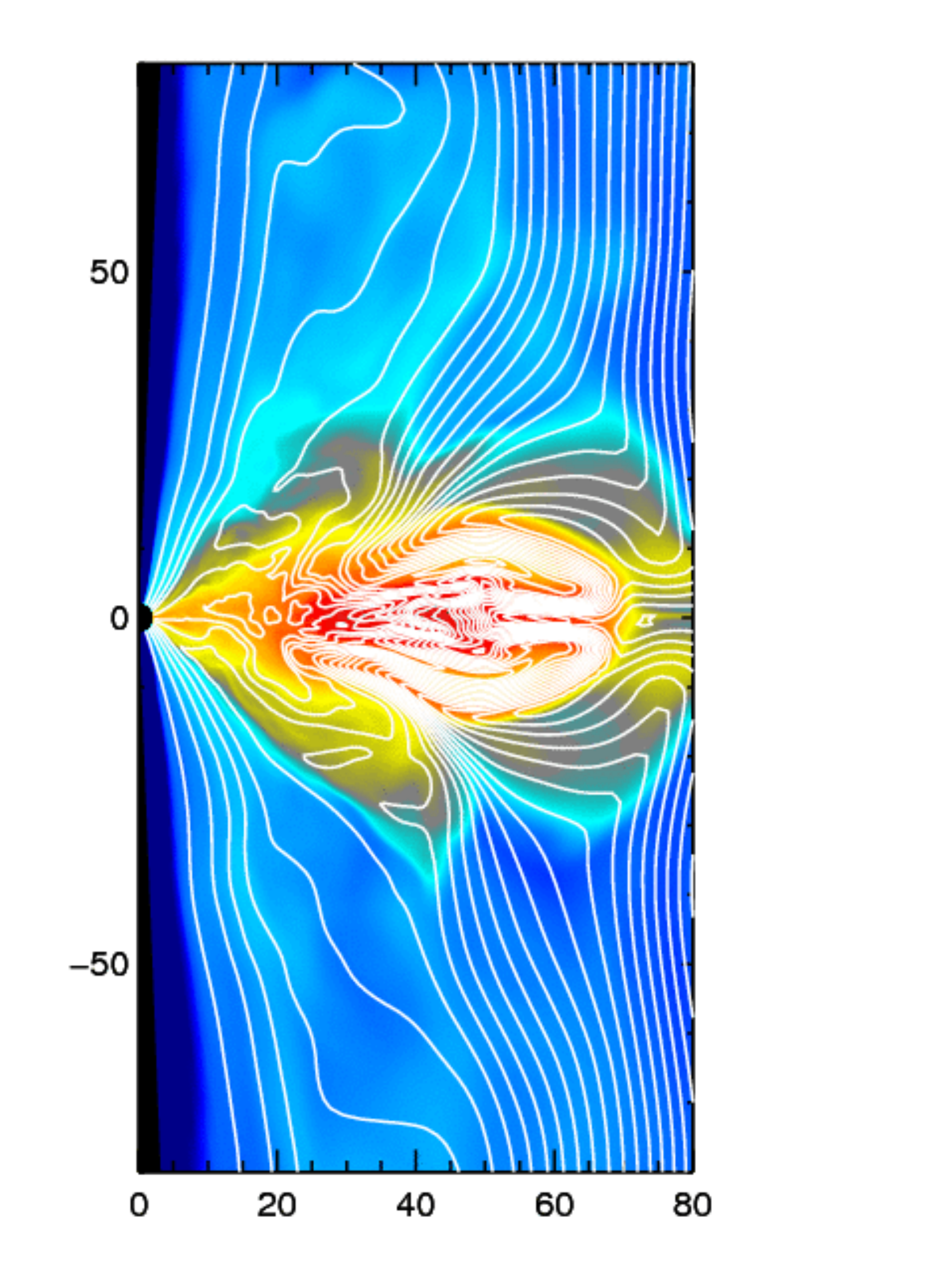}
\includegraphics[width=0.24\textwidth, viewport=40 30 360 620,clip] {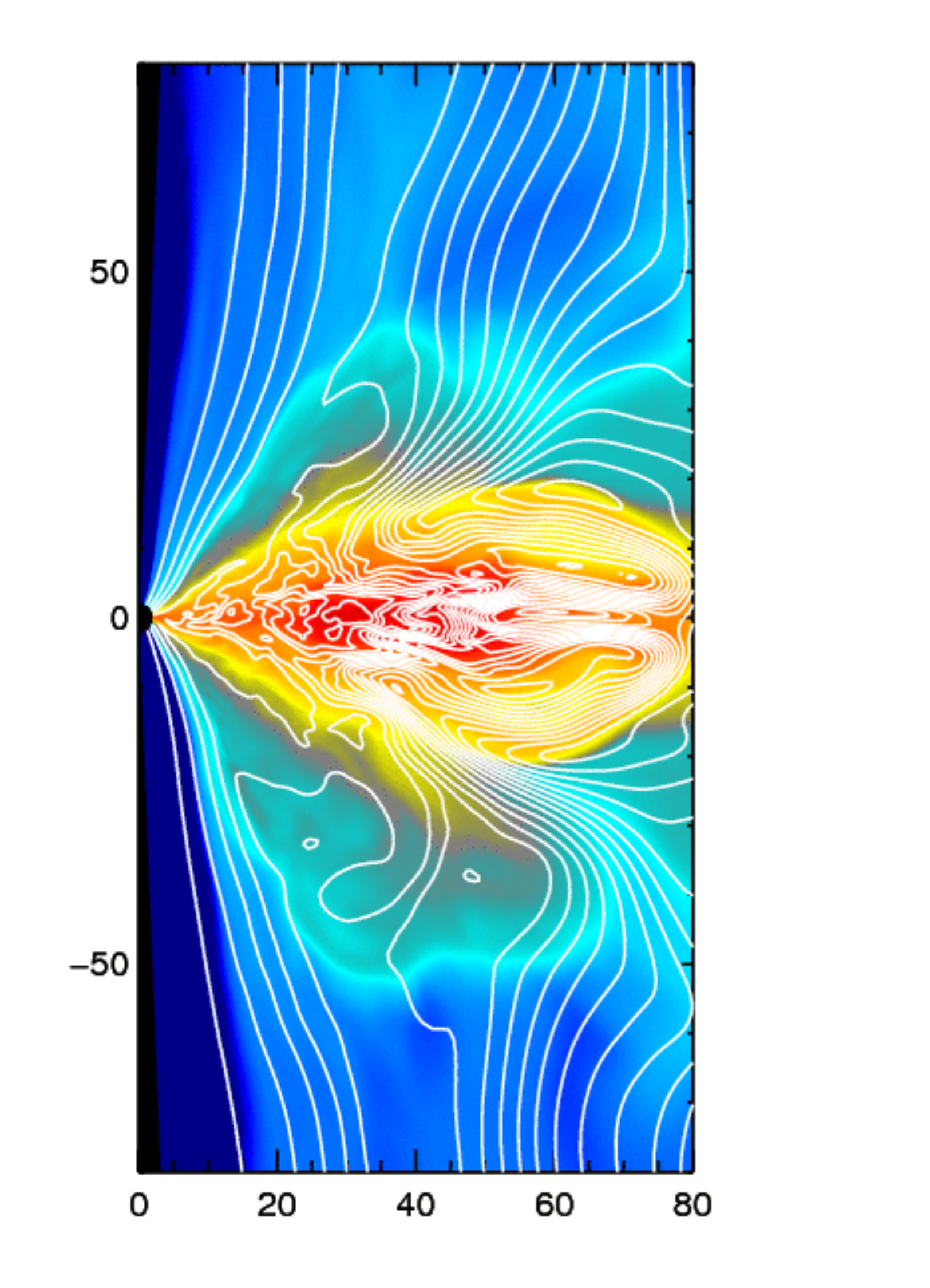}
\includegraphics[width=0.24\textwidth, viewport=40 30 360 620,clip] {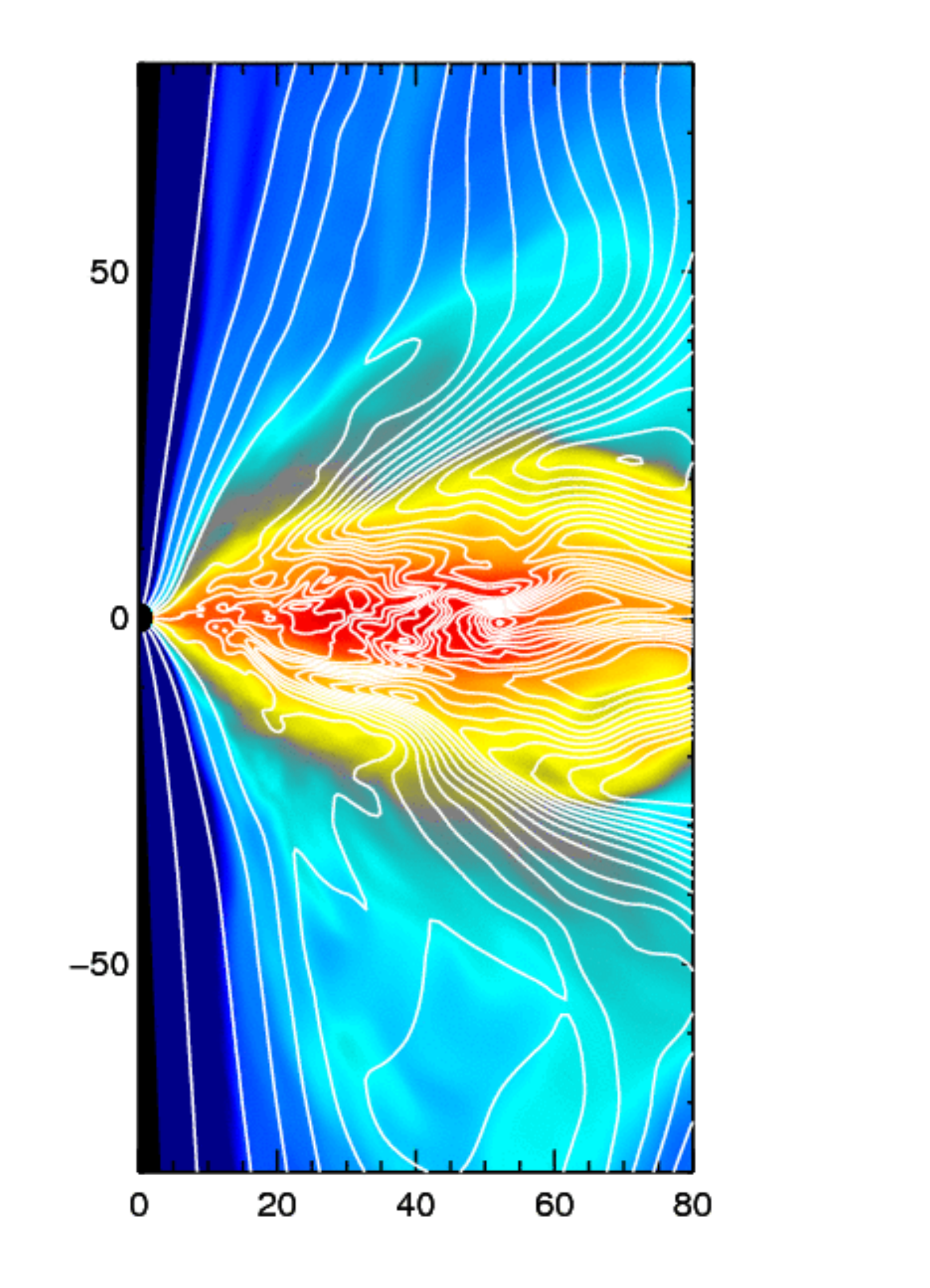}
\includegraphics[width=0.6\textwidth]{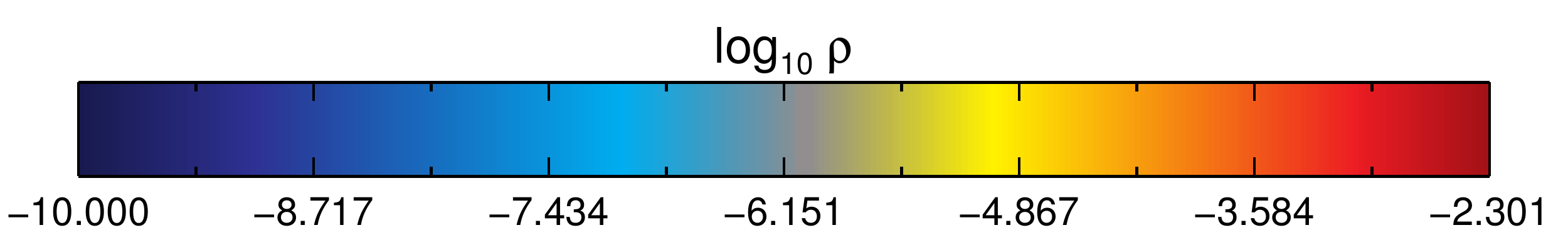}
\end{center}
\caption[]{Growth and subsequent break up of the vertical field MRI 
channel mode in simulation VD0m3d. White contours are magnetic field 
lines, color contours show the density distribution. From left to right, the 
panels correspond to times $t=3000,\,3500,\,4000,\,4500M$. The 
color scale is the same as in Figure \ref{initialstate}. By 
$t=4500M$, the disk has become turbulent;  large scale 
MRI modes persist within the corona.}
\label{channel_evolve} 
\end{figure}

Although low density surface layers participate in the initial avalanche
flow, the dense part of the torus is largely unaffected by these coronal
motions.  By $t=2500M$, however, the MRI reaches a nonlinear amplitude
throughout the torus and the characteristic structures associated
with the ``channel mode'' have become visible within
the torus.  The next stage of evolution is illustrated in
Figure~\ref{channel_evolve} which shows snapshots every $500 M$ in time
from $3000$ to $4500M$.  Gas in the torus moves radially both inward
and outward in the characteristic channels as angular momentum
is transported along field lines, which are themselves increasingly
stretched into long radial filaments.  The channel modes are unstable
to nonaxisymmetric perturbations and rapidly break up, resulting in a
turbulent disk without prominent radial features.  In the axisymmetric
simulations, these relatively coherent large-scale radial flows continue
to dominate throughout the evolution.
This artificial persistence of
coherent motion is a significant limitation
of two-dimensional simulations with vertical field.

Accretion into the black hole does not begin until $1650M$, and the mass
accretion rate at the inner edge of the disk reaches its long-term mean
for the first time shortly after $4500M$.  The mass accretion rate then
continues to vary by factors of $\sim 2$ around this mean through the
remainder of the simulation.  The disk achieves an approximate inflow
equilibrium within $r \simeq 25M$ from time $10^4M$ onward.  Prior to
$10^4M$, the mass in the inner disk ($r\le 25M$) rises steadily, but after
that time, it changes by about $10\%$, and without any prevailing trend.
The mass accretion rate averaged from $10^4M$ until the end of the
simulation at $2\times 10^4 M$ is flat to within $10\%$ for $r \le 25M$.
The total mass on the grid at the end of the simulation is 78\% of its
initial value; of this, $18.8\%$ has gone into the black hole.

\begin{figure}
\leavevmode
\begin{center}
\includegraphics[width=0.7\textwidth]{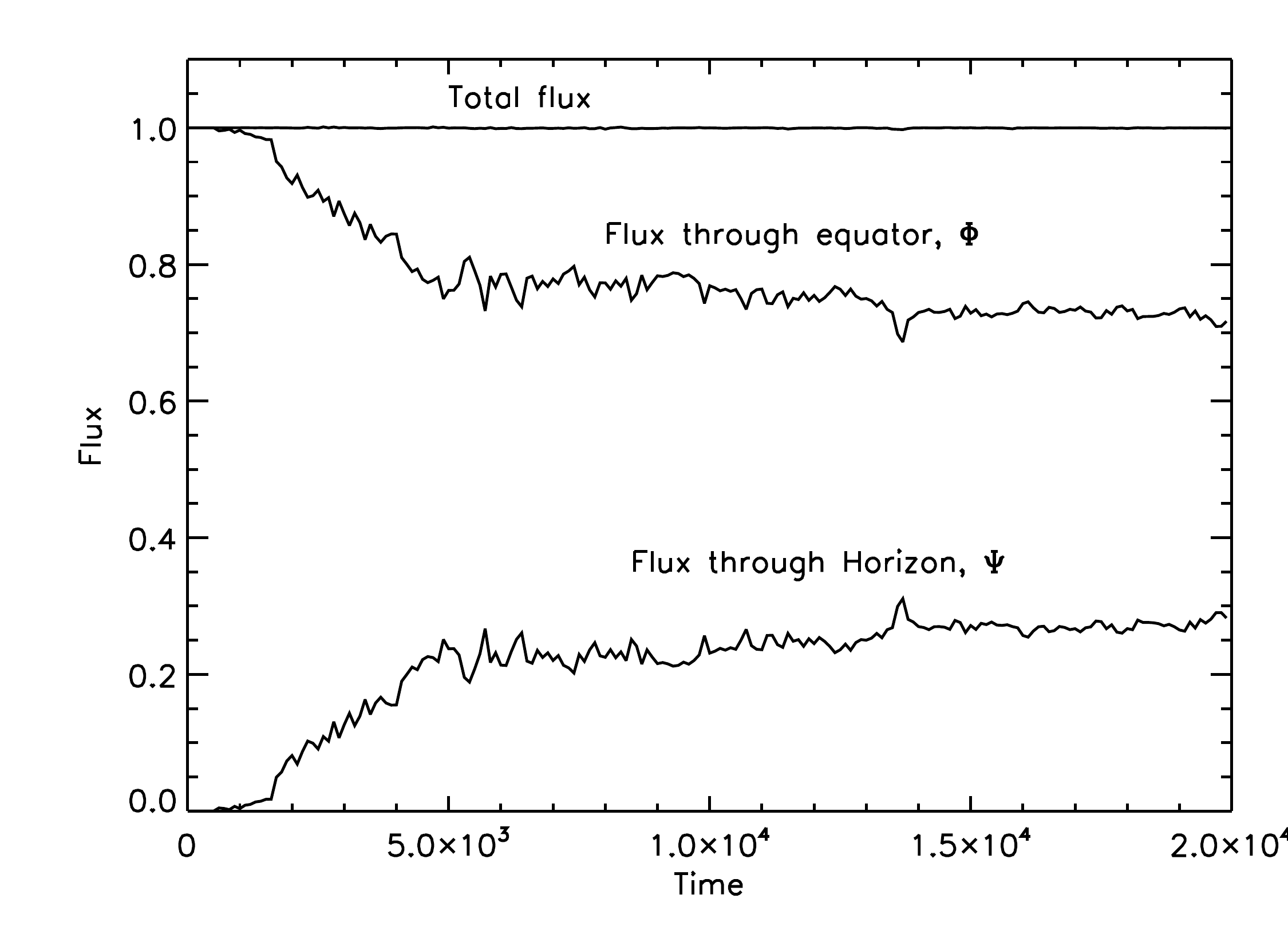}
\end{center}
\caption[]{Total flux $A_\phi$ integrated from the axis along the horizon to
the equator and then out along the equator to the outer boundary.  The
equatorial component, $\Phi$, and the horizon component,
$\Psi$ are shown separately.
The vertical scale is in units of the total initial flux.
The total flux is unchanged through the evolution; 
losses in the equatorial flux are accounted for
in net flux through the horizon.
}
\label{totalflux} 
\end{figure}

The divergence-free nature of the magnetic field means that the total
flux is rigorously conserved except for losses at the boundaries of the
computational domain.  To measure the evolution of the total flux we
integrate along the black hole horizon from the
axis to the midplane and then out along the equator to the outer
boundary.  Figure~\ref{totalflux} shows
this integral as a function of time.
The figure shows that there is no flux lost through the outer radial
boundary during the simulation, but considerable net flux is deposited on
the horizon at the expense of flux through the equator.  The majority of
the flux brought to the horizon arrives by $t\simeq 5000M$, which implies that
the net horizon field is created mostly by coronal motions, not through
disk accretion. During the later stages of more or less steady-state
accretion, the horizon flux grows much more slowly.  Relative to its
magnitude at $t=5000M$, the horizon flux increases by only $28\%$ over
the next $5000M$ and adds only another $11\%$ during the last $10^4 M$.
By contrast, the total mass accreted onto the black hole grows by a
factor of 9 from $5000M$ to $2\times 10^4 M$, with a rate of increase
that is approximately constant from $\simeq 10^{4} M$ onward.

\begin{figure}
\leavevmode
\begin{center}
\includegraphics[width=0.49\textwidth]{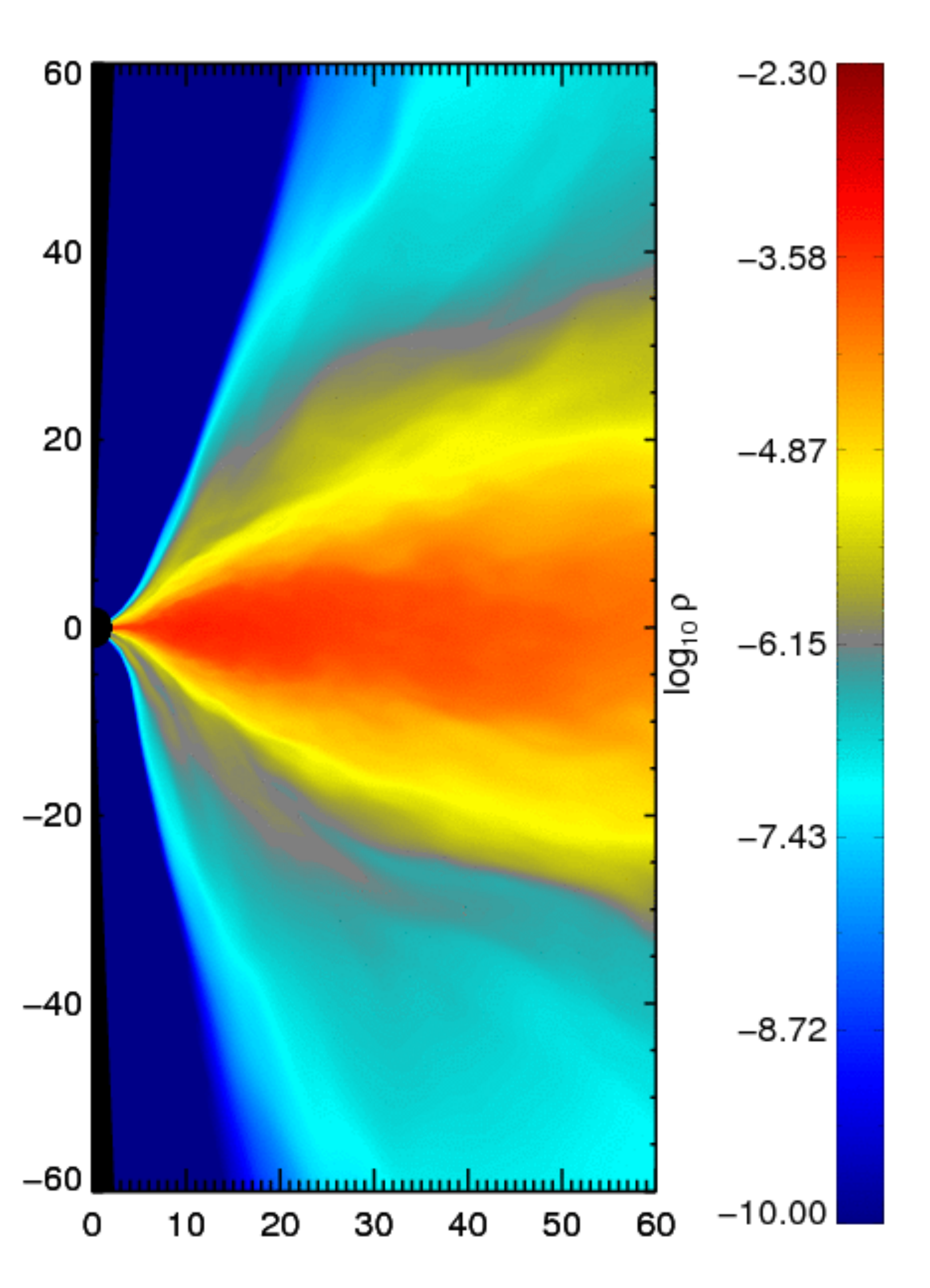}
\includegraphics[width=0.49\textwidth]{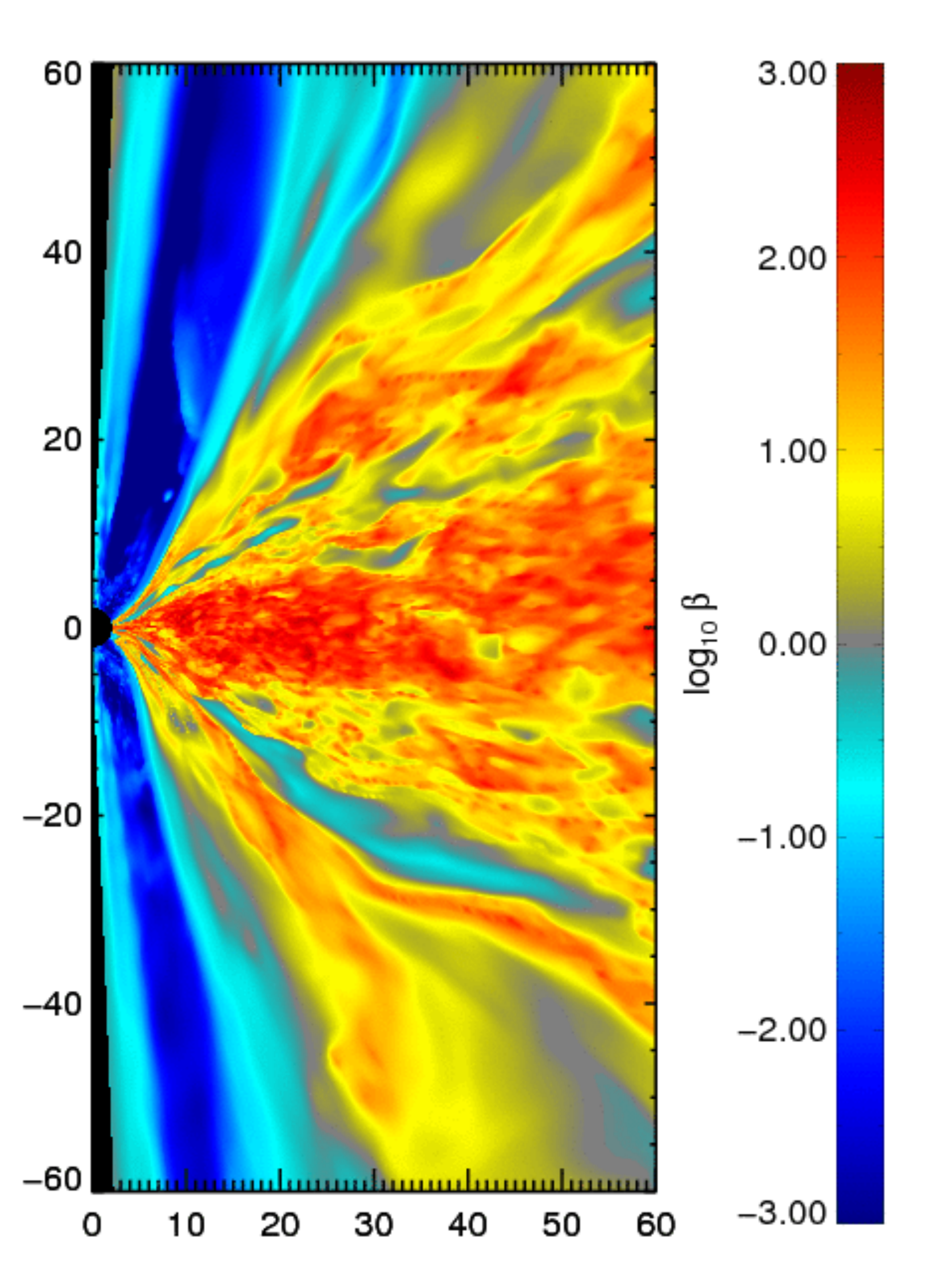}
\end{center}
\caption[]
{Azimuthally-averaged gas density (left panel) and plasma $\beta$ (right
panel) at time $2\times 10^4 M$ in simulation VD0m3d.
}
\label{flow_20000}
\end{figure}

\begin{figure}
\leavevmode
\begin{center}
\includegraphics[width=0.7\textwidth]{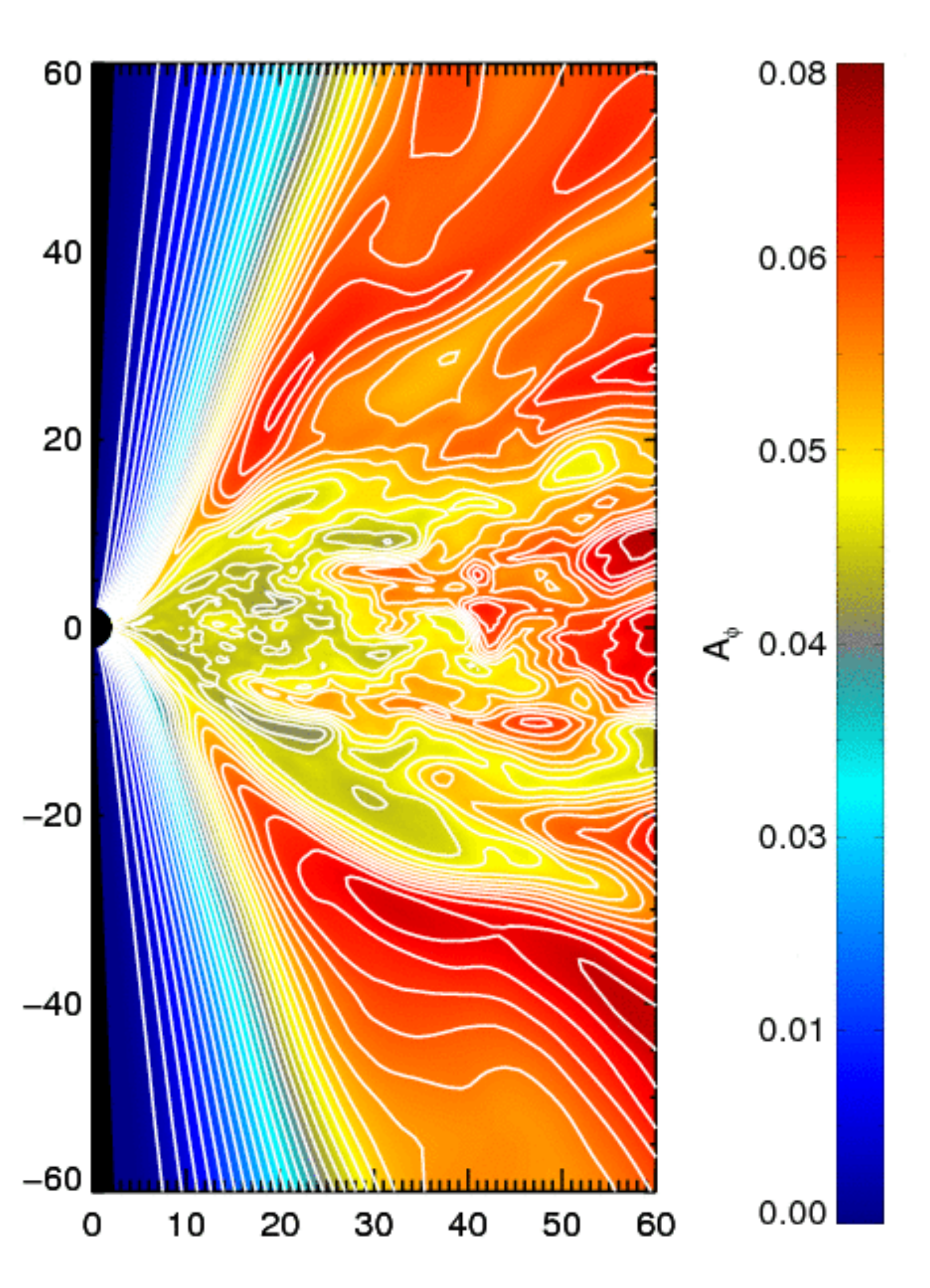}
\end{center}
\caption[]
{Contours of $A_\phi (r,\theta)$, displayed both in color and as poloidal 
field lines (isosurfaces of $A_\phi (r,\theta)$)
at time $2\times 10^4 M$ in simulation VD0m3d.
}
\label{field_20000}
\end{figure}

The appearance of the overall flow at the end of the simulation is
given in Figure~\ref{flow_20000} which shows the azimuthally-averaged
density distribution and the $\beta$ parameter, and in
Figure~\ref{field_20000}, which shows the vector potential
component $A_\phi$ at time $t=2\times 10^4 M$.  There are several
distinct regions.  First, there is a low density, strongly magnetized
funnel in which the field lines are primarily radial.  Second, there is a
corona characterized by large-scale field lines that loop about creating
islands and hairpins, with numerous current sheets separating closely-spaced
regions of oppositely-directed field.
Here there are regions of both relatively strong
and relatively weak magnetization; $\beta$ ranges from $\sim 0.1$ to $\sim 10$.
A thin region of high $\beta$ fluid running along the funnel wall
marks a current sheet between field lines directed in opposite radial
directions as well as the boundary of the region with strong toroidal
field.   Finally, the main disk body remains weakly magnetized, with
$\beta$ ranging from $\sim 10$--$100$.  The disk thickness $H/R$ ranges
from 0.1 to 0.15, using the vertical density moment \citep{Noble:2008}
for the definition of $H$.  As can be seen from
Figures~\ref{channel_evolve} and \ref{field_20000}, substantial parts
of the disk body (most of the disk inside $r \simeq 40M$) are actually
disconnected from the large-scale field, their magnetic connections having
been broken by the reconnection events that transferred field both to
the horizon and to the outside of the torus.  In the outer half of this
region ($20M \lesssim r \lesssim 40M$), however, some field lines do
pass through the equator and out through the disk and into the corona.

\begin{figure}
\leavevmode
\begin{center}
\includegraphics[width=0.45\textwidth,angle=90]{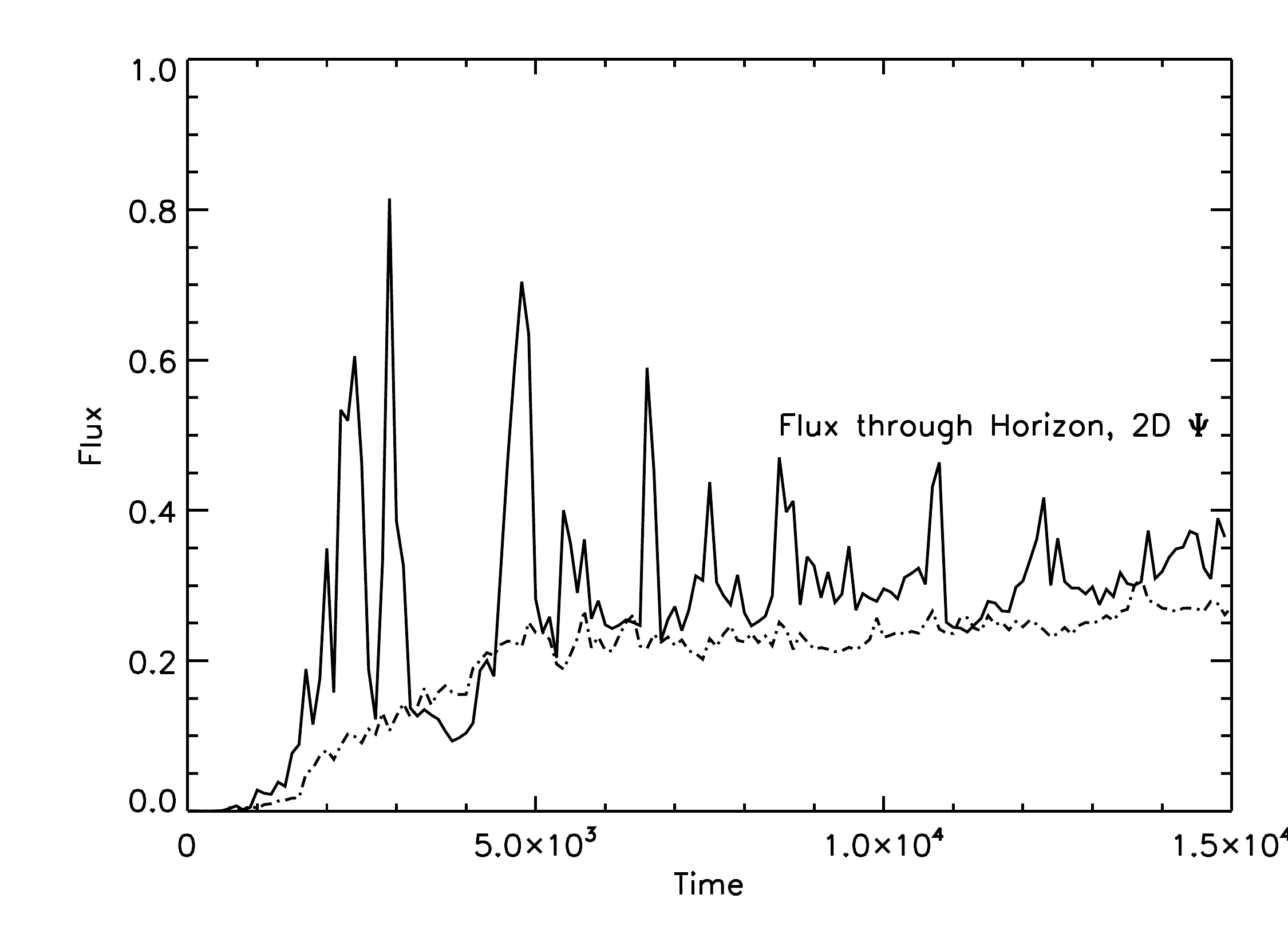}
\includegraphics[width=0.35\textwidth,angle=90]{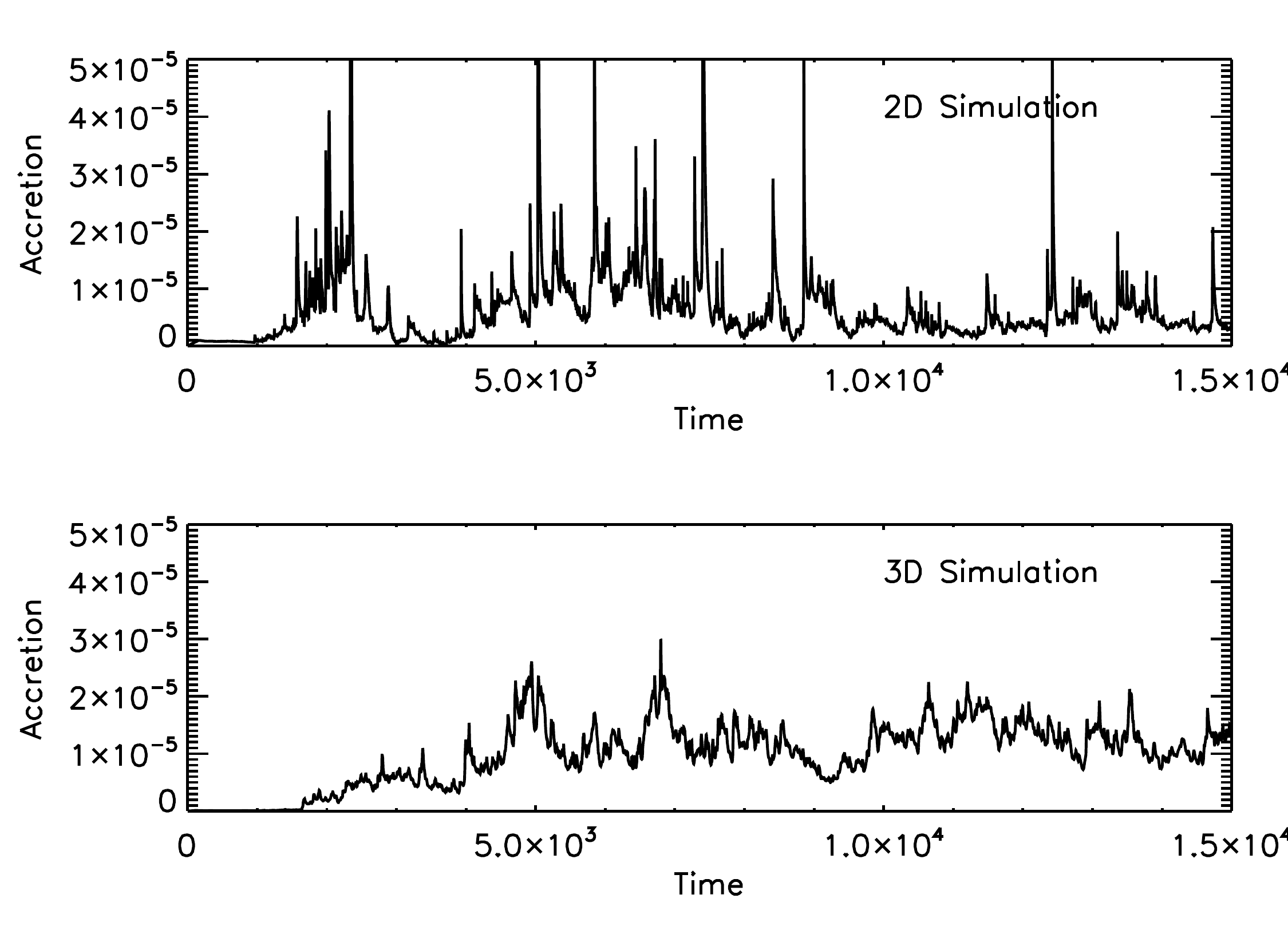}
\includegraphics[width=0.35\textwidth,angle=90]{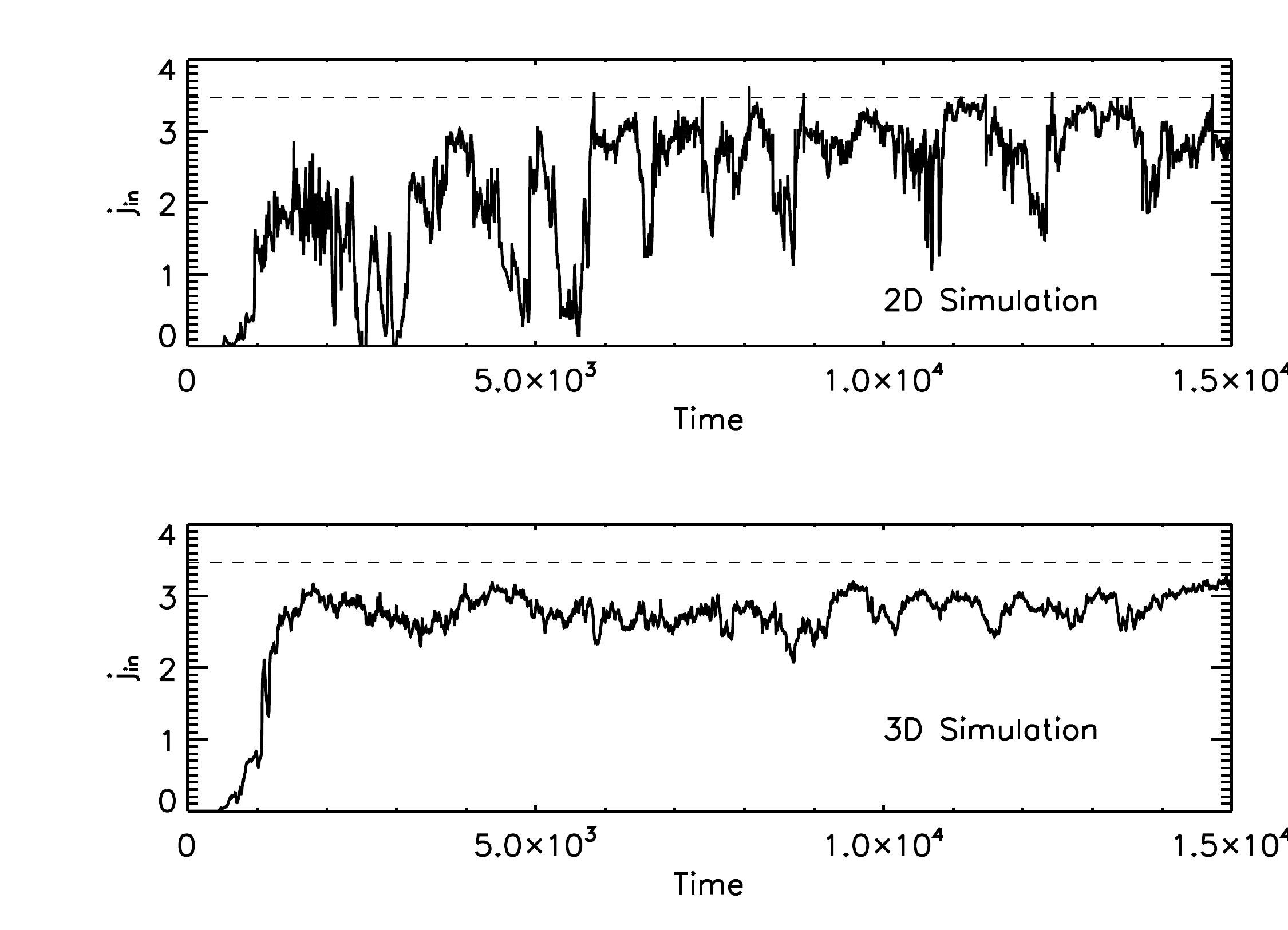}
\end{center}
\caption[]
{Comparison between results from the two-dimensional (VD0m2d) and 
three-dimensional
(VD0m3d) simulation. Top panel shows the flux through the horizon, $\Psi$ for 
VD0m2d (solid line) and VD0m3d (dashed line). The middle panel shows
the net mass accretion rate  through the horizon in units of the 
initial torus mass. 
The bottom panel shows the specific angular momentum carried into the
hole as a function of time for simulations VD0m2d and VD0m3d.  The
strong coherent radial fields in the two-dimensional system create
large fluctations in these quantities.
}
\label{twod_vs_threed} 
\end{figure}

This initial magnetic field and torus configuration was also
simulated in axisymmetry.  While the initial evolutions of the
two- and three-dimensional models are similar, once the MRI sets in
within the torus the subsequent evolutions differ dramatically. In
axisymmetry, the channel modes dominate throughout the simulation,
creating coherent radial field lines that extend over large distances.
These field structures strongly influence the accretion of both mass
and magnetic flux. Figure~\ref{twod_vs_threed} compares the flux through
the horizon, the mass accretion rate and the specific angular momentum
accreted in simulations VD0m3d and VD0m2d.  The axisymmetric simulation
is characterized by very large fluctuations in mass accretion rate.
Similar large fluctuations are seen in the value of the magnetic flux
through the horizon.  Note the contrast between the net specific angular
momentum $j_{in}$ carried into the hole by the accretion flow in the
two- and three-dimensional simulations.  For both plots the dashed line
indicates the specific angular momentum of the innermost stable circular
orbit at $6M$.  The deep minima in the axisymmetric run correspond to
strong torques created by radial field extending through the plunging
region.  These extended radial fields are associated with the coherent
channel flows seen in axisymmetry, and they have an obvious strong effect
on mass, angular momentum and magnetic fluxes carried into the black
hole.  The three-dimensional simulation is qualitatively different.
In three dimensions, the channel modes quickly break up, yielding to
genuine turbulence.  The accretion rate into the hole fluctuates, but not
nearly as strongly as in axisymmetry.
Similarly, only in 3-d can one
speak sensibly about a well-defined net accreted angular momentum per
unit rest-mass, $j_{\rm in}$.  On the basis of these contrasts,
we conclude that axisymmetric simulations, while useful for preliminary
investigations of various initial configurations, are of only limited
utility in investigating the long term behavior of MHD turbulent
disks.

\subsection{The coronal mechanism}
\label{coronal_flux}

The dominant mechanism for bringing flux to the event horizon in
simulation VD0m3d is one that operates primarily in the corona, and not
in the disk accretion flow proper.  Although most of the flux is
brought in during the first $5000M$ in time, there is a slow increase in flux
throughout the remainder of the simulation.  The majority of this flux
appears to be transported through the continuing operation of the
coronal mechanism.  The key aspect is that magnetic field is carried
inward by infalling low density fluid.  Although this coronal mechanism
operates outside the disk where most of the mass accretion happens,
it, too, depends on angular momentum transport to proceed, and the basic
physical mechanism for angular momentum transport is likewise the same:
magnetic torques.  Wherever the vertical field is perturbed so as to
create a radial component, orbital shear acting on the radial field in
turn creates toroidal field.  Moreover, whenever the orbital frequency
decreases outward, the relative signs of the radial and toroidal fields
are always such as to make the $r$-$\phi$ element of the magnetic stress
tensor, $b^r b_\phi/4\pi$, negative.  Such a stress transports angular
momentum outward, tending to accentuate inflow, and therefore growth
in the radial field component; this is, of course, the mechanism of the
magneto-rotational instability.

\begin{figure}
\leavevmode
\begin{center}
\includegraphics[width=0.49\textwidth]{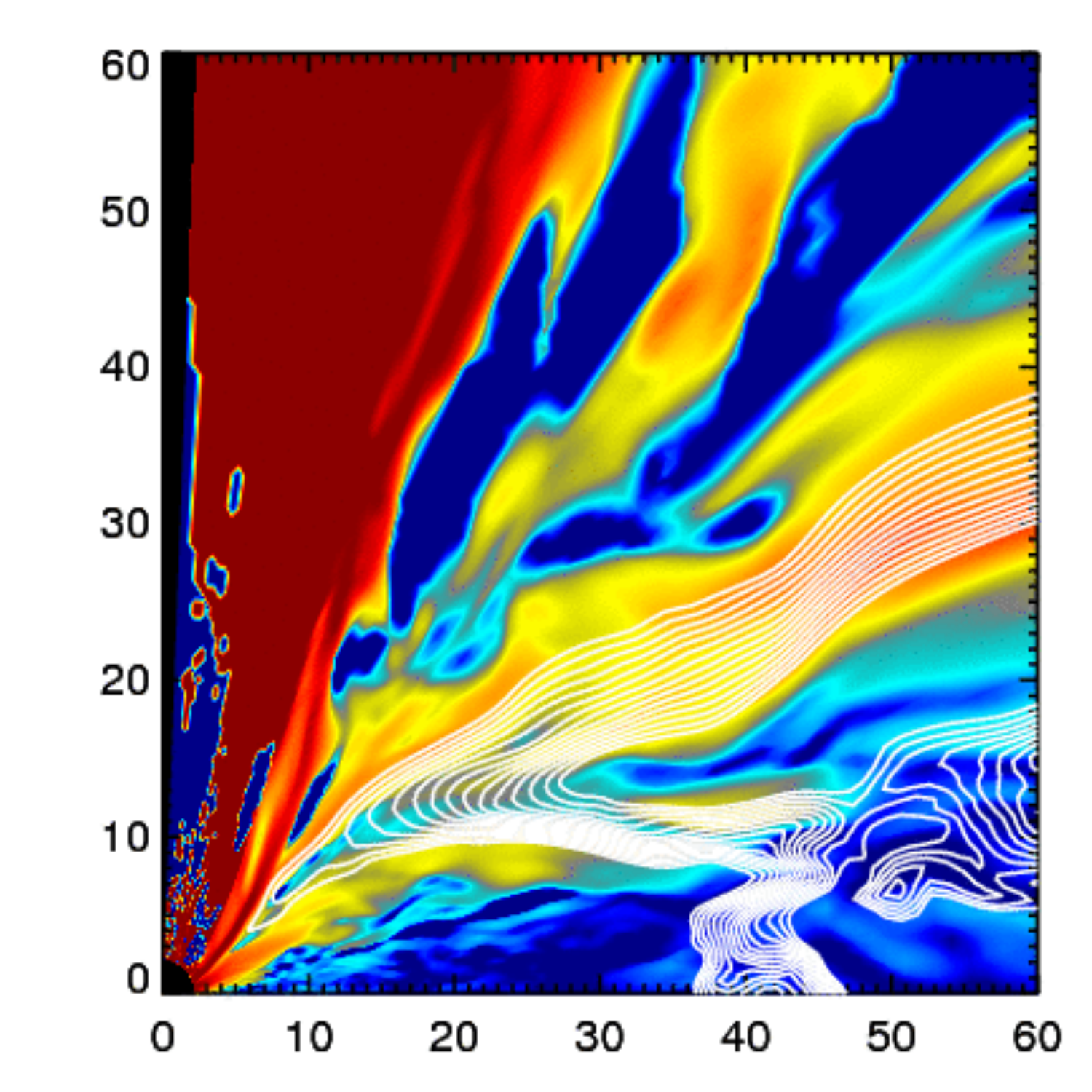}
\includegraphics[width=0.49\textwidth]{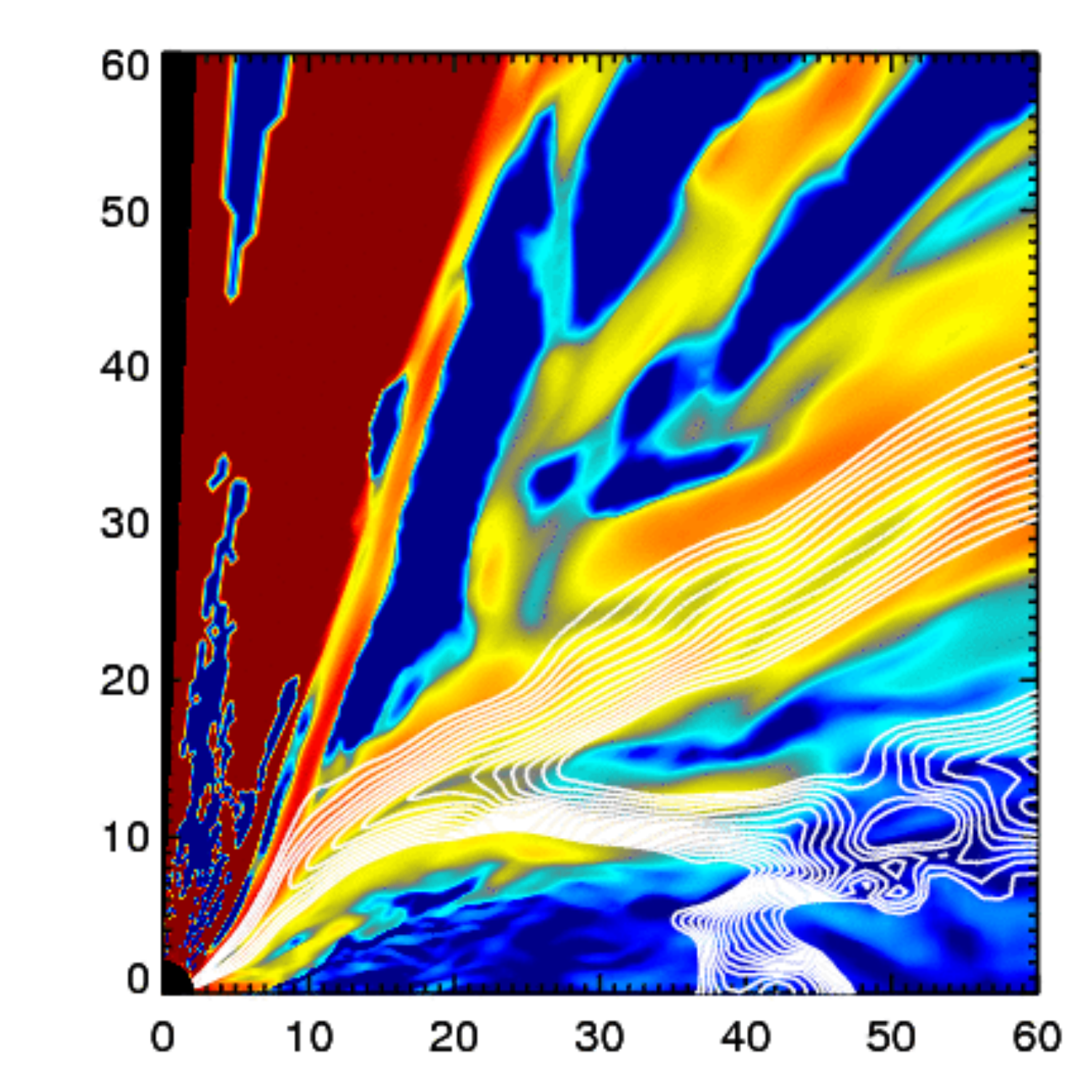}
\includegraphics[width=0.49\textwidth]{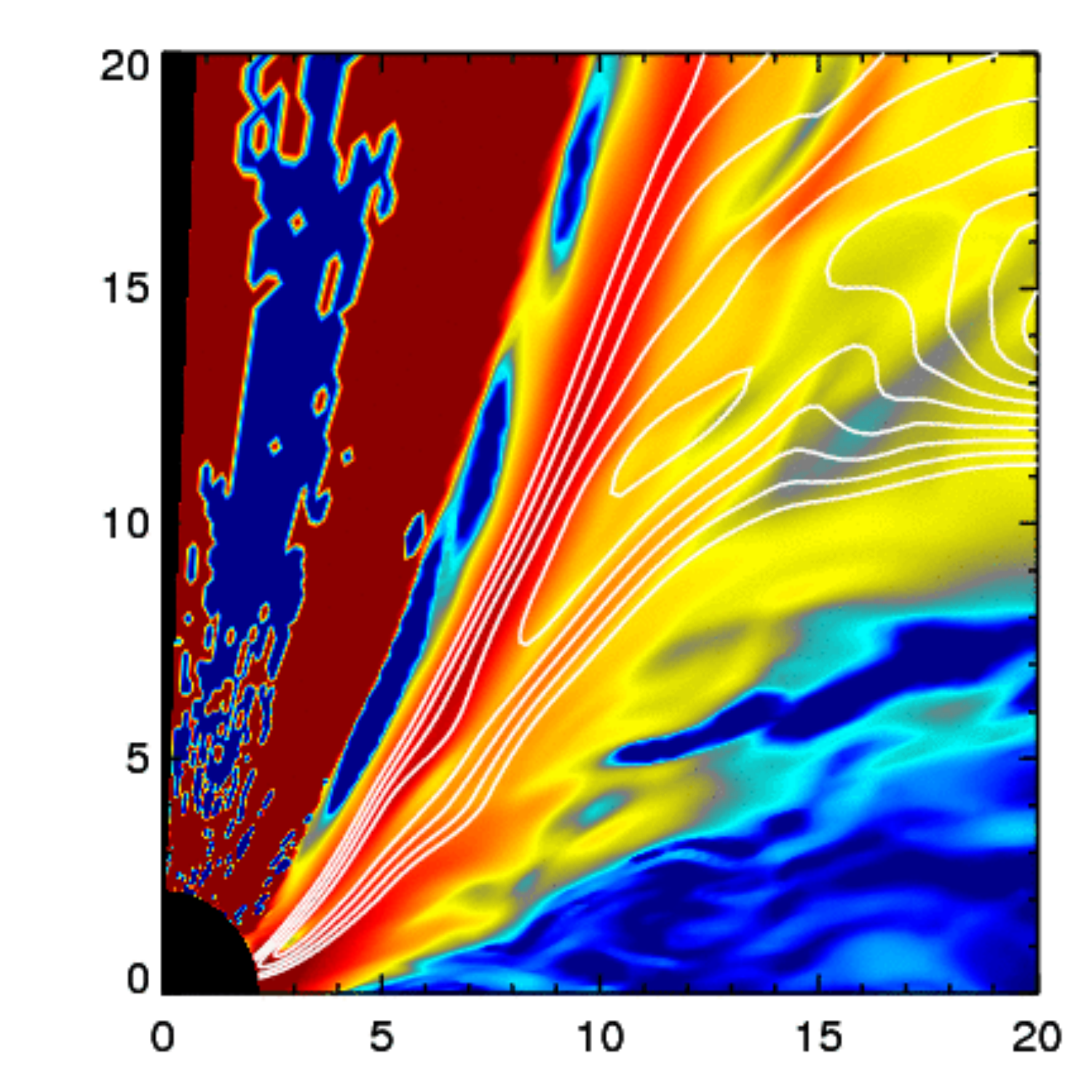}
\includegraphics[width=0.49\textwidth]{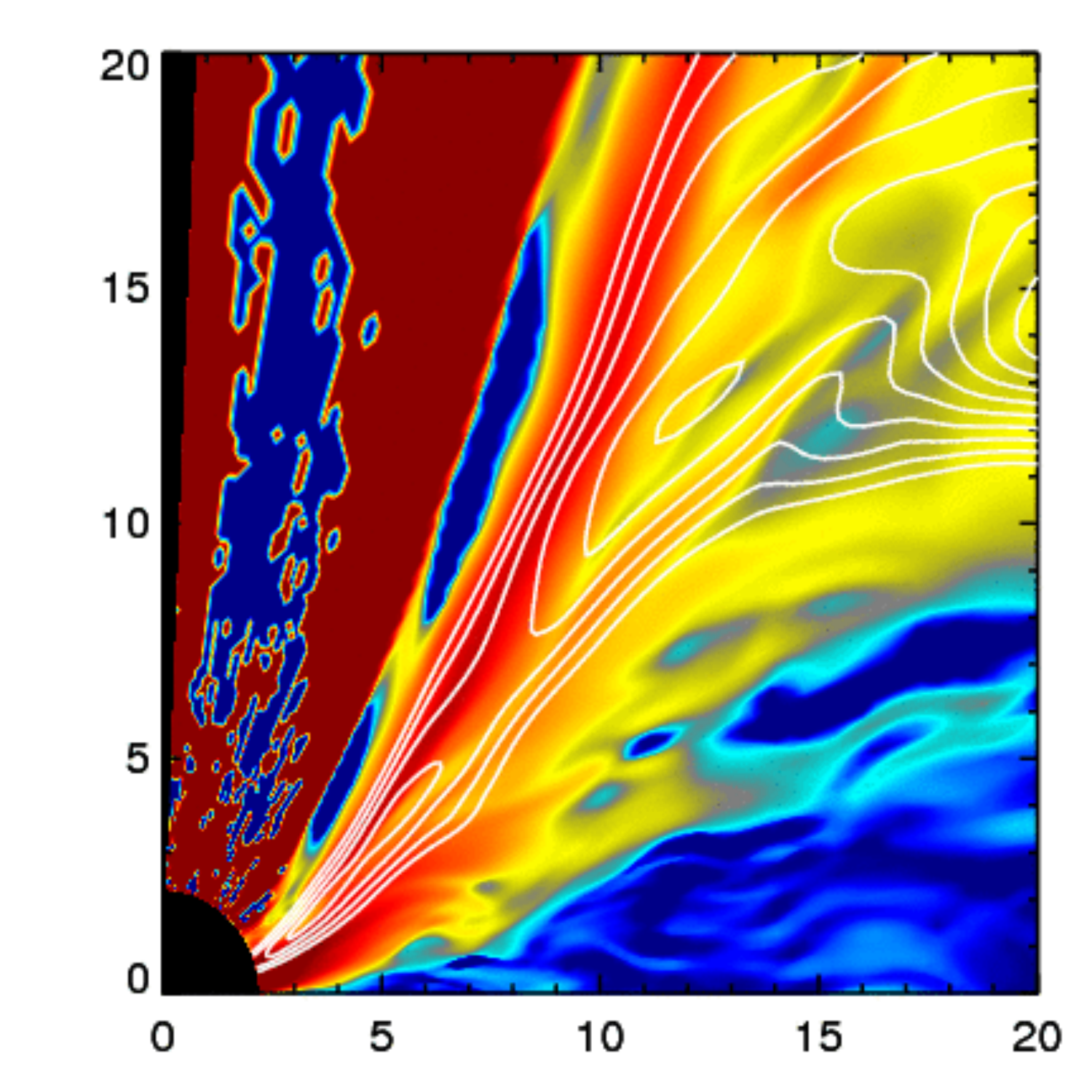}
\includegraphics[width=0.6\textwidth]{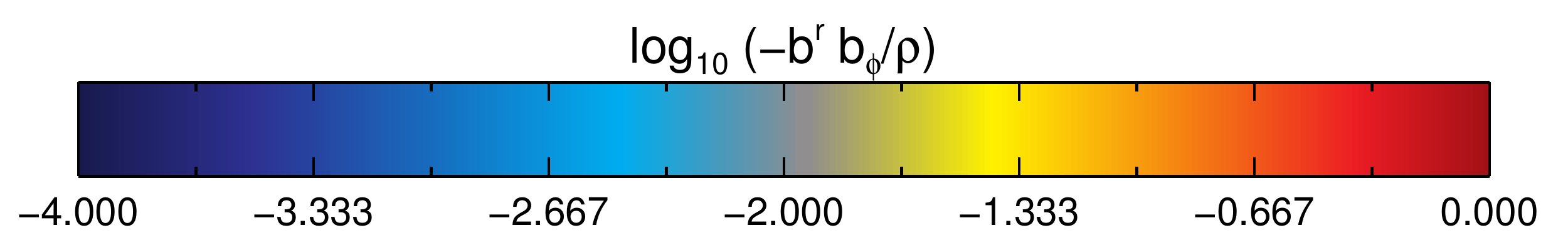}
\end{center}
\caption[]{Evolution of a magnetic ``hairpin" in the corona during the
turbulent steady state. Color contours show the azimuthal average of
stress per unit mass, $-b^r b_\phi/\rho$, and are overlaid with the
poloidal field structure (white lines), which are kept fixed between the
panels.  From left to right, the top panels
show times $t=6000$, and $6200M$ (top row).  The bottom row shows the
near-hole region at times $t=6440$ and $6460M$.  During this time a
reconnection event takes place within the field hairpin.
}
\label{coronalevolve1}
\end{figure}

An example of the coronal mechanism in action is
given in Figure~\ref{coronalevolve1}.  In this figure, color
contours show the strength of the azimuthally averaged Maxwell stress per
unit mass, overlaid with field lines derived from $A_\phi(r,\theta)$.
For clarity we focus on those field lines that still pass through
the equator in the accretion disk.  The first frame, corresponding to
$t=6000M$, features a field that runs from the equator vertically upward
through the disk and then down toward the hole before reversing and moving
radially outward.  The strong stress per unit mass in the vicinity of
the hairpin carries angular momentum away from the associated matter,
enabling it to move inward. In the next frame at $t= 6200$,
the hairpin moves in radially toward the black hole.  The next two
frames show a close up view of the innermost part of the accretion flow
at times $t=6440$ and $6460M$.
By the time of the first one,
the inner portion of the hairpin has reached and crossed the horizon.
Only $20M$ later, the inside hairpin has reconnected, forming a closed
field loop.  Thus, the net result of the coronal mechanism at this
stage is to attach those fieldlines forming the exterior of the hairpin
to the black hole.  These fieldlines then extend out into the corona,
ultimately attaching to the disk at much larger radius.

\begin{figure}
\leavevmode
\begin{center}
\includegraphics[width=0.5\textwidth,angle=90]{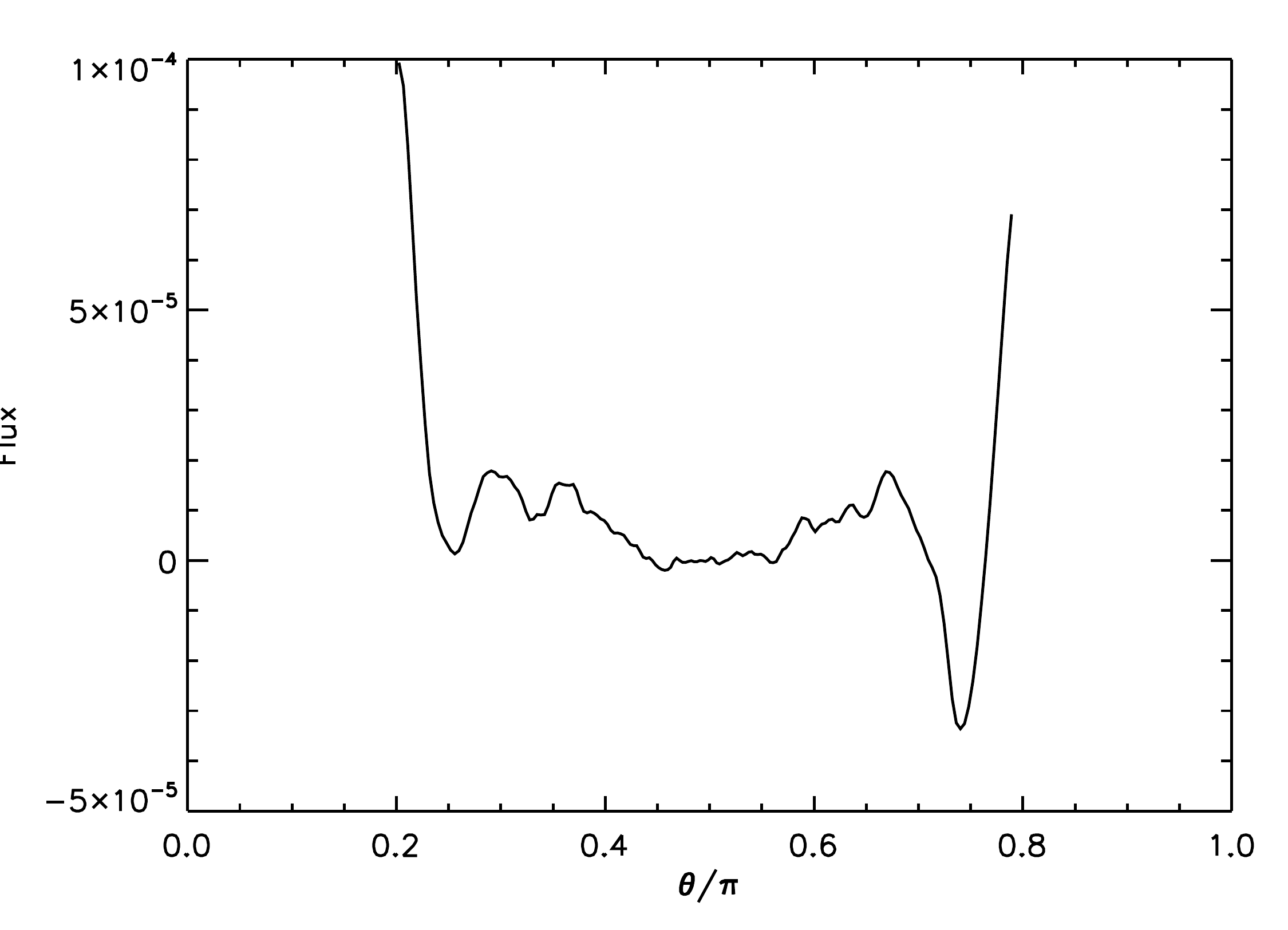}
\includegraphics[width=0.5\textwidth,angle=90]{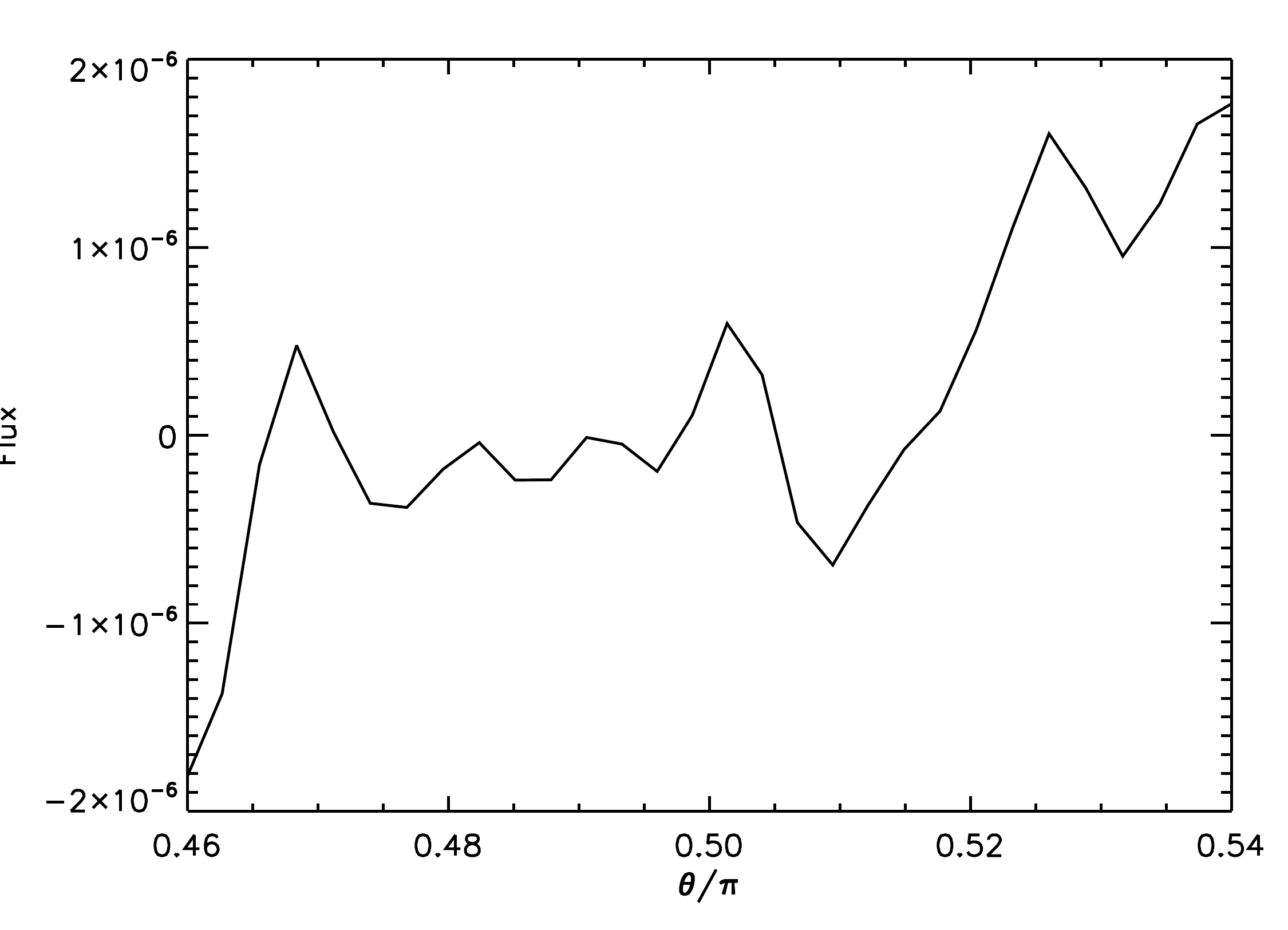}
\end{center}
\caption[]{Azimuthal mean of $-V^r B^\theta(15M)$, averaged over the
time span 5000--$1.5 \times 10^4 M$ at $r=15M$.  Two $\theta$ ranges and
scales are used:  a scale appropriate for the corona and excluding only
the funnel region (top), and only the disk proper plotted on a scale
appropriate to the value there.}
\label{magfluxflow}
\end{figure}

Although net local inflow of magnetic field is a {\it necessary} 
condition for flux accretion, it is not a {\it sufficient} one; 
the net flux changes only when reconnection occurs.  Local motion alone
is insufficient because magnetic flux is an 
essentially {\it global} concept, so local motion is not enough to 
determine net flux transport. This distinction is emphasized by 
contrasting the local magnetic inflow rate with global measures.  We 
define the local rate by
\begin{equation} 
V_{\Phi}(r,\theta) \equiv \frac{1}{2\pi}\int \, d\phi V^r {\cal B}^\theta 
\end{equation} 
averaged over the portion of the simulation during which 
the trapped flux grows slowly (5000--$2\times 10^4 M$). 
This is shown in Figure~\ref{magfluxflow}, for which the radius 
was chosen to be $15M$.  As this figure shows, the local flux inflow 
speed is considerably greater just outside the funnel ($|\pi - \theta| 
\simeq 0.2\pi$), where it is $\simeq 0.5$--$1 \times 10^{-4}$, than in 
the midplane, where it is nearly two orders of magnitude smaller.

\begin{figure}
\leavevmode
\begin{center}
\includegraphics[width=0.6\textwidth]{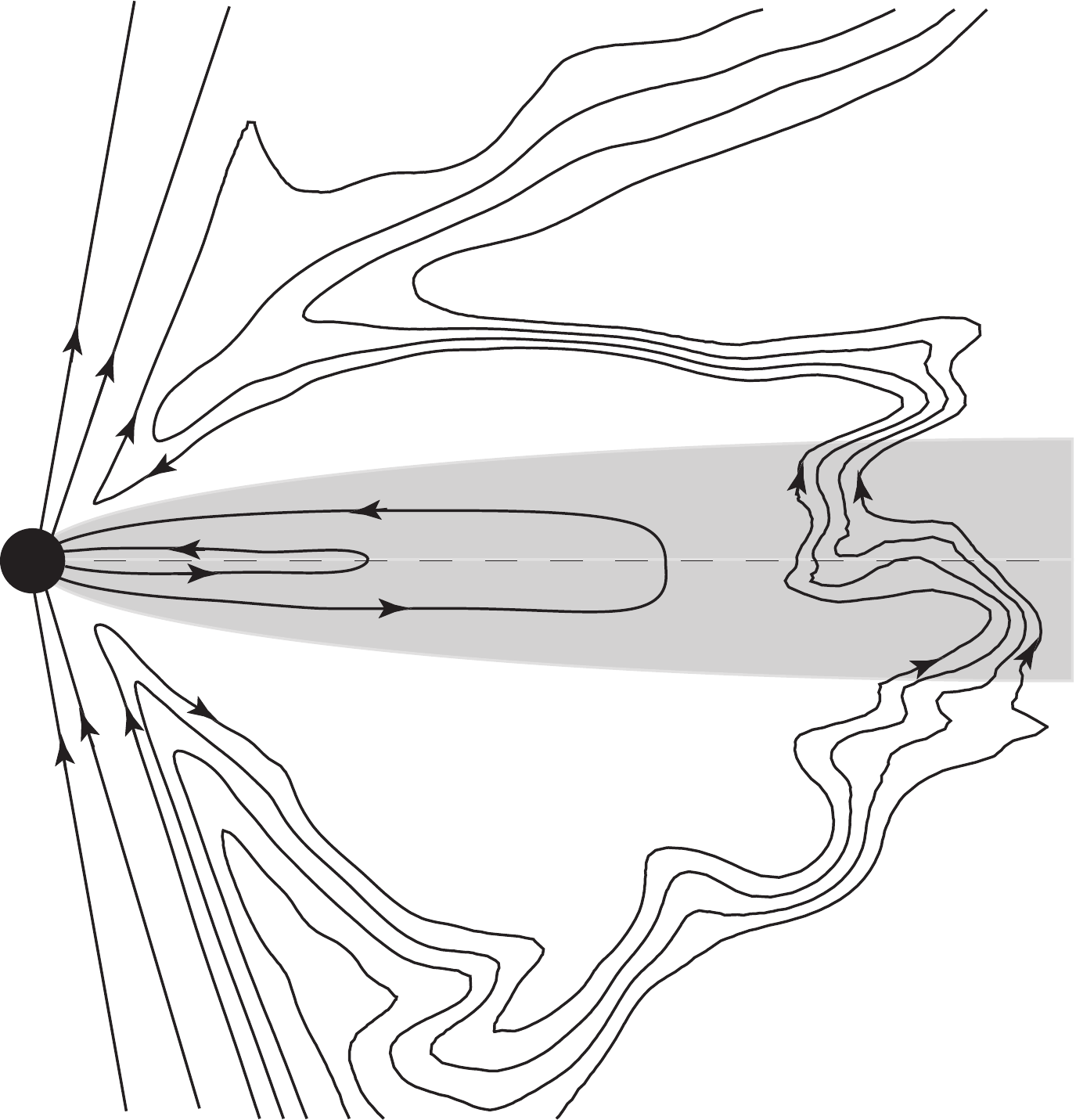}
\end{center}
\caption[]{Schematic of the field line structure in the coronal
mechanism.  Field lines within the coronal are carried in toward the
black hole, forming a hairpin-like structure. When the hairpin connects
to the horizon flux is added to the funnel field and opposite-signed
flux is added to the disk (shaded region).  Reconnection across the
equator (dashed line) allows this field to form loops that accrete.
Accretion of those field loops results in an increase of the net horizon
flux.
}
\label{schematic}
\end{figure}

This local rate of flux inflow is, however, much larger than the
{\it global} rate of flux accumulation.  The global rate is
$\partial A_\phi/\partial t$, which can be inferred from 
Figure~\ref{totalflux}.  For the purpose of this figure, the 
magnetic flux was separated into the flux piercing the midplane, $\Phi$, 
and the flux through the upper hemisphere of the black hole, $\Psi$.  
During the early stages of the simulation (before $5000M$), $\partial 
\Psi/ \partial t \simeq 1 \times 10^{-5}$, but this rate of increase 
diminishes at later times by at least an order of magnitude.  Thus, the 
local inflow rate, $V_\Phi(\theta \simeq 0.2\pi,0.8\pi) \simeq 1 \times 10^{-4}$ 
at $r=15M$ during the latter 3/4 of the simulation, is as much as two
orders of magnitude greater than the net global rate.  In other words,
local field inflow alone does not suffice to create global flux inflow.
Likewise, reconnection on its own is also a {\it necessary} condition
for flux accretion,
but not a {\it sufficient} one. In order for flux accretion
to occur, reconnection must result in a global rearrangement of the
field topology.  This fact is illustrated by the field structures displayed
in Figure~\ref{coronalevolve1}.  In that figure, a narrow current sheet
lies between the two sides of the hairpin, permitting reconnection to happen
there.  In order to transfer magnetic flux inward through the
accretion flow and increase the net flux in the funnel, reconnection
must preferentially destroy the side of the magnetic hairpin that
lies closest to the midplane.  This is possible because reconnection can
also occur across the equator, something that the funnel field cannot
do.  Figure~\ref{schematic} illustrates the
idea.  Coronal hairpin field loops bring field down to the inner disk,
returning along the disk back out to large radius where they eventually
pass through the equator and connect to another hairpin in the opposite hemisphere.
The innermost bend of these hairpins can attach to the horizon.  When this
happens, the field appears to be almost a closed loop, with one end of the
loop inside the horizon (e.g., the two inner-disk field lines in the gray region
of the diagram).  If the two oppositely-directed portions of one of these
loops can meet at intermediate radius, the equatorial flux jumps inward.
This inward jump drastically shortens the time until the entire field loop
can accrete onto the black hole.  When that happens, flux of one sign is
removed from the horizon above the equator and flux of the other sign is
removed from below the equator, leaving a net increase in flux piercing
each horizon hemisphere.

\begin{figure}
\leavevmode
\begin{center}
\includegraphics[width=0.32\textwidth, viewport=20 20 340 620,clip]{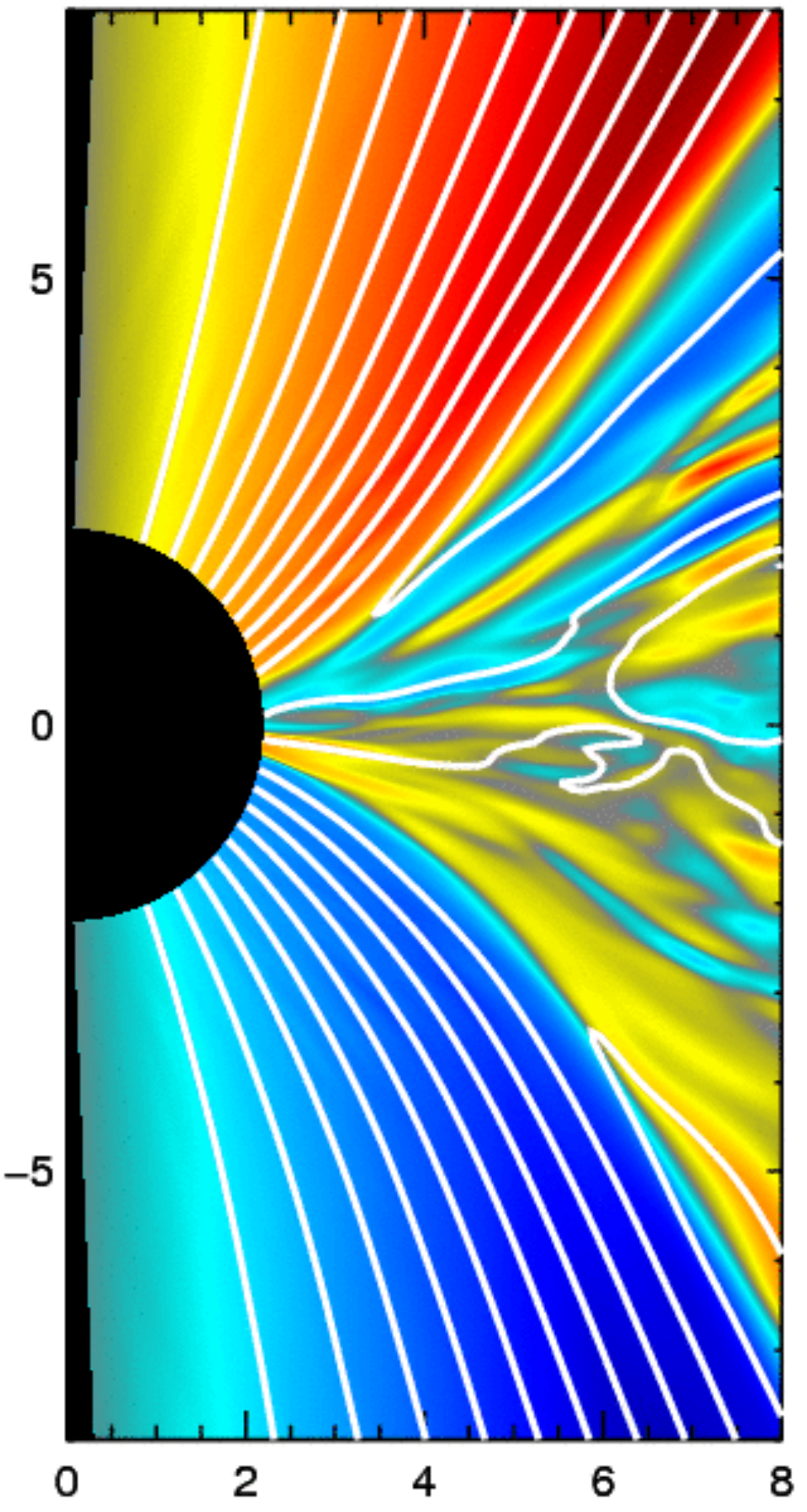}
\includegraphics[width=0.32\textwidth, viewport=20 20 340 620,clip]{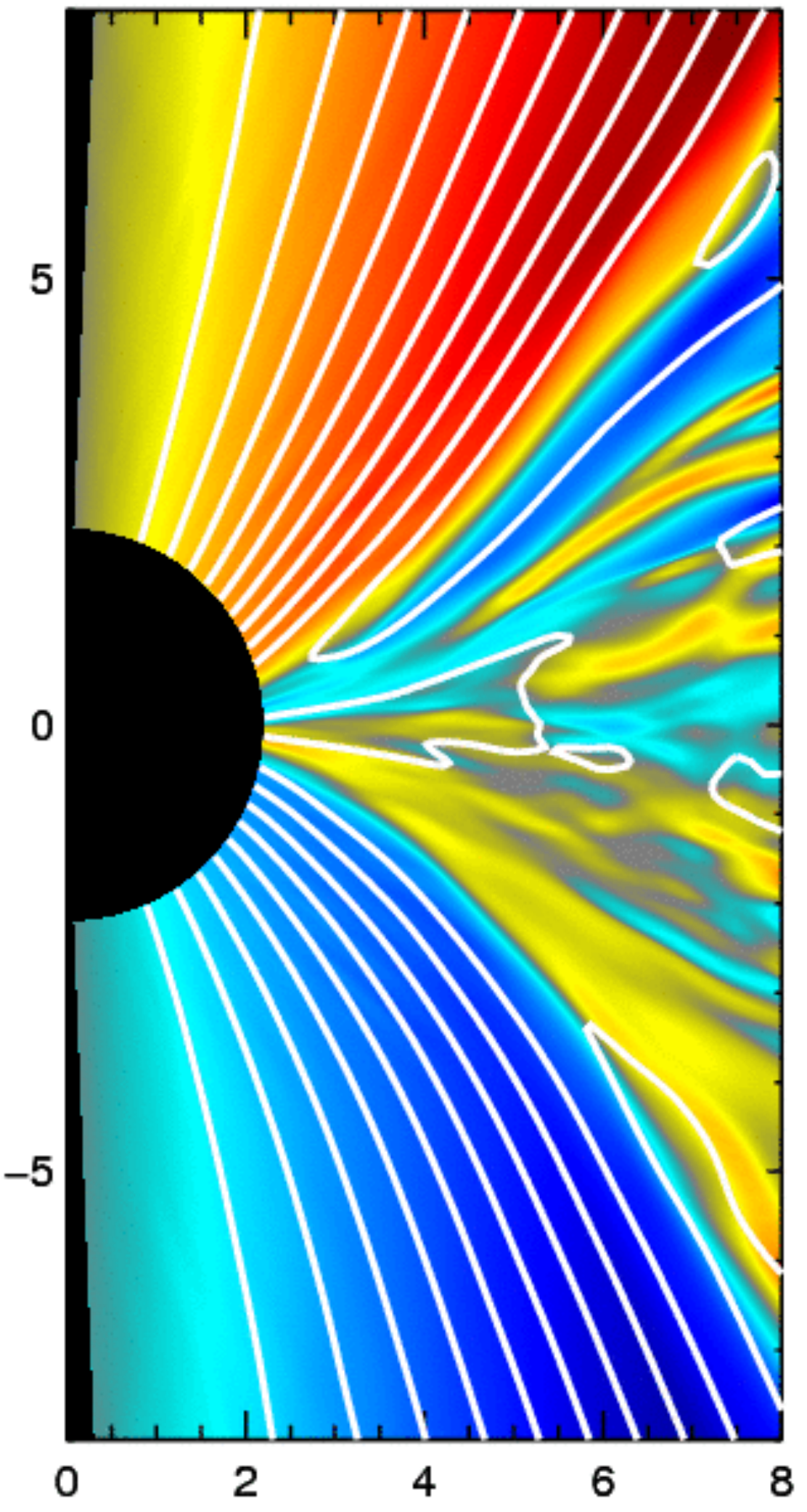}
\includegraphics[width=0.32\textwidth, viewport=20 20 340 620,clip]{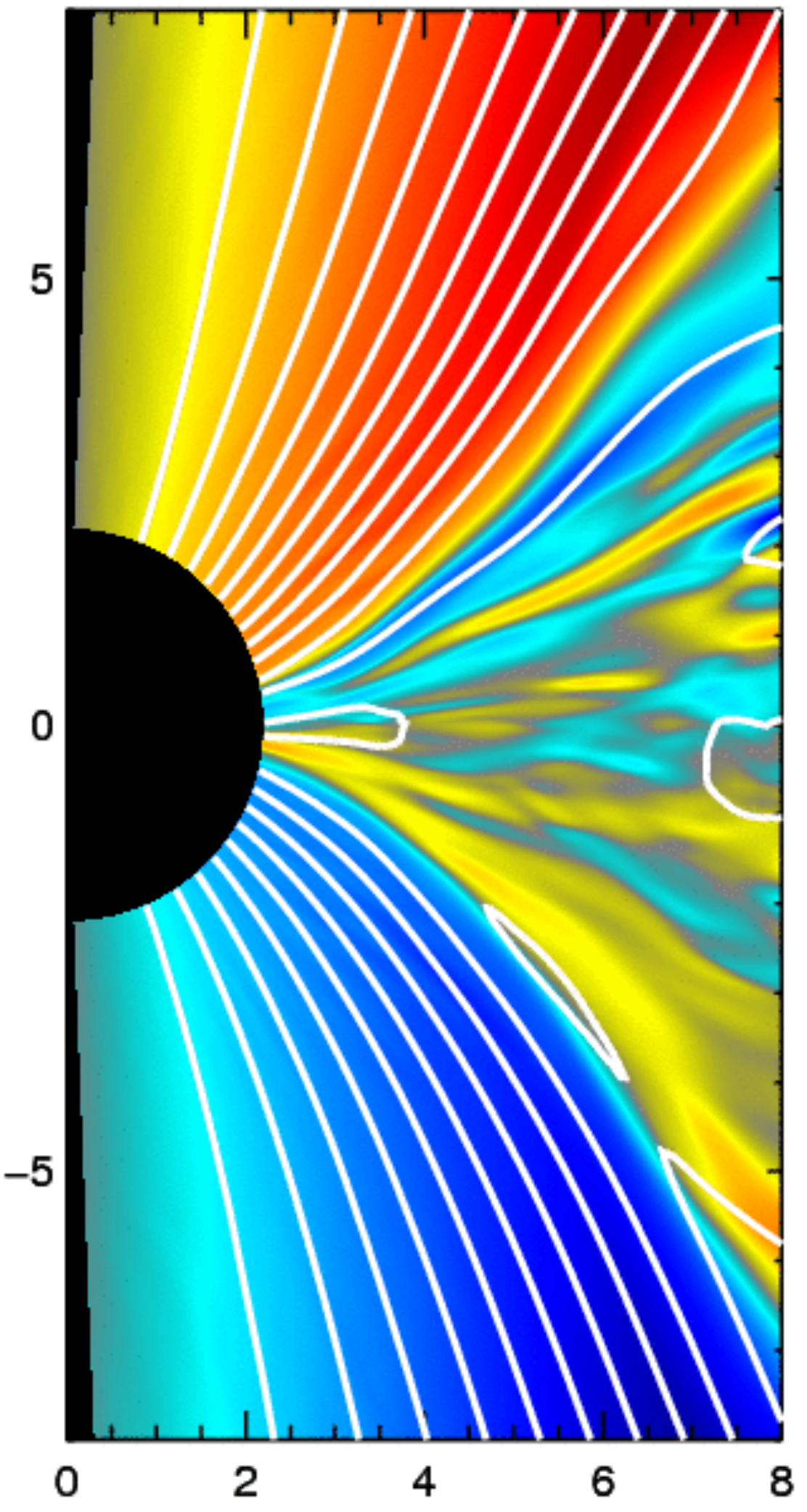}
\includegraphics[width=0.6\textwidth]{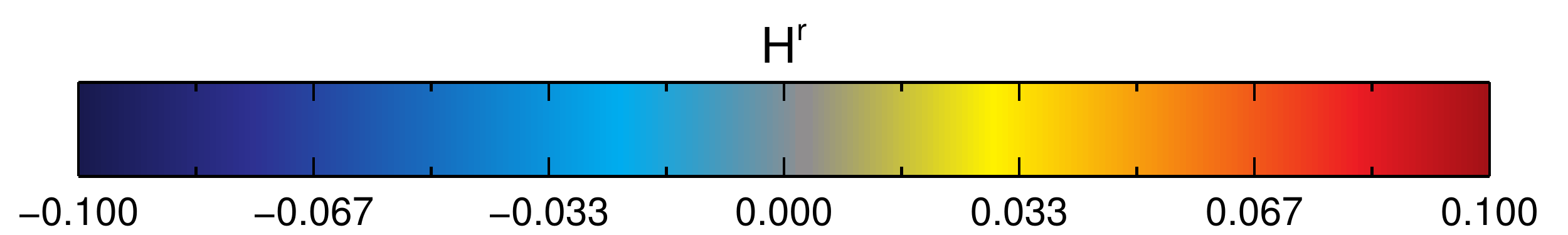}
\end{center}
\caption[]{Reconnection event that occurs in the disk and leads to the rapid
transport of poloidal magnetic flux from the outer disk to the black hole. 
Color contours show ${\cal B}^r$ and are overlaid with the
poloidal field structure (white lines), which are kept fixed between the
panels. From left to right the panels show $t=14600$, $14620$, and $14640M$.}
\label{reconnect}
\end{figure}

\begin{figure}
\leavevmode
\begin{center}
\includegraphics[width=0.4\textwidth, viewport=40 30 360 620,clip]{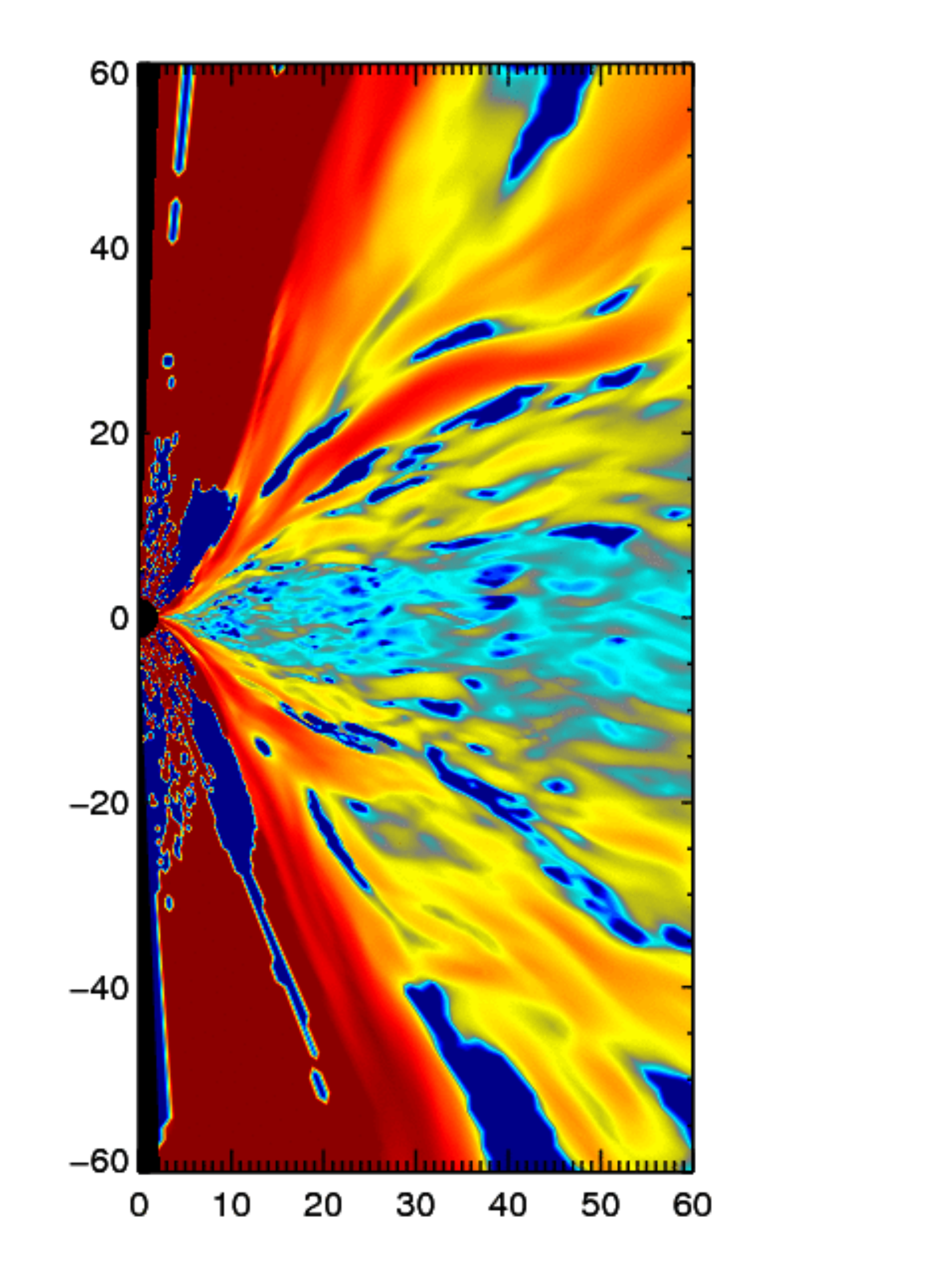}
\includegraphics[width=0.4\textwidth, viewport=40 30 360 620,clip]{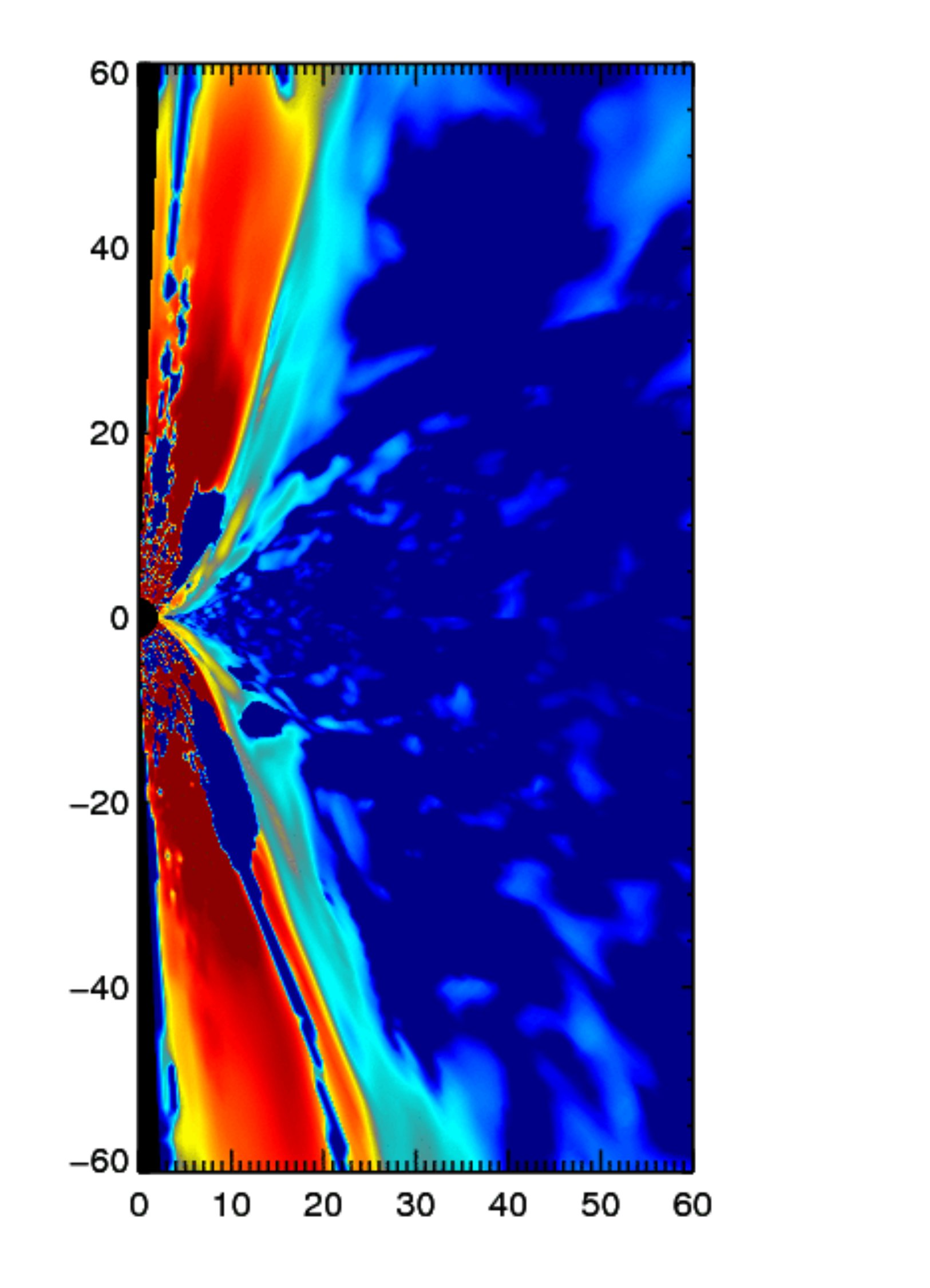}
\includegraphics[width=0.6\textwidth]{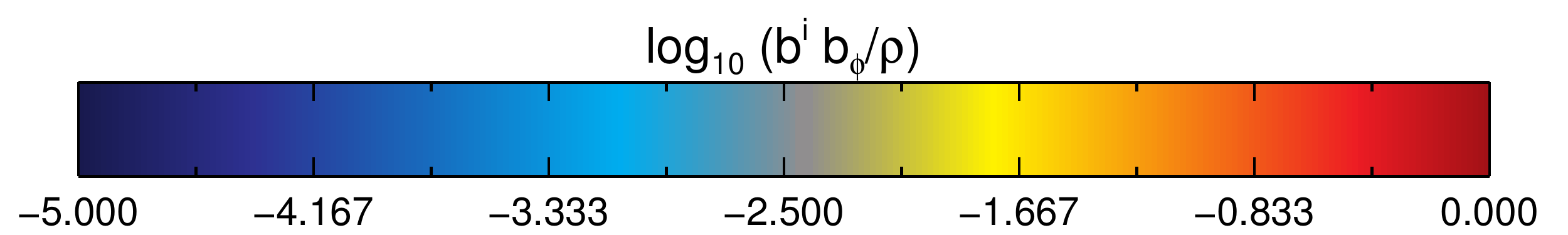}
\end{center}
\caption[]{Azimuthal average of $-b^r b_\phi/\rho$ (left panel) and
$-b^\theta b_\phi/\rho$ (right panel) at the same time as shown in
Fig.~\ref{flow_20000}. Regions with $-b^i b_\phi / \rho > 1$ are
colored dark red, regions with $-b^i b_\phi / \rho < 10^{-5}$
(including $-b^i b_\phi / \rho < 0$) are deep blue. The sign of
$-b^\theta b_\phi/\rho$ is adjusted so that positive values correspond
to electromagnetic angular momentum flux directed away from the disk.
The azimuthal average of $-b^\theta b_\phi$ is much smaller than
that of $-b^r b_\phi$ everywhere outside the funnel.}
\label{stresspermass}
\end{figure}

An example of a reconnection event within VD0m3d that leads to this sort
of global flux
rearrangement is shown in Figure~\ref{reconnect}. In this figure, color
contours denote the radial component of the CT-field, ${\cal B}^r$,
indicating the direction of the fieldline, which are again overlaid
with white contours derived from the azimuthally averaged $A_\phi
(r,\theta)$. Initially ($t=14600M$), fieldlines connect large radii to the
black hole event horizon through the funnel region (deep blue region below
midplane) and then proceed back out into the disk close to the midplane
(yellow region below midplane). These fieldlines cross the midplane at
large radius and proceed back to the black hole north of the midplane
(light blue region north of midplane), before finally proceeding out to
large radius through the funnel region (red region north of midplane). At
time $t=14620M$, the oppositely directly fieldlines that lie just above
and below the midplane have reconnected at $r\sim6M$, and the resulting
closed fieldloop (which is still attached to the black hole) is rapidly
accreted. This process transfers flux from the outer disk to $r\sim6M$
(where the reconnection event occurs) and then down to the black hole
as the field loop is accreted, arriving at the black hole with the outer
edge of the loop.
The coronal mechanism resembles the flux advection mechanism proposed
by \citet{Rothstein:2008} in the sense that most flux transport takes
place outside of the disk body, and that flux in the corona moves
inward because magnetic stresses send angular momentum off to infinity.
They emphasize the contrast between the turbulent disk and non-turbulent
flow outside of the disk.  We likewise observe that large-scale, relatively
laminar flows within the low-density corona easily move field about;
flux-freezing is an excellent approximation there.
On the other hand,
there are also several points of contrast between our results and the
model of \cite{Rothstein:2008}.  They argue that in the disk corona,
defined as the region where $\beta \sim 1$, the $T^z_\phi$ component of
the magnetic stress tensor could transport angular momentum outward as
effectively as the $T^R_\phi$ component to which attention is more often
directed, allowing the disk surface to move inward.  Although this is
similar to what we see in our simulation during the initial avalanche
phase, it is different from the dynamics prevailing at later times.
During the accretion phase of the simulation, the coronal field lines
develop MRI-like bends at high altitudes above the disk ($>>H$), so
that the inflow occurs in the corona rather than along the disk surface.
In addition, as shown by Figure~\ref{stresspermass}, which contrasts the
azimuthal averages of $-b^r b_\phi/\rho$ and $-b^\theta b_\phi/\rho$,
there is very little {\it net} angular momentum flux per unit mass in
the polar angle direction except in the funnel (at latitudes not too far
from the midplane, $b^\theta b_\phi \simeq b^z b_\phi$).  Although the
magnitude of $T^\theta_\phi$ is often comparable to $T^r_\phi$, it has
very nearly the same probability of having either sign, so that its mean
value is small.  On the funnel edge, $T^\theta_\phi$ is more likely to
take angular momentum away from the disk but has magnitude on average
somewhat smaller than $T^r_\phi$, while in the outer corona, on average
it brings angular momentum toward the midplane.

\subsection{Inflow in the disk body}
\label{disk_flux}

The majority of the net flux on the black hole is brought there by
the coronal mechanism during the initial $5000M$ of time.  However, the flux on
the horizon does increase over the last $1.5 \times 10^{4} M$ and, while
the coronal mechanism continues to operate during this interval, it is
possible that part of the flux might be brought in through the accretion
disk itself.  As the right panel of Figure~\ref{magfluxflow} makes
plain, the local magnetic inflow rate $V_\Phi$ is roughly two orders
of magnitude smaller in the midplane than it is at the funnel edge.
Nonetheless, its magnitude is still consistent with the total flux
inflow rate during the latter part of the simulation.  It may therefore
be possible for flux advection directly associated with the accretion
flow in the disk body to account for some part of the flux accumulation
during the quasi-steady accretion phase.  In this section we attempt to
estimate the size of that contribution.

\begin{figure}
\begin{center}
\includegraphics[width=0.5\textwidth,angle=90]{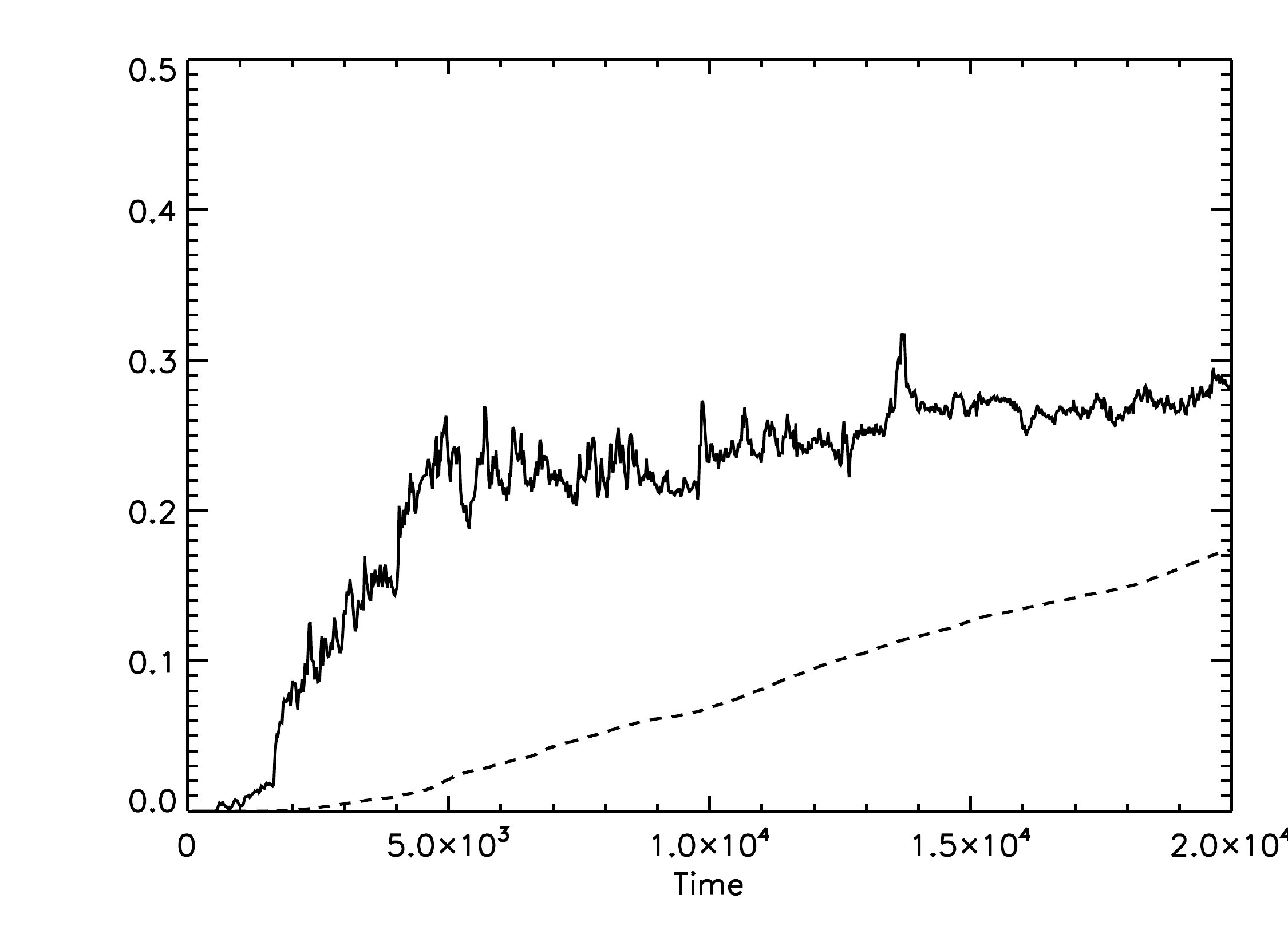}
\includegraphics[width=0.5\textwidth,angle=90]{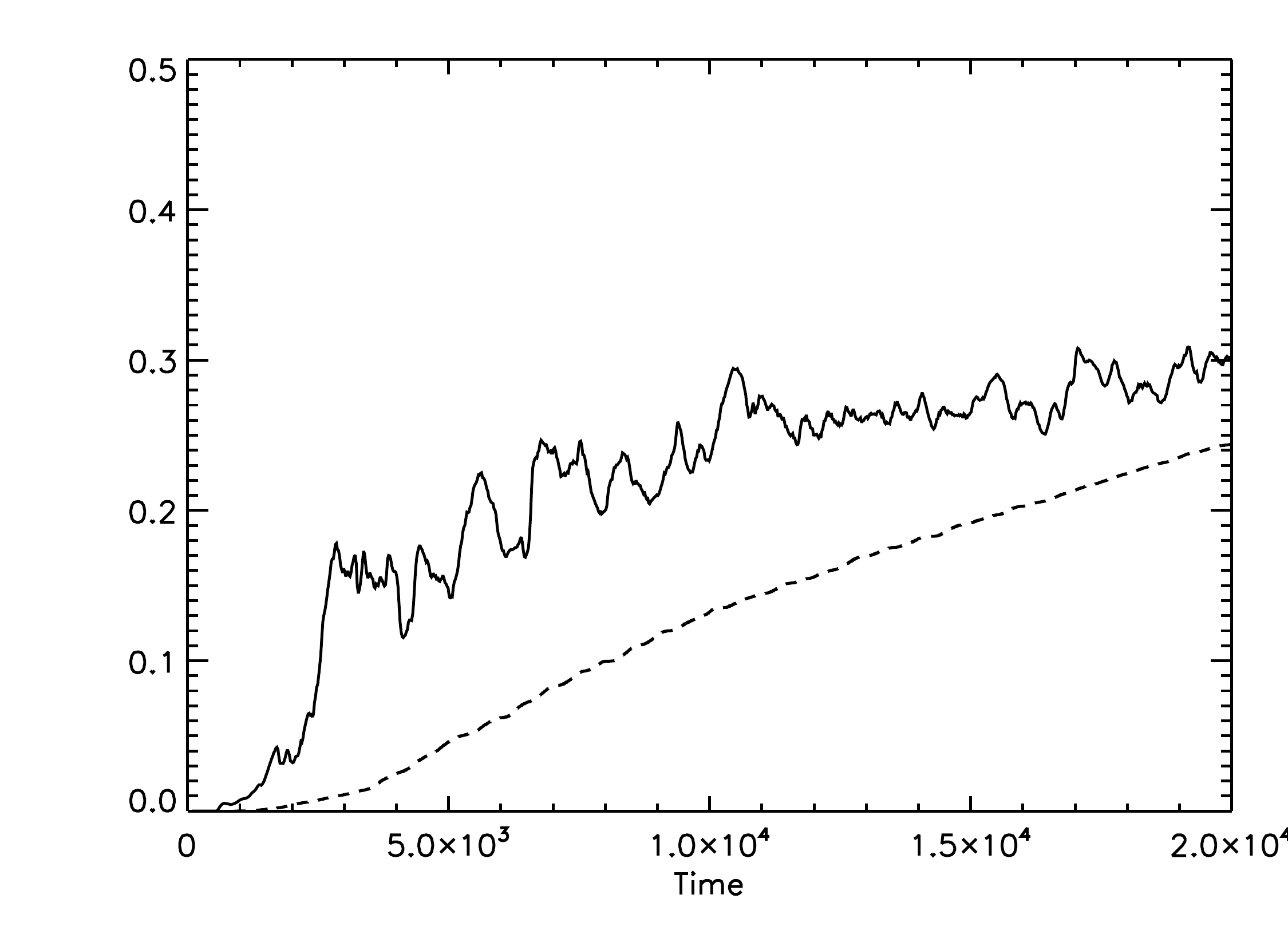}
\end{center}
\caption[]{Fractional accretion of magnetic flux $\cal A$ (solid curve) and 
mass $\cal M$ (dashed
curve) as functions of time.  Because we stored shell-integrated mass accretion
rates every $1M$ in time, but the full 3-d data required to compute the vector
potential only every $10M$, the latter curve has greater time resolution than
the former.  (Top) At $r=2.1M$.  (Bottom) At $r=20M$.}
\label{qinside} 
\end{figure}

We begin by comparing the rate of magnetic flux accretion to
the rate of mass accretion.
If magnetic flux and mass moved inward in lock-step, the functions
\begin{equation}
{\cal M}(r,t) \equiv \int_{0}^{t} \, dt^{\prime} \dot M(r,t^\prime)/M_0
\end{equation}
and
\begin{equation}
{\cal A}(r,t) \equiv A_\phi (r,t,\pi/2)/A_\phi (r_{\rm max},0,\pi/2)
\end{equation}
would be identical.  These quantities are normalized to the initial
mass, $M_0$, and the initial flux; $r_{\rm max}$ is the outer radius of the
initial torus.  In Figure~\ref{qinside}, we show the time-dependence
of both these quantities at two fiducial radii, just outside the event
horizon (left panel) and at $r=20M$ (right panel).  The former, of
course, tells about the flux attached directly to the black hole and
the mass deposited there.  The difference between the two plots
is the mass and flux within the accretion disk between the horizon and
$r=20M$; here we will refer to this region as the ``inner disk.''  

In evaluating Figure~\ref{qinside}, one must allow for an important
distinction between the histories of mass and magnetic flux: the mass
of the black hole {\it must} increase monotonically (and the mass of
the inner disk, although not required to do so, generally does), but
the net magnetic flux can increase or decrease, 
both by magnetic reconnection and by
the radial motion of closed loops.  Thus, the inflow of magnetic flux
is highly intermittent, and the time-derivative of flux within some
radius can easily be negative.  On short timescales, the ratio of the
rate of flux accretion to mass accretion can vary by at least an order
of magnitude, as well as fluctuate in sign.  For this reason, a sensible
comparison of the rate of magnetic flux inflow relative to that of mass
can be made only in regard to long-term time averages.  In addition,
in order to evaluate the efficiency of this process in real disks, we
must restrict these averages to times when the inner disk was in
approximate inflow equilibrium: for this simulation, that means the
latter half of the data, from $t=10^4 M$ to $t=2 \times 10^4 M$.
Within this span, we smooth ${\cal A}(t)$ by taking
a running average $1000M$ wide and then compose the ratio of the change
in ${\cal A}$ to ${\cal M}$ from $10^4 M$ to $2\times 10^4 M$ using the
smoothed data.  We find that $\Delta{\cal A}/\Delta{\cal M}\simeq 0.47$
at the horizon, and $\simeq 0.50$ at $r=20M$.  In other words, relative
to the initial amount embedded in the accretion flow, magnetic flux
moves inward roughly half as fast as mass, and these rates are the same
at both the horizon and $r=20M$.  If the mass is in a state of inflow
equilibrium, the equality of these two ratios shows that the magnetic flux in
the inner disk is also in equilibrium.

This ratio of inflow rates may be contrasted with the local ratio of
magnetic flux to mass.  Looking beyond the short-timescale fluctuations
in the plots of ${\cal A}(t)$ in Figure~\ref{qinside}, one sees that
the average values of the flux at the horizon and the flux contained
within $r=20M$ are very similar.  In other words, the vertical magnetic
flux through the inner disk, $\Phi(r=20M)$, contributes very little
to $\cal{A}$.  Time-averaging over the last $10^4 M$ of data, we find
that it is only $\simeq 0.86\%$ of the initial flux, and it appears
that what flux there is is confined to the outer part of the inner disk.
By contrast, averaged over the same period, $5\%$ of the initial mass can
be found in the inner disk.  Thus, in the inner disk, the time-averaged
flux/mass ratio is only $17\%$ of the initial value.

To reconcile this flux/mass ratio with the accreted flux/mass ratio,
there are only two possibilities: either the magnetic flux moves in at
the same rate (or slower) than the mass and (at least) $2/3$ of the flux
reaching the horizon between $10^{4} M$ and $2\times 10^{4} M$ arrived
there via a different route (most likely the coronal mechanism), or the
magnetic flux on average moved inward three times faster than the mass.
We view the former alternative as much more plausible.  Note, however,
any flux delivery from outside the inner disk must be balanced by flux
losses from the inner disk to the horizon
because the magnetic inflow rate at $r=20M$ is nearly
the same as at the horizon.

\begin{figure}
\begin{center}
\includegraphics[width=0.6\textwidth,angle=90]{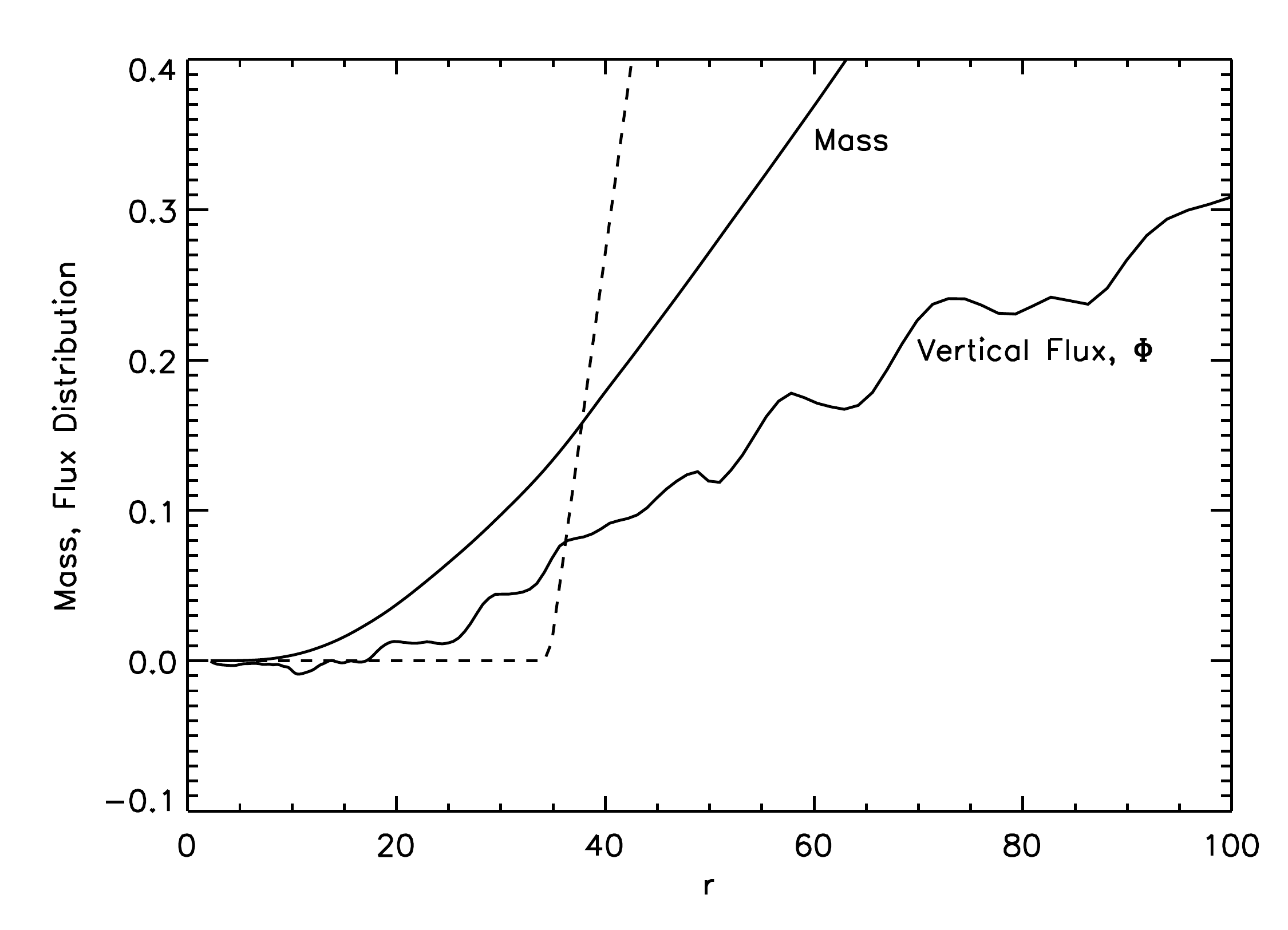}
\end{center}
\caption[]{The net vertical flux through the equator $\Phi(r)$,
integrated from the black hole horizon outward, averaged over the last
$5000M$ of the simulation and normalized to the inital total value
(``Vertical Flux, $\Phi$'').  The initial $\Phi$ distribution is also shown
(dashed line).  The net flux has spread through the disk over the course
of the simulation.  For comparison, the mass distribution ${\cal} M
(r)$ is shown (labeled ``Mass'') for the end time, $t=2\times 10^4~M$,
likewise normalized to its initial total value.  }
\label{distribution} 
\end{figure}

Next we further explore the nature of the magnetic flux equilibrium
in the inner disk, using two different views of this quantity.
Figure~\ref{distribution} shows the distribution of the vertical
flux function $\Phi(r)$ time-averaged over the last $5000M$ of the
simulation and normalized to the initial flux value.  For comparison,
we also plot in that figure both the initial flux distribution and the
final mass distribution in the disk, both likewise normalized to the
initial total.  The radial derivative $d\Phi/dr$ shows the magnitude and
sign of the vertical magnetic field piercing the equator.  We see that,
like the mass, vertical flux has spread away from its initial location,
both in and out, but, as already mentioned, a much smaller part of the
net magnetic flux than the mass resides in the inner disk.

\begin{figure}
\begin{center}
\includegraphics[width=0.6\textwidth,angle=90]{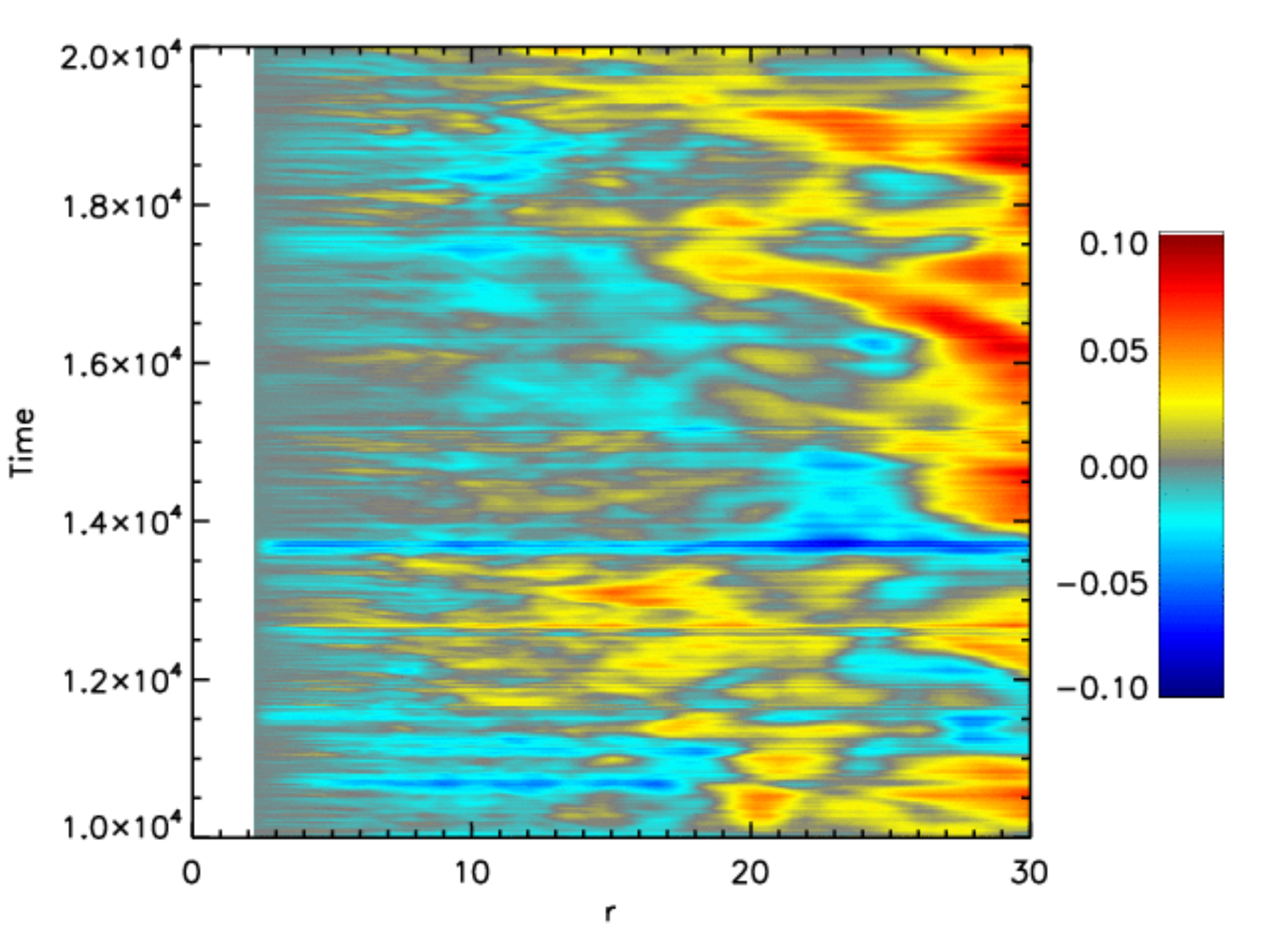}
\end{center}
\caption[]{Space-time contours of the net vertical flux through the
equator $\Phi(r)$, integrated from the black hole horizon outward for
the inner disk during the second half of the simulation, normalized to
the total initial flux.  Never very large in magnitude, the integrated
magnetic flux in this region fluctuates in sign.}
\label{stflux} 
\end{figure}

The time-dependence of the magnetic flux in the inner disk over the
last $10^4 M$ in time is illustrated
in Figure~\ref{stflux}.  As this figure shows, not only is the magnetic
flux contained within the inner disk a rather small part of the total,
it fluctuates in time, frequently going negative.  Little sign of any
long-term trend can be seen, consistent with our contention that the inner
disk magnetic flux has reached a statistical equilibrium.  The frequent
sign changes of the local net flux suggest that the poloidal magnetic
field lines in the inner disk predominantly close within the inner disk.

\begin{figure}
\begin{center}
\includegraphics[width=0.6\textwidth,angle=90]{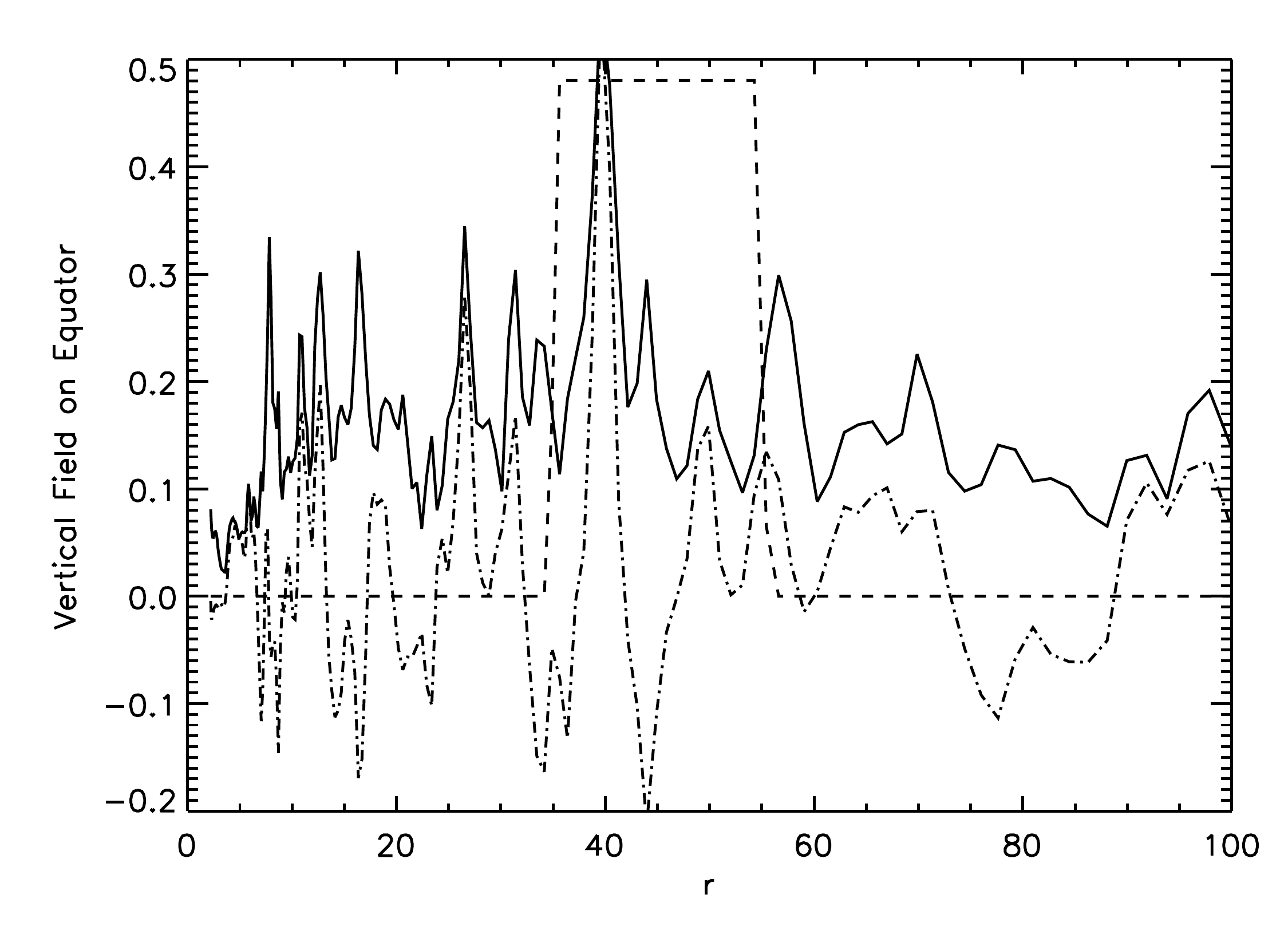}
\end{center}
\caption[]{Azimuthal average of the absolute value of $|{\cal B}^\theta
|$ (solid line) and the average of ${\cal B}^\theta$ (dot-dashed line)
along the equatorial plane at $t=2\times 10^4 ~M$ (solid line) along with
the initial ${\cal B}^\theta$ distribution (dashed line).  The vertical
field remains relatively strong but experiences fluctuations in sign
throughout the disk body.  The {\it net} flux is a small positive bias
in the vertical field value.  }
\label{avgbz} 
\end{figure}

Moreover, the magnetic field corresponding to the net flux is a very small
fraction of the {\it total} field within the disk.  Figure~\ref{avgbz}
shows the azimuthally averaged value of ${\cal B}^\theta$ through the
equator at the end time as a function of radius, along with the initial
value.  The MRI has generated considerable field of both signs throughout
the disk.  Local regions of the disk have a net vertical flux, but their
contrasting signs lead to little contribution to the total disk flux.
The global net flux is present only as a slight positive excess moving
both inward and outward with time.

To reach the state described by these figures, most of the large-scale
magnetic connections to the disk matter must have been reconnected away.
But this should come as no surprise because we have already seen that
a significant part of the original flux has moved {\it by reconnection}
from the disk to the horizon.  The matter in the inner disk has little net
flux because it has been largely bypassed through the coronal mechanism.
Thus, while the MRI has created a large effective turbulent viscosity
in the sense that considerable mass has accreted from the location of
the initial torus, very little net flux has moved with it.  In other
words, within the disk, turbulence created by the MRI transports angular
momentum rapidly and drives accretion, but very little flux moves with
the accreting mass.

\subsection{Strength of the Funnel Field}

\begin{figure}
\leavevmode
\begin{center}
\includegraphics[width=0.6\textwidth,angle=90]{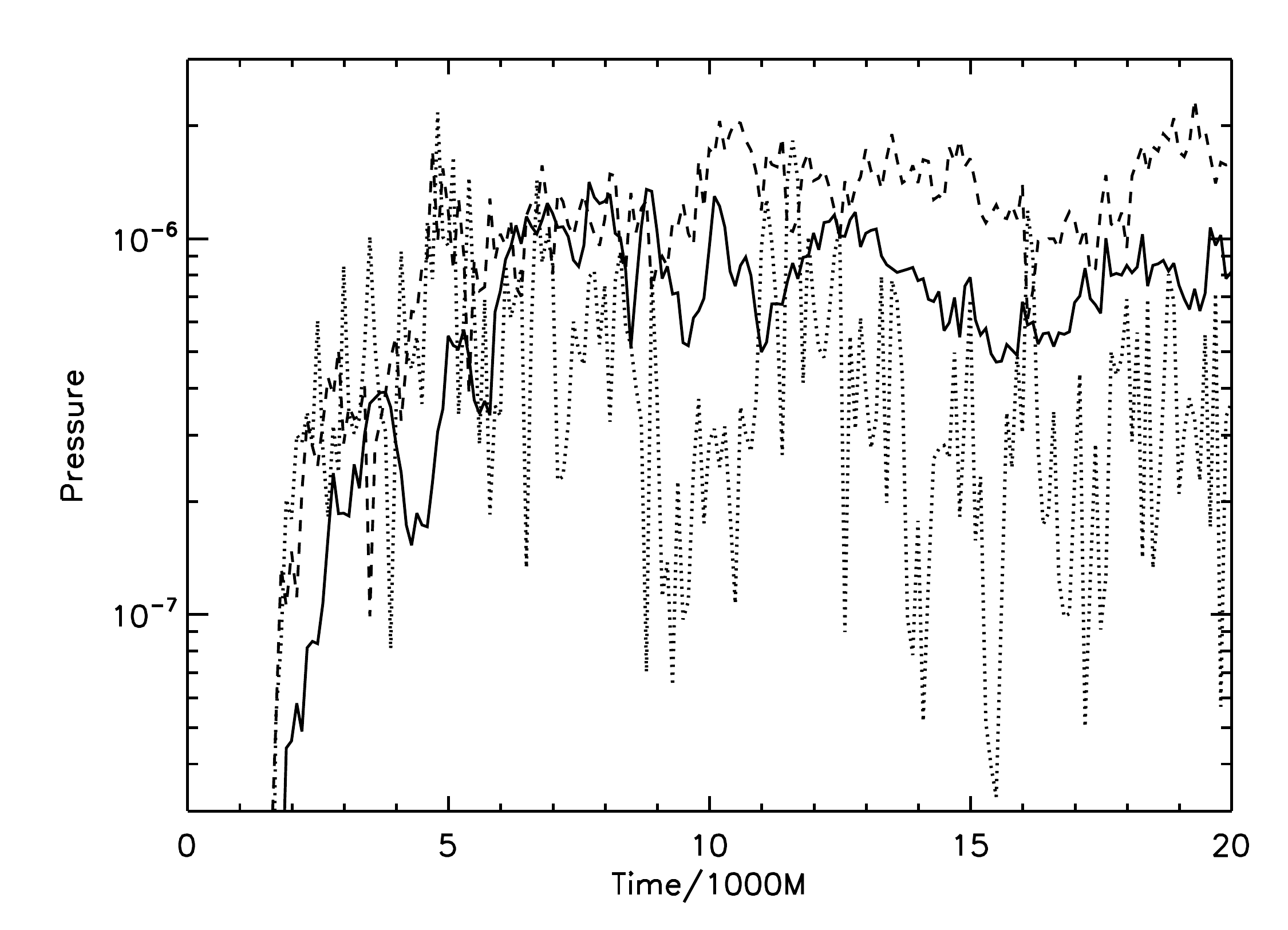}
\end{center}
\caption[]{Azimuthally-averaged magnetic pressure as a function of time
at two locations:
inside the funnel near its base ($r=6M$, $\theta = 0.044\pi$; solid curve);
and in the inner accretion flow ($r=6M$, $\theta=\pi/2$; dotted curve),
and azimuthally-averaged gas pressure in the inner accretion flow (also
$r=6M$, $\theta=\pi/2$; dashed curve).
}
\label{magsaturate}
\end{figure}

Lastly, we raise the question of whether in the very long-term the
magnetic flux on the horizon would continue to grow.  If it is regulated
by a combination of the magnetic and gas pressure near $r=6 M$, and
these are in an approximately time-steady state, one might expect
that further flux accumulation on the horizon would have to stop.
Figure~\ref{magsaturate} supports this idea.  This figure shows that
the magnetic pressure inside the funnel, the magnetic pressure in the
inner accretion flow, and the gas pressure in the inner accretion flow
are well correlated with one another in a temporal sense, even while
all three change by more than two orders of magnitude over the course
of the simulation.  The time-dependent data show fluctuations that are
so large, however, that the much slower accumulation of flux seen in
Figure~\ref{qinside} is well below the noise, so our data on magnetic
and gas pressures cannot answer any questions about long-term saturation
of magnetic flux attached to the horizon.

That the magnetic and gas pressures should be closely related is an old
idea \citep{Rees:1982,MacDonald:1982}, but there is also a long history
of controversy about whether the magnetic field pressure in the funnel
should be closer in magnitude to the (poloidal) magnetic pressure or
the gas pressure in the inner disk. For example, \cite{Ghosh:1997} and
\cite{Livio:1999} argued that as the poloidal magnetic field strength in
the disk is subthermal, the funnel field strength should be regulated by
it rather than the gas pressure.  In the simulation presented here, the
{\it total}, i.e. poloidal plus toroidal magnetic pressure in the disk
is smaller than the gas pressure
for the duration of the simulation (i.e. the field remains subthermal),
and the magnetic pressure from the field in the funnel lies between
these two candidate regulators, sometimes closer to one and sometimes
closer to the other.  Thus, the funnel field is consistently
stronger than the poloidal field in the disk.

The relevance of the inner disk poloidal field to the strength of the
funnel field seems to be rather limited based upon what we observe
about the process carrying flux to the horizon.  The field in the
funnel is nearly radial, and its total intensity is determined by the
poloidal flux that is delivered to the black hole over the course of
the simulation.  This flux delivery system is the coronal mechanism,
and so it should not be surprising that the poloidal flux within the
accretion disk itself plays at most a secondary role (e.g., by determining
the rate of reconnection and accretion of flux loops).  Instead, simple
pressure-matching in the inner regions of the accretion flow---be it gas,
(total, mostly toroidal) magnetic or radiation pressure---appears to be
important in regulating how much flux may be attached to the black hole.
We speculate that when the flux on the horizon has a total field intensity
larger than would be consistent with the gas and magnetic pressure in
the nearby accretion flow, the rate of reconnection along the funnel
field boundary rises, so that the field approaching the black hole is
entirely in closed loops and does not add to the net flux on the horizon.

\section{Summary, Discussion \& Conclusions}\label{conclusion}

In this paper we investigate the evolution of an accretion torus embedded
in a large-scale vertical magnetic field orbiting a Schwarzschild black
hole, with a view toward studying how magnetic flux moves relative to the
accretion flow.  The simulation stretched over $2.0\times10^4 M$ in time,
corresponding to $80 \Omega^{-1}$ at the initial torus pressure maximum, long
enough to establish inflow equilibrium in the inner disk for the second
half of the simulation.  Of particular interest is how the net vertical
field evolves, and whether or not a field distribution consistent with
the formation of jets or winds can develop.  We trace the evolution of
the net poloidal flux distribution with a particular focus on net flux
that becomes attached to the black hole.  Our primary result is that a
significant fraction of the initial flux---$\simeq 27\%$---is brought
to the black hole horizon, even though only a rather smaller fraction
of the initial mass---$\simeq 19\%$ is accreted.  The flux attached to
the horizon supports a coherent poloidal field within the evacuated axial
funnel, a requirement for the creation of a Blandford-Znajek type jet
(if the black hole rotates, which it did not in this simulation).
The mass and flux are carried inward through distinct mechanisms:
the mass by turbulent stresses within the accretion disk, and the flux
by large-scale motions in the low density corona.

Most of the global magnetic flux motion is mediated by a novel
mechanism that we have dubbed the coronal mechanism.  Rather than the
gradual ``diffusive" process that had been the focus of most previous
discussions of magnetic flux motion in accretion flows, in this mechanism
the flux is brought directly down to the horizon as the net flux jumps
discontinuously when large-scale magnetic loops reconnect across the
equator.  The same orbital shear that creates angular momentum transport
in the disk body by correlating radial and azimuthal magnetic field also
acts in the corona; the difference is that in the corona the resulting
stress leads to the formation of large-scale loops stretched rapidly
inward rather than to MRI-driven turbulence.  These stretched loops can
reconnect to oppositely-directed field at much smaller radius than their
footpoint, leading to sudden macroscopic radial changes in the location of
magnetic flux.  Although the coronal mechanism acts particularly rapidly
during the initial transient phase of the simulation, it continues
to dominate flux motion throughout the simulation, including the long
period during which the inner disk is in approximate inflow equilibrium.
We might therefore reasonably expect it to be a property of actual
accretion disks.

The same reconnection that creates global relocation of net flux
simultaneously creates closed field loops within the disk body.  For this
reason, the flux/mass ratio within the accretion flow is suppressed by
a factor of order unity relative to what it is in the initial state.
Earlier global disk simulations have, in most cases, assumed zero
net flux for reasons of simplicity and computational convenience; the
coronal mechanism might, in fact, make this assumption a reasonable
approximation to the magnetic field in typical disks.  The suppression
of flux/mass in the disk body also makes conventional advection of flux
in direction association with mass accretion relatively inefficient.

As seen in previous simulations \citep[particularly][]{Beckwith:2008a},
if the overall field topology at the black hole is dipolar, the funnel
field can be relatively immune to reconnection
and hence long-lived.  We conjecture
that the funnel field amplitude is determined by pressure balance with
the gas and (total, mostly toroidal) magnetic pressures in the inner
disk and coronal regions.
Future experiments could test this hypothesis by running much longer
simulations with more available net field to see if the field strength
levels off or continues to build.

As with any simulation, our results are subject to some uncertainty
due to the limitations inherent in a numerical solution.  One concern
is the importance of reconnection to the overall flux evolution.  In
our numerical simulation, reconnection occurs at the grid scale when
oppositely signed field components are brought together.  Although
leaving the microscopic rate of such an important process to numerical
effects is a concern, the overall motions that lead to the formation of
current sheets and subsequent reconnection area driven by events
on much larger scales.  In this sense, we might argue that the rate
of reconnection seen in the simulation is relatively independent of
our gridscale.  We have not, however,
demonstrated that our results are numerically converged, although the
global field motions in the axisymmetric simulations carried out at two
different resolutions were very similar.  Furthermore, the thermodynamics
of these simulations are incomplete because energy lost to numerical
magnetic reconnection and numerical cancellation of fluid velocities is
not captured; conversely, we do not account for radiation in any way,
either as a cooling agent or as a contributor to the pressure.  One thing
that is clear, regardless of any other numerical limitation, is
that axisymmetric simulations are of limited utility.  In addition to
the constraints associated with the anti-dynamo theorem, the vertical
field channel modes remain dominant throughout the simulation, giving
a qualitatively distinct evolution at late time.  Three dimensional
simulations are essential for studying long term, steady-state behavior.

To place our somewhat surprising results in context, it is worthwhile
contrasting them with previous suggestions about how magnetic flux moves
through accretion flows.  These are quite disparate, for they have in
general been based on approximate analytic arguments.  We examine three
principal approaches, of which one has received much more attention than
the others.

The greatest amount of attention has been given to a picture in
which the matter moves inward by an unspecified angular momentum
transport process (called ``viscosity", but not thought to be
literally that) while the magnetic field diffuses relative
to the matter through another mechanism called ``turbulent
resistivity", but similarly unattached to specific microphysics
\citep[e.g.,][]{van-Ballegooijen:1989,Lubow:1994,Heyvaerts:1996,Ogilvie:2001}.
The ratio between this effective viscosity and resistivity (the nominal
Prandtl number), would then determine whether the flux mostly moves
inward with the mass, or instead diffuses outward relative to the mass
fast enough to avoid much net inward motion.

This approach rests upon two little-examined assumptions: First, that
the overall flow can be regarded as if it were laminar and time-steady,
and second, that the behavior of the underlying turbulence can be reduced
to parameters (an advection rate and a diffusion, i.e., resistivity,
coefficient) whose values are independent of the magnetic field.
With these assumptions, field motion in the disk body governs field
motion far away from the disk, and one may describe the field evolution
in the language of linear diffusion.  Unfortunately, neither assumption
applies to real accretion flows.

It is now well-established that angular momentum transport in accretion
disks is due to turbulent magnetic stresses driven by the MRI.
Consequently, the magnetic field structure is neither laminar nor
time-steady.  Even in the corona, where the MRI {\it per se} does not
occur, the same basic physics leads (as we have earlier discussed) to
irregular field motions.  There is therefore no direct connection between
local field motion deep inside the disk (e.g., resistive diffusion)
and the position of the corresponding field lines far away.  Moreover,
as we have emphasized, flux is a {\it global} concept, not a local one;
the coronal mechanism, for example, could never be described by a local
theory of this sort.

The second assumption, that the evolution of the field can be described
in terms of simple resistive and viscous diffusion operating on a
background field gradient is similarly problematic.  Accretion is driven
by nonlinear magnetic stresses; because the effective inflow velocity
arises only from an average of a strongly fluctuating velocity field,
it is non-local in both time and space.  Similarly, the breakdown
in ideal MHD happens primarily by driven reconnection, the rate of
which depends on the structure and magnitude of both the magnetic and
velocity fields.  As a result, neither the mean rate at which magnetic
field is carried from place to place nor any tendency for its structure
to smooth is either independent of or linear in the field strength.

In fact, the formulation in terms of a competition between ``viscous"
and ``resistive" diffusion breaks down in an even more fundamental way.
Where there is net flux passing through the disk body,
the very same MHD turbulence that transports angular momentum simultaneously
decouples the flux from the matter through turbulence-driven reconnection.
Field lines associated with flux passing through the equator develop bends
that readily break off as closed loops. The matter moves inward with the
closed loops while the flux stays in place.  Thus, in dramatic contrast
to the conventional prediction, rapid ``viscous'' accretion
can, and apparently does, co-exist with largely stationary flux, while
the ``resistivity" that permits this decoupling has no particular impact
on smoothing the field distribution.

Moreover, there is relatively little net flux passing through the disk
because reconnection associated with the coronal mechanism detaches it
efficiently.  In other words, most of the flux motion occurs in an ``end-run"
that bypasses the disk altogether, so that it has rather little to do
with any activity in the disk body, ``viscous", ``resistive", or other.

Local smoothing of field structures can occur, but it is questionable
whether one can define an overall diffusion coefficient to describe its
rate.  For example, \cite{Guan:2009} have recently attempted to quantify
magnetic diffusion in the context of MRI-driven MHD turbulence using
shearing box simulations.  They imposed a sinusoidally varying vertical
field with an amplitude above the background due to the MRI-turbulent
flow and observed the subsequent decrease in this mode's Fourier power.
The peak vertical field amplitudes studied vary from 10--80\% of the
initial toroidal field strength.  They found that the decay time for the
imposed feature is several tens of $\Omega^{-1}$, with the decay rate
an increasing function of the perturbation amplitude.  These results
indicate that local field gradients can be smoothed within the turbulent
flow, but it is unclear to what degree the inferred average diffusion
coefficient depends on the specific situation studied.  In particular,
in accretion disks, the {\it net} field, the part determining the flux,
is, especially in the disk body, small in magnitude compared to the
fluctuating turbulent field.  When the dynamics are strongly nonlinear,
spreading of structures in this small net field may proceed in a way
very different from spreading of larger amplitude structures; in fact,
the motion of the net field may have much more to do with the dynamics
of the fluctuating field than to its own disposition.

For all these reasons, therefore, we see no useful way to describe
the results of the present simulation in the traditional language of
arbitrarily specified transport coefficients applied to gradients of
the time-averaged magnetic field.

Two other concepts have recently been studied.  One of these
\citep{Rothstein:2008,Rothstein:2009} in some respects is a blend
of the ``advection/diffusion" picture and the coronal mechanism.
On the one hand, it uses the conventional formalism of fixed transport
coefficients; on the other hand, it relies heavily on field line motions
in the corona.  These two approaches are united by supposing that the
MRI and its resulting turbulence are suppressed in the corona, so that
it suffers neither ``turbulent viscosity" nor ``turbulent resistivity".
Consequently, coronal motion is dominated by {\it vertical} transport of
angular momentum, in a magneto-centrifugal wind when the effective Prandtl
number is large, and in a Poynting flux-dominated outflow otherwise.
The result is that (for most parameters), the upper surface of the disk
moves inward, carrying the footpoints of the large-scale flux with it,
while the flow in the midplane is {\it outward}.

As we have previously discussed, the fundamental physical element of the
MRI, the creation of substantial magnetic stresses by orbital shear,
does operate within the corona, and strongly---this is the foundation
of the coronal mechanism.  Despite these coronal motions,
this is not a significant source of angular momentum transport for
the main accretion disk.  Although the instantaneous amplitude of the
magnetic $T^\theta_\phi$ stress is typically sizable, outside the funnel
its sign fluctuates so that we see no significant net vertical transport
of angular momentum from the disk.  In addition, the net fluid radial
velocity in the midplane, although small compared to its {\it rms} value,
is inward.  Nonetheless, our results are consistent with their suggestion
that magnetic stresses carrying angular momentum outward through the
corona can be effective in moving field lines inward.  Because the mass
density in the corona is so low, removing a comparatively small amount
of angular momentum can lead to large-scale field motions.  Because
the subsequent motions are not dominated by turbulence, flux freezing
remains effective within the coronal plasma.

In the second of the two recent suggestions for how flux moves in
accretion flows, \cite{Spruit:2005} propose that the flux motion is
not controlled by a simple gradient, but rather by the dynamics of
intermittent field bundles.  These authors suggest that magnetized
patches could accrete relatively rapidly from angular momentum losses
through a wind.  Further, if such fields were strong enough to suppress
the MRI, then turbulence, and subsequently magnetic reconnection,
would be greatly reduced for those patches.  While such magnetized
patches would not themselves lead to a significant net magnetic flux
passing through the disk {\it per se}, they could accumulate at the
central black hole until the field strength itself suppressed further
accretion.  Such a scenario is similar to that of the magnetically arrested
accretion model of \cite{Narayan:2003}.  Strong nonaxisymmetric field
structures appear in a simulation of \cite{Igumenshchev:2008} where a
strong field is continually injected at the outer boundary, providing
some support for the \cite{Spruit:2005} concept.  In our simulation, we
do see nonaxisymmetric variations in the vertical flux throughout the
disk, but not in the form of local pockets of intense vertical flux,
nor do we see either any significant disk wind
or any interruption of the accretion flow due to ``magnetic arrest.''
Nonetheless, we agree with
\cite{Spruit:2005} in emphasizing the intermittency of the flow and the
potential importance of nonaxisymmetries.

The study carried out here is, of course, only a first step, and the
results presented here point to several avenues of investigation with
future numerical experiments.  For example, simulations that continue for
longer times and begin with net field at larger radii can better explore the
efficacy of the coronal mechanism.  In particular longer simulations with
more available net field are required to see if the field strength in
the funnel levels off or continues to build over time.  Simulations in
the Kerr metric with non-zero spin parameters could also study directly
how magnetic flux evolution and the coronal mechanism relate to jet
launching and collimation.  The requirements of three spatial dimensions
and increased spatial resolution will continue to make these
investigations challenging.

\acknowledgements{ This work was supported by NSF grant PHY-0205155 and
NASA grant NNX09AD14G
(JFH), and by NSF grant AST-0507455 (JHK). We thank Sean Matt, Jim
Pringle, Chris Reynolds, Charles Gammie and Scott Noble for useful
discussions.  We acknowledge Jean-Pierre De~Villiers for improvements to
the algorithms used in the GRMHD code.  The simulations described
were carried out on the Teragrid Ranger system at TACC, supported by 
the National Science Foundation.  }


\begin{thebibliography}{40}
\expandafter\ifx\csname natexlab\endcsname\relax\def\natexlab#1{#1}\fi

\bibitem[{{Balbus} \& {Hawley}(1991)}]{Balbus:1991}
{Balbus}, S.~A., \& {Hawley}, J.~F. 1991, \apj, 376, 214

\bibitem[{{Balbus} \& {Hawley}(1998)}]{Balbus:1998}
------. 1998, Reviews of Modern Physics, 70, 1

\bibitem[{{Beckwith} {et~al.}(2008{\natexlab{a}}){Beckwith}, {Hawley}, \&
  {Krolik}}]{Beckwith:2008a}
{Beckwith}, K., {Hawley}, J.~F., \& {Krolik}, J.~H. 2008{\natexlab{a}}, \apj,
  678, 1180, arXiv:0709.3833

\bibitem[{{Beckwith} {et~al.}(2008{\natexlab{b}}){Beckwith}, {Hawley}, \&
  {Krolik}}]{Beckwith:2008b}
------. 2008{\natexlab{b}}, \mnras, 390, 21, 0801.2974

\bibitem[{{Blandford} \& {Payne}(1982)}]{Blandford:1982}
{Blandford}, R.~D., \& {Payne}, D.~G. 1982, \mnras, 199, 883

\bibitem[{{Blandford} \& {Znajek}(1977)}]{Blandford:1977}
{Blandford}, R.~D., \& {Znajek}, R.~L. 1977, \mnras, 179, 433

\bibitem[{{De Villiers} \& {Hawley}(2003)}]{De-Villiers:2003}
{De Villiers}, J., \& {Hawley}, J.~F. 2003, \apj, 589, 458

\bibitem[{{De Villiers} {et~al.}(2003){De Villiers}, {Hawley}, \&
  {Krolik}}]{De-Villiers:2003b}
{De Villiers}, J., {Hawley}, J.~F., \& {Krolik}, J.~H. 2003, \apj, 599, 1238

\bibitem[{{De Villiers} {et~al.}(2005){De Villiers}, {Hawley}, {Krolik}, \&
  {Hirose}}]{De-Villiers:2005}
{De Villiers}, J., {Hawley}, J.~F., {Krolik}, J.~H., \& {Hirose}, S. 2005,
  \apj, 620, 878

\bibitem[{{De Villiers}(2006)}]{De-Villiers:2006}
{De Villiers}, J.-P. 2006, ArXiv Astrophysics e-prints, astro-ph,
  astro-ph/0605744

\bibitem[{{Evans} \& {Hawley}(1988)}]{Evans:1988}
{Evans}, C.~R., \& {Hawley}, J.~F. 1988, \apj, 332, 659

\bibitem[{{Ghosh} \& {Abramowicz}(1997)}]{Ghosh:1997}
{Ghosh}, P., \& {Abramowicz}, M.~A. 1997, \mnras, 292, 887

\bibitem[{{Guan} \& {Gammie}(2009)}]{Guan:2009}
{Guan}, X., \& {Gammie}, C.~F. 2009, ArXiv e-prints, 0903.3757

\bibitem[{{Hawley} \& {Krolik}(2006)}]{Hawley:2006}
{Hawley}, J.~F., \& {Krolik}, J.~H. 2006, \apj, 641, 103, astro-ph/0512227

\bibitem[{{Heyvaerts} {et~al.}(1996){Heyvaerts}, {Priest}, \&
  {Bardou}}]{Heyvaerts:1996}
{Heyvaerts}, J., {Priest}, E.~R., \& {Bardou}, A. 1996, \apj, 473, 403

\bibitem[{{Hirose} {et~al.}(2004){Hirose}, {Krolik}, {De Villiers}, \&
  {Hawley}}]{Hirose:2004}
{Hirose}, S., {Krolik}, J.~H., {De Villiers}, J., \& {Hawley}, J.~F. 2004,
  \apj, 606, 1083

\bibitem[{{Igumenshchev}(2008)}]{Igumenshchev:2008}
{Igumenshchev}, I.~V. 2008, \apj, 677, 317, arXiv:0711.4391

\bibitem[{{Koide} {et~al.}(1999){Koide}, {Shibata}, \& {Kudoh}}]{Koide:1999}
{Koide}, S., {Shibata}, K., \& {Kudoh}, T. 1999, \apj, 522, 727

\bibitem[{{Krolik} {et~al.}(2005){Krolik}, {Hawley}, \& {Hirose}}]{Krolik:2005}
{Krolik}, J.~H., {Hawley}, J.~F., \& {Hirose}, S. 2005, \apj, 622, 1008

\bibitem[{{Livio}(2000)}]{Livio:2000}
{Livio}, M. 2000, in American Institute of Physics Conference Series, Vol. 522,
  American Institute of Physics Conference Series, ed. S.~S. {Holt} \& W.~W.
  {Zhang}, 275--297

\bibitem[{{Livio} {et~al.}(1999){Livio}, {Ogilvie}, \& {Pringle}}]{Livio:1999}
{Livio}, M., {Ogilvie}, G.~I., \& {Pringle}, J.~E. 1999, \apj, 512, 100,
  arXiv:astro-ph/9809093

\bibitem[{{Lovelace} {et~al.}(2009){Lovelace}, {Rothstein}, \&
  {Bisnovatyi-Kogan}}]{Rothstein:2009}
{Lovelace}, R.~V.~E., {Rothstein}, D.~M., \& {Bisnovatyi-Kogan}, G.~S. 2009,
  ArXiv e-prints, 0906.0345

\bibitem[{{Lubow} {et~al.}(1994){Lubow}, {Papaloizou}, \&
  {Pringle}}]{Lubow:1994}
{Lubow}, S.~H., {Papaloizou}, J.~C.~B., \& {Pringle}, J.~E. 1994, \mnras, 267,
  235

\bibitem[{{MacDonald} \& {Thorne}(1982)}]{MacDonald:1982}
{MacDonald}, D., \& {Thorne}, K.~S. 1982, \mnras, 198, 345

\bibitem[{{Matsumoto} {et~al.}(1996){Matsumoto}, {Uchida}, {Hirose}, {Shibata},
  {Hayashi}, {Ferrari}, {Bodo}, \& {Norman}}]{Matsumoto:1996}
{Matsumoto}, R., {Uchida}, Y., {Hirose}, S., {Shibata}, K., {Hayashi}, M.~R.,
  {Ferrari}, A., {Bodo}, G., \& {Norman}, C. 1996, \apj, 461, 115

\bibitem[{{McKinney} \& {Blandford}(2009)}]{McKinney:2009}
{McKinney}, J.~C., \& {Blandford}, R.~D. 2009, \mnras, 394, L126, 0812.1060

\bibitem[{{McKinney} \& {Gammie}(2004)}]{McKinney:2004}
{McKinney}, J.~C., \& {Gammie}, C.~F. 2004, \apj, 611, 977, astro-ph/0404512

\bibitem[{{Narayan} {et~al.}(2003){Narayan}, {Igumenshchev}, \&
  {Abramowicz}}]{Narayan:2003}
{Narayan}, R., {Igumenshchev}, I.~V., \& {Abramowicz}, M.~A. 2003, \pasj, 55,
  L69, arXiv:astro-ph/0305029

\bibitem[{{Noble} {et~al.}(2008){Noble}, {Krolik}, \& {Hawley}}]{Noble:2008}
{Noble}, S.~C., {Krolik}, J.~H., \& {Hawley}, J.~F. 2008, ArXiv e-prints,
  0808.3140

\bibitem[{{Ogilvie} \& {Livio}(2001)}]{Ogilvie:2001}
{Ogilvie}, G.~I., \& {Livio}, M. 2001, \apj, 553, 158, arXiv:astro-ph/0007474

\bibitem[{{Pudritz} {et~al.}(2007){Pudritz}, {Ouyed}, {Fendt}, \&
  {Brandenburg}}]{Pudritz:2007}
{Pudritz}, R.~E., {Ouyed}, R., {Fendt}, C., \& {Brandenburg}, A. 2007, in
  Protostars and Planets V, ed. B.~{Reipurth}, D.~{Jewitt}, \& K.~{Keil},
  277--294

\bibitem[{{Rees} {et~al.}(1982){Rees}, {Phinney}, {Begelman}, \&
  {Blandford}}]{Rees:1982}
{Rees}, M.~J., {Phinney}, E.~S., {Begelman}, M.~C., \& {Blandford}, R.~D. 1982,
  \nat, 295, 17

\bibitem[{{Reynolds} {et~al.}(2006){Reynolds}, {Garofalo}, \&
  {Begelman}}]{Reynolds:2006}
{Reynolds}, C.~S., {Garofalo}, D., \& {Begelman}, M.~C. 2006, \apj, 651, 1023,
  arXiv:astro-ph/0607381

\bibitem[{{Rothstein} \& {Lovelace}(2008)}]{Rothstein:2008}
{Rothstein}, D.~M., \& {Lovelace}, R.~V.~E. 2008, \apj, 677, 1221,
  arXiv:0801.2158

\bibitem[{{Schnittman} {et~al.}(2006){Schnittman}, {Krolik}, \&
  {Hawley}}]{Schnittman:2006}
{Schnittman}, J.~D., {Krolik}, J.~H., \& {Hawley}, J.~F. 2006, \apj, 651, 1031,
  arXiv:astro-ph/0606615

\bibitem[{{Spruit} \& {Uzdensky}(2005)}]{Spruit:2005}
{Spruit}, H.~C., \& {Uzdensky}, D.~A. 2005, \apj, 629, 960,
  arXiv:astro-ph/0504429

\bibitem[{{Thorne} {et~al.}(1986){Thorne}, {Price}, \&
  {MacDonald}}]{Thorne:1986}
{Thorne}, K.~S., {Price}, R.~H., \& {MacDonald}, D.~A. 1986, {Black holes: The
  membrane paradigm} (Black Holes: The Membrane Paradigm)

\bibitem[{{Tout} \& {Pringle}(1996)}]{Tout:1996}
{Tout}, C.~A., \& {Pringle}, J.~E. 1996, \mnras, 281, 219

\bibitem[{{Uzdensky} \& {Goodman}(2008)}]{Uzdensky:2008}
{Uzdensky}, D.~A., \& {Goodman}, J. 2008, \apj, 682, 608, 0803.0337

\bibitem[{{van Ballegooijen}(1989)}]{van-Ballegooijen:1989}
{van Ballegooijen}, A.~A. 1989, in Astrophysics and Space Science Library, Vol.
  156, Accretion Disks and Magnetic Fields in Astrophysics, ed. G.~{Belvedere},
  99--106

\end{thebibliography}

\newpage

\end{document}